\documentclass[a4paper,11pt]{article}
\pdfoutput=1 
\usepackage[table]{xcolor}
\usepackage{jheppub} 
\usepackage[export]{adjustbox}
\usepackage[utf8]{inputenc}
\usepackage[compat=1.1.0]{tikz-feynman}
\usepackage{multirow,array}
\usepackage{soul}
\usepackage{slashed}
\usepackage{bm}
\usepackage{fancyvrb}
\usepackage{fvextra}
\usepackage{comment}
\usepackage{rotating}
\usepackage{tikz}
\usepackage{tkz-euclide}
\usepackage{booktabs}
\usepackage{epsfig,amsfonts,mathrsfs,amsmath,amssymb,graphicx,color,slashed,multirow}
\usepackage{amsmath,latexsym,amssymb,graphicx,slashed,hyperref,color,enumerate,url,cancel,gensymb}
\usepackage{textcomp}
\usepackage{physics}
\usepackage{enumitem}
\usepackage{textcomp}
\usepackage{subcaption}
\usepackage{multirow}
\usepackage{bbm}
\usepackage[normalem]{ulem}

\newcommand{\be}{\begin{equation}}
	\newcommand{\ee}{\end{equation}}
\newcommand{\bea}{\begin{equation} \begin{aligned}}
	\newcommand{\eea}{\end{aligned} \end{equation}}	
	
\newcommand{\beq}{\begin{equation}}
\newcommand{\eeq}{\end{equation}}
\renewcommand{\O}{\mathcal{O}}
\newcommand{\R}{\mathcal{R}}

\newcommand{\nn}{\nonumber}
\def\equationautorefname~#1\null{Eq.~(#1)\null}
\def\sectionautorefname~#1\null{Sec.\,#1\null}
\def\subsectionautorefname~#1\null{Sec.\,#1\null}
\def\figureautorefname~#1\null{Fig.~#1\null}
\def\appendixautorefname~#1\null{App.\,#1\null}

\newcommand{\del}{\partial}
\newcommand{\La}{\mathscr{L}}
\newcommand{\Amp}{\mathcal{A}}	
	
\newcommand{\B}{\mathcal{B}}
\newcommand{\W}{\mathcal{W}}

\newcommand{\A}{\mathcal{A}}

\newcommand{\s}{\mathcal{S}}
\newcommand{\Q}{\mathcal{Q}}
\newcommand{\e}{\mathfrak{e}}
\newcommand{\h}{\mathfrak{h}}

\newcommand{\ba}{\begin{array}}
\newcommand{\ea}{\end{array}}
\def\bV{\begin{Verbatim}[breaklines=true, breakanywhere=true]}
\def\eV{\end{Verbatim}}

\def \ie{{\it i.e. }}

\def \muH{\mu_\phi}
\def \VV{VV}

\newcommand{\al}[1]{\begin{align}\begin{aligned} #1 \end{aligned}\end{align}}

\renewcommand{\h}{k}

\renewcommand{\order}[1]{O\!\left(#1\right)}

\title{
A log story\hspace{0.1em}\scalebox{0.01}{\textcolor{white}{(not}}\hspace{0.1em}\scalebox{0.01}{\textcolor{white}{so)}}\hspace{0.1em}short: running contributions to radiative Higgs decays in the SMEFT
}

\author[a,b,c]{Christophe~Grojean,}
\author[a]{Guilherme~Guedes,}
\author[a,b]{Jasper~Roosmale~Nepveu,}
\author[a,d]{Gabriel~M.~Salla}

\affiliation[a]{Deutsches Elektronen-Synchrotron DESY, Notkestr. 85, 22607 Hamburg, Germany}
\affiliation[b]{Institut für Physik, Humboldt-Universität zu Berlin, 12489 Berlin, Germany}
\affiliation[c]{Theoretical Physics Department, CERN, 1211 Geneva 23, Switzerland}
\affiliation[d]{Departamento de F\'{i}sica Matem\'{a}tica, Instituto de F\'{i}sica Universidade de S\~ao Paulo, C. P. 66.318, 05315-970 S\~ao Paulo, Brazil}

\emailAdd{christophe.grojean@desy.de}
\emailAdd{guilherme.guedes@desy.de}
\emailAdd{jasper.roosmalenepveu@desy.de}
\emailAdd{gabriel.massoni.salla@usp.br}

\abstract{
We investigate the renormalization of the radiative decays of the Higgs to two gauge bosons in the Standard Model Effective Field Theory at mass dimension eight. Given that these are loop-level processes, their one-loop renormalization can be phenomenologically important when triggered by operators generated through the tree-level exchange of heavy particles (assuming a weakly coupled UV model).
By computing the tree-level matching conditions of all relevant extensions of the Standard Model, we demonstrate that this effect is indeed present in the
$h\to \gamma Z$ decay at dimension eight, even though it is absent at dimension six. 
In contrast, the $h\to gg$ and $h\to \gamma\gamma$ decays can only be renormalized by operators generated by one-loop processes. 
For UV models with heavy vectors, this conclusion hinges on the specific form of their interaction with massless gauge bosons which is required for 
perturbative unitarity. 
We study the quantitative impact of the possible logarithmic enhancement of $h\to \gamma Z$, and we propose an observable to boost the sensitivity to this effect. Given the expected increased precision of next-generation high-energy experiments, this dimension-eight contribution could be large enough 
to be probed and could therefore give valuable clues about new physics by revealing some of its 
structural features manifesting first at dimension eight.
}

\begin{document}
\begin{flushright}
CERN-TH-2024-075\\
DESY-24-077\\
HU-EP-24/15-RTG
\end{flushright}
\maketitle

\section{Introduction}

Higgs interactions with gauge bosons are of crucial phenomenological importance. The di-photon decay  ($h\to\gamma\gamma$) was one of the channels in which the Higgs was first observed~\cite{ATLAS:2012yve,CMS:2012qbp}, while gluon fusion ($gg\to h$) is the most important production mechanism for the Higgs at hadron colliders. Furthermore, the rarer $h\to \gamma Z$ decay has recently been observed by the ATLAS and CMS collaborations~\cite{ATLAS:2023wqy,ATLAS:2023yqk}.
All of these processes occur at loop level in the Standard Model (SM) and they are promising directions to look for beyond the Standard Model (BSM) physics. 

Assuming this new physics to be heavy -- as suggested by the continuous survival of the SM under the extreme scrutiny from the past years -- the Standard Model Effective Field Theory (SMEFT) is an ideal tool to probe deviations from the SM predictions in a mostly model-independent way. 
In the SMEFT, the loop and heavy mass expansions of the theory suggest a hierarchy on the Wilson coefficients (WCs). 
Higher-order terms in the expansion in the UV scale of new physics ($\Lambda$) are generally expected to be suppressed.
Notwithstanding, effects at dimension eight, \textit{i.e.}~$\mathcal{O}(1/\Lambda^4)$,
can sometimes give relevant or even the leading contributions to an observable~\cite{Adams:2006sv,Hays:2018zze,Panico:2018hal,Zhang:2018shp,Bi:2019phv,Alioli:2020kez,Bonnefoy:2020yee,Boughezal:2021tih,Ardu:2021koz,Ellis:2022zdw,Kim:2022amu,Dawson:2022cmu,Asteriadis:2022ras,Degrande:2023iob,Chala:2023jyx,Corbett:2023qtg,Chala:2023xjy}. 

Up to dimension eight, the SMEFT contribution to the amplitudes for the $h\to \VV$ decays, with 
$\VV = gg,~\gamma\gamma$ or $\gamma Z$, can schematically be written as (suppressing an overall kinematic structure)%
    \footnote{
       We focus on the scaling of different terms in the amplitude because this determines the behavior  of the observable decay rate, which involves the square of the amplitude. 
       There are no non-interference selection rules between the SM and SMEFT amplitudes and therefore the scaling derived in Eq.~\eqref{1.1} will dictate the scaling of the decay rate in a straightforward way.
        }
\al{\label{1.1}
\Amp[h\to \VV] \times v &\simeq \frac{c_{\VV}^{(4)}}{16\pi^2} + c_{\VV}^{(6)}\frac{v^2}{\Lambda^2} 
+ \frac{c_{\VV}^{(6)'}}{16\pi^2}\frac{v^2}{\Lambda^2} 
\,\log\frac{v}{\Lambda}+c_{\VV}^{(8)}\frac{v^4}{\Lambda^4} 
+ \frac{c_{\VV}^{(8)'}}{16\pi^2}\frac{v^4}{\Lambda^4}
\,\log\frac{v}{\Lambda}\,,
}
where $v$ is the Higgs vacuum expectation value (vev) and the coefficients $c_{\VV}^{(d)},\,c_{\VV}^{(d)'}$ represent combinations of SM and EFT parameters at mass dimension $d$. For example, 
$c^{(8)}_{\VV}$ includes terms quadratic in the dimension-six WCs and other terms
linear in the dimension-eight WCs. The SM contribution to the decay amplitude is denoted by
$c_{\VV}^{(4)}/(16\pi^2)$, 
where $c_{\VV}^{(4)} = \order{g_\text{SM}^2}$ in terms of the SM (\textit{e.g.}~gauge) couplings and
we factored out the known loop suppression.
In the SMEFT, there are higher-dimensional contact interactions between the Higgs and gauge bosons, which can generate contributions to the decay amplitudes already at tree level, such that the terms proportional to $c_{\VV}^{(6,8)}$ do not contain a loop suppression at this stage. 
The logarithms stem from the renormalization group equations (RGEs) that  describe the evolution of the WCs from the matching scale $\Lambda$ to the electroweak scale $v$. 
This is a loop effect, hence the explicit $1/(16\pi^2)$ scaling. (We have suppressed the running contribution at dimension four, which is a two-loop effect.)

The fact that the $c^{(6,8)}_{\VV}$ 
terms \emph{a priori} contribute at lower loop order than the SM makes it important to understand their actual scaling by considering their UV origin. 
The coefficients $c_{\VV}^{(d)}$ and $c_{\VV}^{(d)'}$ at dimension six and eight
can be determined in terms of the UV parameters by a matching calculation at the scale of new physics.
Throughout this paper, we will restrict our analysis to UV scenarios that are weakly coupled and renormalizable, and which respect the SM symmetries. In this case,
the leading new physics contribution to $h\to \VV$ is at loop level, because the $SU(3)_c$ and $U(1)_{\rm EM}$ gauge symmetries remain unbroken. 
In the SMEFT parameterization~\eqref{1.1}, this translates to a loop suppression of the unprimed coefficients:
$c_{\VV}^{(6)}= \order{g_\text{BSM}^2g_\text{SM}^2/(16\pi^2)}$ and 
$c_{\VV}^{(8)} = \order{g_{\text{BSM}}^2g_\text{SM}^4/(16\pi^2)}$, 
where 
 the interaction with coupling $g_{\text{BSM}}$ involves new heavy particles.
This illustrates the more general fact that, under the stated assumptions on the UV, some physical effects cannot be generated by tree-level exchange of heavy states~\cite{Arzt:1994gp, Einhorn:2013kja}.
Therefore, the leading order computation of $c_{\VV}^{(6,8)}$ in terms of the UV parameters
requires a matching computation performed at the one-loop level.\footnote{
There will also be contributions from a one-loop computation in the EFT that is obtained from a tree-level matched UV, \textit{e.g.} a modified Yukawa interaction. 
These do not affect the conclusions on the $1/(16\pi^2)$ scaling made here, but we do include their quantitative effect in the rest of this paper (see Section~\ref{sec:FullModel}).}

Even though there has been an intense effort to develop tools that automatize  matching calculations at loop level~\cite{DasBakshi:2018vni,Fuentes-Martin:2020udw,Cohen:2020qvb,Carmona:2021xtq,Fuentes-Martin:2022jrf,Guedes:2023azv}, these remain more cumbersome than their tree-level counterparts, particularly at dimension eight. They are more time consuming; the projection to a physical basis is more involved; and the set of relevant UV models is not cataloged, in contrast to the tree-level extensions~\cite{deBlas:2017xtg}. Furthermore, dealing with spontaneously broken symmetries in loop-level matching calculations is still an active area of research, see 
\textit{e.g.}~Ref.~\cite{Thomsen:2024abg}.

In contrast to the non-logarithmic terms, the primed coefficients in Eq.~\eqref{1.1}, $c_{\VV}^{(6,8)'}$, 
can in principle receive non-vanishing contributions by a matching computation already at tree level, together with the application of known RGEs~\cite{Jenkins:2013zja,Jenkins:2013wua,Alonso:2013hga,Chala:2021pll,AccettulliHuber:2021uoa,DasBakshi:2022mwk}. This would result 
in the same loop suppression as $c_{\VV}^{(6,8)}$ given the loop factor from the RGEs. 
If the appropriate tree-level matching result is non-zero, this could therefore lead to a significant correction at the same perturbative order with a logarithmic enhancement~\cite{Henning:2014wua}. 
This amounts to a tree-level effect mixing into a loop-level process.

In the SMEFT, the possibility of leading-order contributions through the RGEs -- when tree-level effects mix into loop-suppressed processes -- has been studied in detail at dimension six. 
This effect was confirmed and found to be phenomenologically important in calculations of dipole moments~\cite{Panico:2018hal,Buttazzo:2020ibd,Aebischer:2021uvt}.
On the other hand, it was found that the Higgs decays to gauge bosons at dimension six are not renormalized at one loop, 
\textit{i.e.}~in Eq.~\eqref{1.1}, ${c^{(6)'}_{\VV}= 
\order{ g_\text{BSM}^2 g_\text{SM}^4 /(16\pi^2)}}$ instead of ${c^{(6)'}_{\VV}= 
\order{ g_\text{BSM}^2 g_\text{SM}^2 }}$, such that the logarithm at dimension six is a two-loop effect~\cite{Grojean:2013kd,Elias-Miro:2013gya}. 
This is also implied by the non-renormalization theorem of Ref.~\cite{Cheung:2015aba} in combination with a classification of operators according to the perturbative order at which they can be generated by matching~\cite{Craig:2019wmo}.
Even though the same arguments extend to dimension eight, they do not exclude the renormalization of the Higgs decays at this order.
This motivates the investigation of whether the logarithmic enhancement is realised in the SMEFT for $h\to \VV$ at dimension eight, and thereby to determine if $c^{(8)'}_{\VV}$ in Eq.~\eqref{1.1} should be included in future phenomenological analyses.

In this paper, we perform explicit tree-level matching calculations in weakly coupled UV models consisting of heavy scalars and vectors, to demonstrate the presence or absence of the renormalization of the $h\to \VV$ amplitudes, while highlighting its phenomenological impact. 
In the limit of vanishing Yukawa and CP-violating couplings, we consider all bosonic extensions of the SM which match onto the SMEFT and potentially generate operators (through tree-level matching) that can renormalize the operators responsible for Higgs decays at one loop.
At dimension six, the matching results are transparent: 
the fact that 
the Higgs decays are loop-level processes,
\textit{i.e.}~${c_{\VV}^{(6)} = \order{g_\text{BSM}^2g_\text{SM}^2/(16\pi^2)}}$ in Eq.~\eqref{1.1},
and that they are not affected by renormalization group (RG) mixing from operators which are generated at tree level,
$c_{\VV}^{(6)'} = \order{
g_\text{BSM}^2g_\text{SM}^4/(16\pi^2)}$, 
is implied by the vanishing result of the appropriate WCs 
in the complete tree-level matching results of Ref.~\cite{deBlas:2017xtg}.

At dimension eight, in the commonly used basis of Ref.~\cite{Murphy:2020rsh}, our results show that the loop-level order of 
$c_{\VV}^{(8)} = \order{g_\text{BSM}^2g_\text{SM}^4/(16\pi^2)}$
is encoded through intricate correlations between multiple WCs which are individually generated by tree-level matching calculations. 
Furthermore, to obtain the expected loop suppression factor in the heavy vector extensions, an additional interaction involving the heavy fields had to be included. 
This ``magnetic dipole term'' is necessary to ensure tree-level perturbative unitarity~\cite{Brandeis1970,Ferrara:1992yc,Henning:2014wua}.
Dimension eight is the leading order 
at which the SMEFT is sensitive to this interaction in tree-level matching calculations.

To determine the scaling of $c_{\VV}^{(8)'}$, we supplement the RGEs of Refs.~\cite{Chala:2021pll,
AccettulliHuber:2021uoa,DasBakshi:2022mwk} with the obtained tree-level matching results. We show that the 
Higgs decays $h\to gg$ and $h\to \gamma\gamma$ are not renormalized at one loop. 
For the $h\to\gamma\gamma$ amplitude, this is a non-trival consequence of cancellations between the RGEs of various WCs. 
Knowledge of such correlations and cancellations within the EFT is important for phenomenological studies, as they motivate the restriction to a smaller set of directions in parameter space. 
To expose the correlations that are present in the matching results and the RGEs, we propose a new basis for a subset of dimension-eight operators that describe the Higgs decays. 
Similarly to the SILH basis~\cite{Giudice:2007fh, Contino:2013kra}, the Higgs decay into two photons is parameterized by a single coefficient in this basis. Moreover, the $h\to\gamma Z$ decay requires only one additional parameter (at tree level) and
also the running (and mixing) of these two coefficients is made more transparent.

Remarkably, we do find that the $h\to \gamma Z$ amplitude is logarithmically enhanced in some of the UV models, 
\textit{i.e.}~$c_{\gamma Z}^{(8)'}= \order{g_\text{BSM}^2g_\text{SM}^4}$. 
This non-trivial renormalization of $h\to\gamma Z$ by potentially tree-level generated operators
is the main result of this work. It is a novel effect starting at dimension eight and it suggests that the RGEs could provide sizable corrections to phenomenologically relevant observables.

The results presented in this paper help understanding the conditions (or UV models) under which 
a logarithmic enhancement can occur in the $h\to \gamma Z$ decay at dimension eight. 
We consider both custodial $SU(2)_L$ conserving and violating UV models, and we find that the renormalization of the $h\to \gamma Z$ decay is correlated with custodial symmetry breaking for pure scalar extensions of the SM. That is,
$c_{\gamma Z}^{(8)'}$ (in Eq.~\eqref{1.1}) is in fact not generated by tree-level matching when custodial symmetry is preserved in UV models with only heavy scalars. Given the tight experimental bounds on custodial breaking, this renders the effects of the logarithm very constrained. This is reminiscent of the fact that custodial symmetry breaking also protects $h\to \gamma Z$ in Composite Higgs Models~\cite{Azatov:2013ura}. However, if one includes heavy vectors, this correlation is broken and the dimension-eight renormalization can arise in several new physics scenarios while respecting custodial symmetry at tree level.  

As precision increases in Higgs physics with the High-Luminosity program at the LHC and at possible future colliders such as the FCC, the detailed understanding of the role of the dimension-eight contributions becomes fundamental, particularly after the observation of the $h\to\gamma Z$ decay~\cite{ATLAS:2023wqy,ATLAS:2023yqk}. We 
identify observables in which
these higher-order contributions can be relevant to disentangle possible UV scenarios in case a signal is observed by future experiments.

We start this paper by reviewing the tree-level parameterization of the Higgs decays in the SMEFT up to dimension eight in
Section~\ref{sec:review}. 
We also discuss the non-renormalization result of Ref.~\cite{Cheung:2015aba}, which implies that the Higgs decays are not renormalized at one loop and dimension six in the SMEFT. 
In Section~\ref{sec:matching}, we introduce 
the set of candidate weakly coupled UV models with heavy scalars and vectors that may result in logarithmically enhanced Higgs decays at one loop.
We also calculate their tree-level matching onto the SMEFT. These results are combined with the RGEs from the literature in Section~\ref{sec:running} to determine whether the Higgs decays can be renormalized at dimension eight.
Upon confirming that the $h\to\gamma Z$ decay amplitude does indeed receive contributions from the RG mixing of tree-level generated operators, we study the phenomenological implications of this finding in Section~\ref{sec:FullModel}. We demonstrate that there exist UV models in which the dimension eight RGEs generate sizable corrections, while evading bounds from custodial symmetry breaking.
We then conclude in Section~\ref{sec:conclusion}.
In addition, we provide more details on the computations in three appendices.

\section{Renormalization of Higgs decays in the SMEFT}
\label{sec:review}

In this section, we will review the relevant background material for the rest of the paper. We will first write the Higgs decay amplitudes in terms of SMEFT parameters at tree level up to dimension eight. The considered Higgs decays are loop-level processes in weakly coupled renormalizable theories. 
We will therefore also review the general classification of SMEFT operators that are necessarily loop-level generated, 
\textit{i.e.}~these operators are never generated in the matching at tree level of any weakly coupled renormalizable model. Such operators may receive important contributions through the RGEs, when induced by potentially tree-level generated operators~\cite{Henning:2014wua}. 
We will review this possibility on general grounds, before focusing on the Higgs decays in more detail in subsequent sections.

\subsection{Parameterization of Higgs decays in the SMEFT}\label{sec:higgsdecays}

Unlike the SM, in which the leading contribution to the processes $h\to gg$, ${h\to \gamma\gamma}$ and $h\to \gamma Z$ arises at loop level, the SMEFT introduces corrections to these decays at tree level in the effective theory, which we denote by \emph{direct} contributions. 
We focus on the parameterization of these effects in this section.
In addition, the SMEFT also contributes \emph{indirectly}, through redefinitions of the SM couplings in the broken phase, involving the WCs. The latter only affect loop diagrams and 
are consequently not considered in this section. They may however be relevant when computing the SMEFT contributions to the decay widths, and as such we discuss these indirect effects in Appendix~\ref{app:EFT}.

We define the bosonic sector of the SM Lagrangian as
\al{\label{eq:La_SM}
\La_\text{SM} \supset 
-\frac14 G^A_{\mu\nu}G^{\mu\nu A} -\frac14 W^a_{\mu\nu}W^{\mu\nu a} -\frac14 B_{\mu\nu}B^{\mu\nu} + D_\mu\phi^\dagger D^\mu\phi + \muH^2 |\phi|^2 - \lambda |\phi|^4 ,
}
where $\phi$ is the Higgs doublet with mass parameter $\muH^2>0$, $B_{\mu\nu}$ and $W_{\mu\nu}^a$ are the electroweak gauge boson field strengths, and $G^A_{\mu\nu}$ is the gluon field strength.
The SMEFT operators that contribute directly to the considered Higgs decays up to dimension eight are%
    \footnote{We omit CP-odd operators involving the dual field-strength tensors.
    Our analysis can be extended to include CP-odd operators without affecting the conclusions.}
{\allowdisplaybreaks
\begin{align}
\La_\text{SMEFT} & \supset 
\frac{C_{\phi G}^{\phantom{1}}}{\Lambda^2} |\phi|^2G_{\mu\nu}^AG^{A\mu\nu} +
\frac{C_{\phi W}}{\Lambda^2}|\phi|^2W^a_{\mu\nu}W^{a\mu\nu} + 
\frac{C_{\phi B}}{\Lambda^2}|\phi|^2B_{\mu\nu}B^{\mu\nu} 
\nn\\&
+ \frac{C_{\phi WB}}{\Lambda^2}(\phi^\dagger \sigma^a\phi)B_{\mu\nu}W^{a\mu\nu} +
 \frac{C_{\phi D}}{\Lambda^2}|\phi^\dagger D_\mu \phi|^2 
 + \frac{C_{\phi\Box}}{\Lambda^2}|\phi|^2\Box |\phi|^2  \nn\\&
 +\frac{C_{\phi^4 G^2}^{(1)}}{\Lambda^4} |\phi|^4G_{\mu\nu}^AG^{A\mu\nu}+
 \frac{C_{\phi^4 W^2}^{(1)}}{\Lambda^4}|\phi|^4W^a_{\mu\nu}W^{a\mu\nu} 
 + \frac{C_{\phi^4 W^2}^{(3)}}{\Lambda^4}(\phi^\dagger \sigma^a\phi)(\phi^\dagger \sigma^b\phi)W^a_{\mu\nu}W^{b\mu\nu}\nn\\
& + \frac{C_{\phi^4 B^2}^{(1)}}{\Lambda^4}|\phi|^4B_{\mu\nu}B^{\mu\nu} + \frac{C_{\phi^4 WB}^{(1)}}{\Lambda^4}|\phi|^2(\phi^\dagger\sigma^a\phi)B_{\mu\nu}W^{a\mu\nu}\nn\\
& + \frac{iC_{W\phi^4 D^2}^{(1)}}{\Lambda^4}|\phi|^2(D_\mu\phi^\dagger \sigma^a D_\nu \phi)W^{a\mu\nu} + \frac{iC_{B\phi^4 D^2}^{(1)}}{\Lambda^4}|\phi|^2(D_\mu\phi^\dagger D_\nu \phi )B^{\mu\nu},
\label{eq:La_EFT}
\end{align}
where} the $C_i$ are the WCs. The first two lines contain dimension-six operators from the Warsaw basis~\cite{Grzadkowski:2010es}, and the last three lines the dimension-eight operators in Murphy's basis~\cite{Murphy:2020rsh}. Our notation for the labelling of the WCs follows these works.

Let us begin with the decay of the Higgs boson into two gluons. 
When the Higgs is expanded around its vev, $\phi = (0,(h+v)/\sqrt{2})^T$, the operator $\O_{\phi G}$ corrects the kinetic term of the gluon. Similarly, $\O_{\phi D}$ and $\O_{\phi\Box}$ modify the kinetic term of the Higgs. 
After field redefinitions to canonically normalize both kinetic terms, 
see Appendix~\ref{app:EFT},
this effect gets distributed over the remaining operators. In particular, it results in dimension-eight contributions at quadratic order in the dimension-six couplings, which we denote by $6\times6$.
Separating the full tree-level amplitude into a dimension-six contribution, $\Amp\left[hgg\right]^{(6)}$, and the two different dimension-eight terms, $\Amp\left[hgg\right]^{(6\times 6)}$ and $\Amp\left[hgg\right]^{(8)}$, we find
\al{\label{eq:h_gg}
&\frac{\Amp\left[hgg\right]^{(6)}}{v/\Lambda^2}  = C_{\phi G},\\
&\frac{\Amp\left[hgg\right]^{(6\times 6)}}{v^3/\Lambda^4} 
= 2C_{\phi G}^2 + \left(C_{\phi\Box}-\frac{C_{\phi D}}{4}\right) 
C_{\phi G},\\
&\frac{\Amp\left[hgg\right]^{(8)}}{v^3/\Lambda^4} = C_{\phi^4 G^2}^{(1)},
}
where we have stripped the amplitudes from a common factor $4(p_{g_1}^\mu p_{g_2}^\nu - \eta^{\mu\nu}(p_{g_1}\cdot p_{g_2}))$, with $p_{g_{1,2}}$ the momenta of the gluons.
This factor corresponds to the Feynman rule of the operator $hG_{\mu\nu}^AG^{A\mu\nu}$,
\be\label{eq:FR_hgg}
\La\supset 
c_{hgg}\,hG_{\mu\nu}^AG^{A\mu\nu} 
\ \implies \ 
i\Amp[hgg] = 4ic_{hgg}(p_{g_1}^\mu p_{g_2}^\nu - \eta^{\mu\nu}(p_{g_1}\cdot p_{g_2}))\,.
\ee
The amplitudes in Eq.~\eqref{eq:h_gg} were previously obtained in Ref.~\cite{Corbett:2021cil}.

In order to compute the amplitudes for the decays to the electroweak gauge bosons, we need to consider the electroweak symmetry breaking mechanism and rotate $B_\mu^{\phantom{3}}$ and $W^3_\mu$ to obtain the physical photon and $Z$ boson. 
The extra field redefinitions and diagonalizations to normalize the kinetic terms in the presence of the higher-dimensional operators are more involved than with the gluonic operators.
More details can be found in Appendix~\ref{app:EFT}.

The tree-level amplitudes at dimension six and dimension eight for $h\to \gamma\gamma$ are given by
{\allowdisplaybreaks
\begin{align}
\label{eq:h_gammagamma}
\frac{\Amp\left[h\gamma\gamma\right]^{(6)}}{v/\Lambda^2} & = e^2\left(\frac{C_{\phi W}}{g^2}+\frac{C_{\phi B}}{{g'}^2} - \frac{C_{\phi WB}}{g'g}\right),\nonumber\\[3mm]
\frac{\Amp\left[h\gamma\gamma\right]^{(6\times 6)}}{v^3/\Lambda^4} & =\left(C_{\phi W} -s_Wc_W C_{\phi WB}\right)\left(2s_W^2C_{\phi W}-s_Wc_W C_{\phi WB}\right)\nn\\
&\quad + \left(C_{\phi B} -s_Wc_W C_{\phi WB}\right)\left(2c_W^2C_{\phi B}-s_Wc_W C_{\phi WB}\right) \nn\\
&\quad + \left(C_{\phi\Box}-\frac{C_{\phi D}}{4}\right)\frac{\Amp\left[h\gamma\gamma\right]^{(6)}}{v/\Lambda^2}\,,\\[3mm]
\frac{\Amp\left[h\gamma\gamma\right]^{(8)}}{v^3/\Lambda^4} & = e^2 \left(
\frac{C_{\phi^4 W^2}^{(1)}}{g^2}
+\frac{C_{\phi^4 W^2}^{(3)}}{g^2}
+\frac{C_{\phi^4 B^2}^{(1)}}{{g'}^2} 
-\frac{C_{\phi^4 WB}^{(1)}}{g'g}\right)\,\nonumber.
\end{align}
Analogous} to the Feynman rule for $h\to gg$ in Eq.~\eqref{eq:FR_hgg}, for $h\to \gamma\gamma$ we obtain an overall kinematic structure $4(p^\mu_{\gamma_2}p^\nu_{\gamma_1}-\eta^{\mu\nu}(p_{\gamma_1}\cdot p_{\gamma_2}))$, with $p_{\gamma_{1,2}}$ the incoming momenta of the photons, which we suppress when writing the amplitudes above. The couplings $e$, $g$ and $g'$ are the $U(1)_{\rm EM}$, $SU(2)_L$ and $U(1)_Y$ charges, respectively. In the presence of higher-dimensional operators, we define these charges as the ones that preserve the form of the covariant derivatives.
Moreover, $s_{W}\equiv \sin \theta_W=g'/\sqrt{g^2+{g'}^2}$, $c_{W}\equiv \cos \theta_W=g/\sqrt{g^2+{g'}^2}$ are the sine and cosine of the weak angle. We refer to Appendix~\ref{app:EFT} for more details on our conventions.
We note the similarity in structure in the contributions from single dimension-six and dimension-eight WCs.
The expressions in Eq.~\eqref{eq:h_gammagamma} agree with Ref.~\cite{Hays:2020scx}.

Suppressing the common factor of $2(p_Z^\mu p_\gamma^\nu - (p_Z\cdot p_\gamma)\eta^{\mu\nu})$, with $p_{\gamma,Z}$ being the 4-momenta of the photon and of the $Z$ boson, the results for $h\to \gamma Z$ read
{\allowdisplaybreaks
\begin{align}
&\frac{\Amp\left[h\gamma Z\right]^{(6)}}{v/\Lambda^2}  = \frac{g^2-{g'}^2}{g'g}\frac{\Amp\left[h\gamma \gamma\right]^{(6)}}{v/\Lambda^2} + gg'\left(\frac{C_{\phi W}}{g^2}-\frac{C_{\phi B}}{{g'}^2}\right)\,,
\nn\\[3mm]
&\frac{\Amp\left[h\gamma Z\right]^{(6\times 6)}}{v^3/\Lambda^4} 
= 2s_{2W}\left(C_{\phi W}^2 - C_{\phi B}^2\right)+\left(C_{\phi\Box}-\frac{C_{\phi D}}{4}\right) \frac{\Amp\left[h\gamma Z\right]^{(6)}}{v/\Lambda^2} 
\nn\\
\label{eq:h_gammaZ}
&\qquad\qquad\qquad - c_{2W}C_{\phi WB}\left[(2s_W^2+1)C_{\phi W} + (2c_W^2+1)C_{\phi B} -s_{2W}C_{\phi WB}\right],\\[3mm]
&\frac{\Amp\left[h\gamma Z\right]^{(8)}}{v^3/\Lambda^4}  = \frac{g^2-{g'}^2}{g'g}\frac{\Amp\left[h\gamma \gamma\right]^{(8)}}{v^3/\Lambda^4} + gg'\left(\frac{C_{\phi^4 W^2}^{(1)}+C_{\phi^4 W^2}^{(3)}}{g^2}-\frac{C_{\phi^4 B^2}^{(1)}}{{g'}^2}\right)\nn\\
&\qquad\qquad\qquad+\frac{gg'}{8}\left(\frac{C_{B\phi^4D^2}^{(1)}}{g'}-\frac{C_{W\phi^4D^2}^{(1)}}{g}\right),\nn
\end{align}
where} we used the short-hand notation 
$s_{2W}\equiv \sin 2\theta_W$, $c_{2W}\equiv \cos 2\theta_W$.
Compared to the case of $h \to \gamma \gamma$,
the amplitude $\Amp\left[h\gamma Z\right]^{(8)}$ receives contributions from an additional operator class, $X\phi^4 D^2$, where $X$ corresponds to a field-strength tensor and $D$ to a covariant derivative.
Equation\,\eqref{eq:h_gammaZ} agrees with Ref.~\cite{Corbett:2021iob}.

\subsection{Tree- and loop-level generated operators and their RG mixing}\label{sec:tlo_llo}

In this section, we review the arguments that can be used to obtain information on the perturbative origin of the WCs and their RG mixing structure, with particular interest in the operators discussed in the previous section, responsible for the decays of the Higgs.
To this end, our starting assumption is that the UV completion is a weakly coupled renormalizable theory. 
 In this case, in a chosen basis, some operators are not generated by any tree-level matching calculation, simply because there exist no contributing diagrams. For this reason they are called ``loop-level generated operators'' (LLOs)~\cite{Arzt:1994gp, Einhorn:2013kja}. In contrast, we will call operators for which this cannot be argued on general grounds ``potentially tree-level generated operators'' (TLOs).
 Given the loop suppression, TLOs are expected to give the leading contribution to observables when present in their parameterization.

The classification of EFT effects as loop-level generated happens most transparently at the amplitude level. 
For this reason, it is convenient to choose a basis in which all operators have a direct correspondence  to amplitudes with the same field content. This is for instance not the case in the SILH basis~\cite{Giudice:2007fh,Contino:2013kra}, where for example the operator $\left(i \phi^\dagger D_\nu \phi \partial_\mu B^{\mu\nu}+\mathrm{ h.c.}\right)$ does not contribute to the $\phi^\dagger\phi B$ amplitude at tree level, but only to amplitudes with more external particles. 
In the Warsaw basis~\cite{Grzadkowski:2010es} and the dimension-eight basis of Ref.~\cite{Murphy:2020rsh}, 
such operators are replaced by operators with more fields. 
In these bases, the classification of an amplitude as occurring only at loop level directly translates to a statement at the level of operators with the same field content. 
We will therefore work in these bases unless otherwise stated.
We emphasize, however, that statements about amplitudes are basis independent.

Within these bases, the general argument for the perturbative origin of the WCs is that 
there only exist tree-level diagrams in renormalizable UV models to generate operators with at least four particles that are not SM gauge bosons, \textit{i.e.} Higgs bosons or fermions.
External Higgs bosons or fermions are necessary because SM gauge bosons couple diagonally to heavy BSM particles, as the SMEFT assumes the SM gauge symmetries to be unbroken. 
We thus classify operators of Refs.~\cite{Grzadkowski:2010es,Murphy:2020rsh} that do not have four Higgs boson or fermions fields as necessarily loop-level generated.
Without further information, any other operator could be considered to be potentially tree-level generated. However, as we will discuss below, this classification can be further refined.

Besides classifying operators according to their potential perturbative origin, for LLOs it is important to account for 
RG mixing from TLOs, because such contributions may come with the
same loop suppression (but with a further logarithmic enhancement). Prime examples are the electric or magnetic dipole moments of light fermions. 
Corrections to the dipole moments in the dimension-six SMEFT come from LLOs.
However, explicit computation of the dimension-six RGEs has shown that TLOs induce the dipole operators through renormalization. References~\cite{Panico:2018hal,Buttazzo:2020ibd,Aebischer:2021uvt} showed that this effect is important and should be taken into account when studying possible heavy scenarios behind a BSM contribution to the anomalous magnetic moment of the muon.
Therefore, knowledge of the structure of the anomalous dimension matrix of an EFT is a crucial ingredient when classifying operators as TLOs and LLOs. 

For the study of Higgs decays, following the parameterization introduced in Section~\ref{sec:higgsdecays}, we will 
investigate whether the operators appearing in 
Eqs.~\eqref{eq:h_gg}-\eqref{eq:h_gammaZ} can be generated at tree level and how they mix in their RGEs. The relevant operator classes are $X^2\phi^2$ and $\phi^4 D^2$ at dimension six, and $X^2\phi^4$ and $X\phi^4 D^2$ at dimension eight.

\subsubsection{Dimension six}

All operator classes of the Warsaw basis are listed in Fig.~\ref{fig:tabledim6} -- which reproduces the results from Refs.~\cite{Cheung:2015aba,Craig:2019wmo} -- with color-coding to distinguish TLOs from LLOs. 
Even though the classification of TLOs is not definitive (potentially tree-level generated operators could in fact be loop-level generated) it was explicitly found that all TLOs at dimension six can be generated with independent coefficients at tree level~\cite{deBlas:2017xtg}. This means that this classification cannot be further refined (without making more restrictive assumptions on the UV).

The operators in Fig.~\ref{fig:tabledim6} are also separated according to their so-called holomorphic weights, $w(\O)$,
and anti-holomorphic weights, $\overline w(\O)$. These weights were introduced in Ref.~\cite{Cheung:2015aba} because they determine part of the renormalization structure within EFTs, as we will discuss below. The weights are defined
by minimizing over the values for all (non-zero) tree-level amplitudes resulting from the insertion of an operator,
\begin{align}
\label{eq:weights}
    &w(\O) \equiv \text{min}\big(w(\A)\big) 
    \equiv \text{min}\big(n(\A) - h(\A)\big)\,, \nn\\&
    \overline w(\O) \equiv \text{min}\big(\overline w(\A)\big) 
    \equiv \text{min}\big(n(\A) + h(\A)\big)\,,
\end{align}
where $n(\A)$ is the number of external particles of the amplitude, $\A$, and $h(\A)$ is the sum over helicities of the external states. 
To this end, we write the operators in terms of fields that generate SM states of definite helicity.
For example, field-strength tensors $X$
are decomposed into 
$F$ with helicity $+1$ and $\bar F$ with helicity $-1$.
These definitions can be applied in basis-independent way. 
In the Warsaw basis, the amplitude with minimal weight has the external particle content corresponding to the term in $\O$ with the smallest number of fields (when expanding the covariant derivatives and field strengths).

\newcommand{\TLO}[1]{\textcolor{red}{#1}}
\newcommand{\mat}[1]{\begin{matrix}#1\end{matrix}}
\def\vr{\vrule width 1pt height 2mm}
\def\hr{\hspace{1.6mm} \rule[1mm]{3.8mm}{1pt} \hspace{-6.4mm}\phantom{.}}
\newcommand{\shade}{\cellcolor{gray!20}}

\def\arrowpicture{
\resizebox{1.7cm}{!}{
\begin{tikzpicture}
\draw[->, line width=1pt] (0,0) -- (1,0);
\draw[->, line width=1pt] (0,0) -- (0,1);
\filldraw[black] (0,0) circle (2pt);
\node[below,font=\small] at (current bounding box.south) 
    {RG mixing};
\end{tikzpicture}
}
}

\begin{figure}[t]
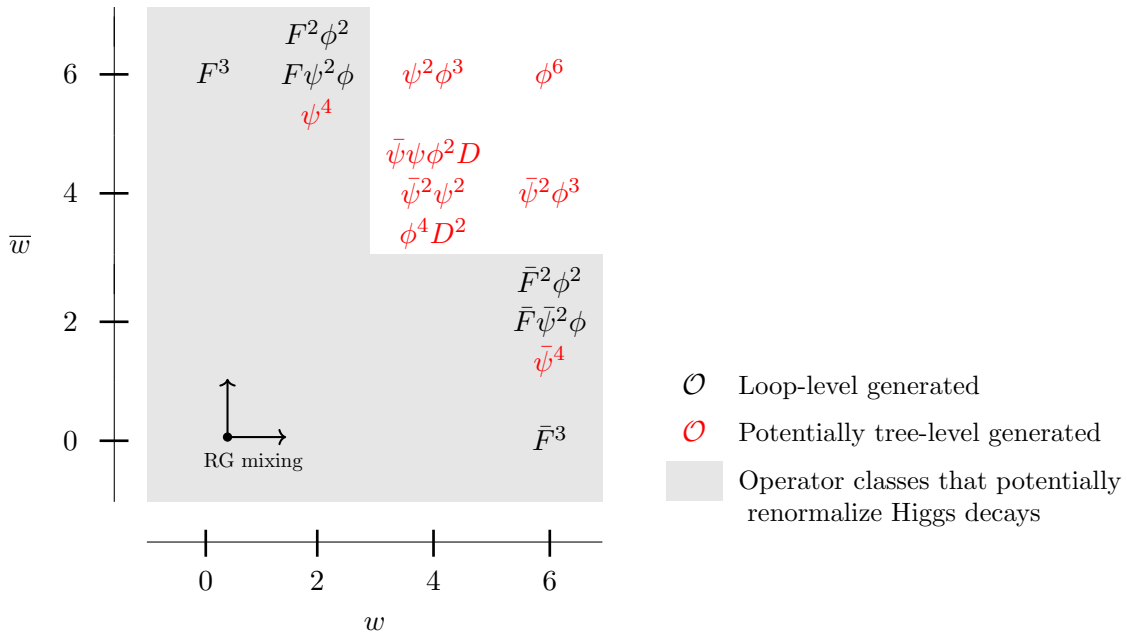

\centering
\hspace{-4cm}
\begin{minipage}{0.7\textwidth} 
\centering
\begin{tabular}{ c c|ccccc } 
    \multirow{12}{*}{$\overline w$}
            &&&\shade&\shade&&
            \\[-4mm]
    &6\hr &&\shade $\phantom{\psi^2}F^3\phantom{\phi}$&
        \shade $\begin{matrix} F^2\phi^2\\
                        F\psi^2\phi\\
                        \TLO{\psi^4}
         \end{matrix}$
         &$\TLO{\psi^2\phi^3}$&$\TLO{\phi^6}$\\
    &4\hr&&\shade&\shade&$\begin{matrix} \TLO{\bar\psi \psi\phi^2 D}\\
                         \TLO{\bar\psi^2 \psi^2}\\
                         \TLO{\phi^4D^2}
            \end{matrix}$&$\TLO{\bar\psi^2\phi^3}$\\
            &&&\shade&\shade&\shade&\shade
            \\[-4mm]
    &2\hr&&\shade&\shade&\shade&\shade$\begin{matrix} \bar F^2\phi^2\\
                          \bar F\bar \psi^2\phi\\
                          \TLO{\bar\psi^4}
            \end{matrix}$\\
    &0\hr&&\shade 
    &\shade&\shade&\shade$\begin{matrix}\\\bar F^3\\\phantom{x}\end{matrix}$\\\\\cline{4-7}\\[-8.5mm]
    &\multicolumn{1}{c}{}&&
                          $\mat{\vr \\\\[-4.5mm]0}$
                         &$\mat{\vr \\\\[-4.5mm]2}$
                         &$\mat{\vr \\\\[-4.5mm]4}$
                         &$\mat{\vr \\\\[-4.5mm]6}$\\
    &\multicolumn{2}{c}{}&\multicolumn{4}{c}{$w$}\\[-3.5cm]
    \multicolumn{7}{c}{\hspace{-1.3cm} \raisebox{5cm}{\arrowpicture}}\\[-3cm]
\end{tabular}
\end{minipage}
\hspace{-0.7cm}
\begin{minipage}{0.2\textwidth} 
\centering
\begin{tabular}{l l}
&\\[2.8cm]
\textcolor{black}{$\O$} 
    & {\small Loop-level generated}\\[1mm]
\textcolor{red}{$\O$} & 
    {\small Potentially tree-level generated}\\[1mm]
\shade & {\small Operator classes that potentially}\\[-1mm]
        & {\small \ renormalize Higgs decays} 
\end{tabular}
\end{minipage}
\caption{Operator classes in the Warsaw basis of the SMEFT at dimension six, distributed according to their holomorphic weights. 
Following Ref.~\cite{Cheung:2015aba}, the arrows indicate the direction in which one-loop RG mixing is possible. That is, operators do not mix into other operators that are on their left or below them in the diagram.
The operators in red can be tree-level generated, while LLOs are written in black. 
The shaded area contains all operator classes which could potentially mix into the Higgs decays, captured by the operator classes $F^2\phi^2$ and $\bar F^2\phi^2$.
}
\label{fig:tabledim6}
\end{figure}

Following the introduction of the weights in Eq.~\eqref{eq:weights}, Ref.~\cite{Cheung:2015aba} formulated a non-renormalization theorem based on generalized unitarity which fully explains the near absence of TLOs mixing into LLOs in the SMEFT at mass dimension six, 
while leaving room for the single exception to this rule -- see Refs.~\cite{Grojean:2013kd, Elias-Miro:2013mua, Elias-Miro:2013gya,Alonso:2014rga,Cheung:2015aba} for other studies of the renormalization structure of the SMEFT.

The one-loop RGEs of the WCs, $C_i$ associated to the operators $\mathcal{O}_i$, are described by the anomalous dimension matrix $\gamma$:
\begin{equation}
    16\pi^2\mu\frac{\dd C_i}{\dd \mu} = \sum_j \gamma_{ij} C_j\,,
\end{equation}
where $\mu$ is the renormalization scale and we omitted higher-order terms in the WCs.%
    \footnote{Terms in the RGEs at non-linear orders in the Wilson coefficients also exhibit remarkable patterns of zeros, some of which are captured by the non-renomalization theorem of Ref.~\cite{Cao:2023adc}.}
The non-renormalization theorem of Ref.~\cite{Cheung:2015aba} is given by%
    \footnote{Reference~\cite{Cheung:2015aba} identified one counterexample to this rule, which occurs for so-called exceptional amplitudes in the SM with 
    $\omega(\A_\text{SM})=2$. 
    This results in mixing between the operator classes 
    $\psi^2\phi^3 \leftrightarrow \bar\psi^2\phi^3$ and 
    $\psi^4 \leftrightarrow \bar \psi^2 \psi^2 \leftrightarrow 
    \bar\psi^4$ and does not affect the conclusions on the mixing of TLOs into LLOs in this section.} 
\begin{equation}\label{eq:nonrenResult}
    \gamma_{ij} = 0 \qquad\mathrm{if}\qquad \omega(\O_i) < \omega(\O_j) \qquad\mathrm{or} \qquad\overline{\omega}(\O_i) < \overline{\omega}(\O_j)\,.
\end{equation}
In terms of Fig.~\ref{fig:tabledim6}, this implies that operators cannot renormalize those from classes that appear on their left or below them~\cite{Cheung:2015aba}.
Since $F^2\phi^2$ ($\bar F^2\phi^2$) is not renormalized at one loop by $\psi^4$ ($\bar\psi^4$) simply because no candidate Feynman diagrams exist, it follows that the operators contributing to the relevant Higgs decays at dimension six -- see Eqs.~\eqref{eq:h_gg}, \eqref{eq:h_gammagamma} and \eqref{eq:h_gammaZ} -- are not renormalized by TLOs.
The only mixing of a TLO into an LLO allowed by the non-renormalization theorem is $\psi^4$ mixing into $F \psi^2 \phi$ (and similarly $\bar\psi^4$ into $\bar F \bar\psi^2\phi$). 
This possibility is in fact realized in the SMEFT, as was verified in the renormalization of the dipole operators by the four-fermion operators 
$\mathcal{O}_{lequ}$ and
$\mathcal{O}_{quqd}$~\cite{Jenkins:2013wua}.

\begin{figure}[t]
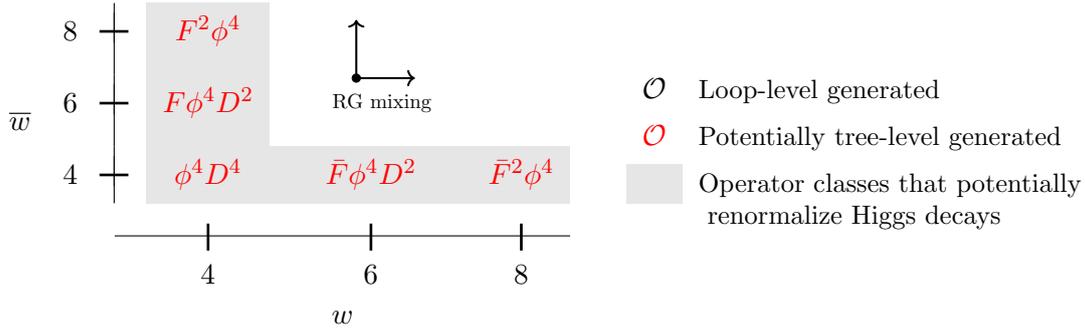

\centering 
\hspace{-1.3cm}
\begin{minipage}{0.6\textwidth}
\centering
\begin{tabular}{ c c|cccc } 
    \multirow{6}{*}{$\overline w$}&&&\shade&&\\[-4mm]
    &8\hr &&\shade$\TLO{F^2\phi^4}$&&\\[3mm]
        &&&\shade&&\\[-4mm]
    &6\hr&&\shade$\TLO{F\phi^4D^2}$&&\\[3mm]
        &&&\shade&\shade&\shade\\[-4mm]
    &4\hr&&\shade$\TLO{\phi^4D^4}$&\shade$\TLO{\bar F \phi^4D^2}$
        &\shade$\TLO{\bar F^2 \phi^4}$\\[1mm]
    &\multicolumn{1}{c}{}&&&&\\[-1mm]\cline{3-6}\\[-8.5mm]
    &\multicolumn{1}{c}{}&
                         &$\mat{\vr \\\\[-4.5mm]4}$
                         &$\mat{\vr \\\\[-4.5mm]6}$
                         &$\mat{\vr \\\\[-4.5mm]8}$\\
    &\multicolumn{1}{c}{}&\multicolumn{4}{c}{$w$}\\[-4.2cm]
    \multicolumn{5}{c}{\hspace{3.8cm} \raisebox{1.3cm}{\arrowpicture}}\\[1.3cm]
\end{tabular}
\end{minipage}%
\begin{minipage}{0.4\textwidth} 
\centering
\begin{tabular}{l l}
\textcolor{black}{$\O$} 
    & {\small Loop-level generated}\\[1mm]
\textcolor{red}{$\O$} & 
    {\small Potentially tree-level generated}\\[1mm]
\shade & {\small Operator classes that potentially}\\[-1mm]
        & {\small \ renormalize Higgs decays} \\\\[-0.2cm]
\end{tabular}
\end{minipage}
\caption{Subset of dimension-eight operators in the SMEFT with exactly four Higgs fields and without fermions, distributed according to their holomorphic weights~\cite{Cheung:2015aba}.
The shaded area contains all bosonic operator classes which could potentially mix into the Higgs decays, captured by the operator classes $F^2\phi^4$,
$\bar F^2\phi^4$,
$F\phi^4D^2$,
and $\bar F\phi^4D^2$.
This figure should be contrasted to Fig.~\ref{fig:tabledim6} at dimension six,
in which all operator classes that can renormalize the Higgs decays are loop-level generated.
A complete classification of all operator classes at dimension eight can be found in Refs.~\cite{Craig:2019wmo, Murphy:2020rsh}.
}
\label{fig:tabledim8}
\end{figure}

\subsubsection{Dimension eight}

The TLO/LLO classification and the non-renormalization theorem of Ref.~\cite{Cheung:2015aba} are general and also apply to the SMEFT at dimension eight, which was explicitly considered in Refs.~\cite{Craig:2019wmo,Murphy:2020rsh}. First, assuming the basis of Ref.~\cite{Murphy:2020rsh}, operators can be classified as necessarily generated at loop level unless they have four fields that are Higgs bosons or fermions~\cite{Craig:2019wmo}. 
Second, as opposed to dimension six, the non-renormalization result of Eq.~\eqref{eq:nonrenResult} now allows for several cases of TLOs mixing into LLOs. 
Explicit results have confirmed this richer structure of TLOs mixing with LLOs at dimension eight. For example, 
$\phi^4D^4$ mixes into $\bar FF \phi^2 D^2$,
and $\bar \psi \psi \phi^2 D^3$ induces 
$\bar FF \phi^2 D^2$~\cite{DasBakshi:2022mwk}.
It is also interesting to note that $\phi^4D^4$ mixes into the dimension-six LLOs of the class $X^2\phi^2$, with terms proportional to the Higgs mass parameter 
$\muH^2$~\cite{DasBakshi:2022mwk}. 

In Fig.~\ref{fig:tabledim8} we show the weight diagram for the operators relevant to the Higgs decays at dimension eight. These operators are classified as potentially tree-level generated and they can receive mixing contributions from other TLOs. 
However, it should be emphasized again that the minimal requirement for the tree-level generation of some operators, namely the presence of four Higgs or fermion fields, is a necessary, but insufficient condition for an operator to be generated at tree level. Indeed,  in any renormalizable weakly coupled UV theory,
the Higgs decays to two gluons, to two photons, or to one photon and one $Z$ boson can only arise at loop level, because of the unbroken $SU(3)_c\times U(1)_{\mathrm{EM}}$ gauge symmetries. This implies that operators which contribute to these Higgs decays (through tree-level diagrams in the EFT) must be generated in a correlated way by tree-level matching, such that their combined contribution to the amplitude cancels. In other words, the classification of the parameter space of TLOs can be further refined by identifying loop-level generated directions, corresponding to linear combinations of the Wilson coefficients (in a generic basis).

In the remainder of this paper, we will make the TLO/LLO classification of the operators related to the Higgs decays more precise, following explicit matching calculations in Section~\ref{sec:matching}.
The next step is then to explore whether TLOs can mix into the LLOs responsible for these Higgs decays. The study of the RGE mixing structure among these operators focusing on TLOs mixing into LLOs will be performed in Section~\ref{sec:running}.

\section{Tree-level matching results}
\label{sec:matching}

The amplitudes in Eqs.~\eqref{eq:h_gg}, \eqref{eq:h_gammagamma} and \eqref{eq:h_gammaZ} were derived from a bottom-up approach, 
\textit{i.e.}~starting from the most general EFT Lagrangian
without any considerations on the UV. 
As was noted in Section~\ref{sec:tlo_llo}, gauge invariance in any underlying weakly coupled UV completion implies that the combinations of WCs related to $h\to gg$, $h\to\gamma\gamma$ and $h\to\gamma Z$ in Eqs.~\eqref{eq:h_gg}, \eqref{eq:h_gammagamma} and \eqref{eq:h_gammaZ}  vanish at tree level.
For all considered decays, $\Amp^{(6)}$ and $\Amp^{(6\times6)}$ are loop-level generated because they are proportional to WCs in the $X^2\phi^2$ class. In contrast, the dimension-eight amplitudes $\Amp^{(8)}$ depend on operators from potentially tree-level generated classes.

For the decay into two gluons, $\Amp[h gg]^{(8)}$ depends only on $C_{\phi^4 G^2}^{(1)}$.
The fact that $h\to gg$ is a loop-level process thus leads to the conclusion that 
$C_{\phi^4 G^2}^{(1)}$ is itself loop-level generated.
This also follows from the fact that any heavy field linearly coupled to two Higgs fields cannot be charged under $SU(3)_c$. It is therefore impossible to attach gluons to a diagram representing the exchange of said heavy particle, which implies that 
$C_{\phi^4 G^2}^{(1)}$ cannot be generated at tree level.

In contrast, the Higgs decays to photons or to a photon and a $Z$ boson depend on multiple WCs.
Considered separately, these WCs can each potentially be generated at tree level. However, there should exist correlations between them such that
$\Amp[h\gamma\gamma]^{(8)}=\Amp[h\gamma Z]^{(8)}=0$ at tree level.
To make these correlations manifest, we analyse all relevant (CP-even) UV models which 
generate operators in the classes $\phi^4D^4$, $X\phi^4D^2$ and $X^2\phi^4$ at tree level.
In particular, we consider UV models with a single additional heavy particle in the following representations of the $SU(3)_c\times SU(2)_L\times U(1)_Y$ gauge group:%
    \footnote{We do not consider models with two or more BSM particles, because tree-level contributions to the considered operators result from the exchange of a single heavy field. 
    For theories with multiple heavy fields, for example be relevant in UV complete models with heavy vectors, the contributions from each particle in the single-field results can be summed.}
\begin{itemize}[itemsep=5pt, topsep=5pt, partopsep=0pt, parsep=0pt, leftmargin=1cm,
before=\vspace{1mm},after=\vspace{2mm}]
    \item a real singlet scalar $S \sim(1,1,0)$\,; 
    \item a real $SU(2)_L$ triplet scalar $\Xi \sim (1,3,0)$\,; 
    \item a complex $SU(2)_L$ triplet scalar $\Xi_1\sim (1,3,1)$\,; 
    \item a real Abelian vector boson $\mathcal{B} \sim (1,1,0)$\,; 
    \item a complex $SU(2)_L$ singlet vector boson $\mathcal{B}_1 \sim (1,1,1)$\,; 
    \item a real $SU(2)_L$ triplet vector boson $\mathcal{W} \sim (1,3,0)$\,; 
    \item and a complex $SU(2)_L$ triplet vector boson $\mathcal{W}_1 \sim (1,3,1)$\,. 
\end{itemize}   
Particles with higher representations of $SU(2)_L$ cannot couple to just two Higgs doublets, 
since only singlet and triplet currents of $SU(2)_L$ can be constructed with two Higgs fields. We also do not consider the complex scalar singlet $S_1\sim  (1,1,1)$, 
because the interaction with two Higgs fields vanishes, 
$S_1\phi^\dagger i\sigma_2 \phi^*=0$. For these reasons, the above list considers all possible UV models that potentially generate the relevant classes of operators at tree level. 
The Lagrangians for these UV models are listed in Appendix~\ref{app:UVmodels}, where we restrict to real couplings between the heavy particles and two Higgs fields.

In the heavy scalar extensions we will assume that the new scalar potential  does not introduce extra sources of electroweak symmetry breaking (EWSB), so that the matching can be performed to the SMEFT instead of the more general Higgs Effective Field Theory (HEFT)~\cite{Cohen:2020xca}.
When considering massive vectors, amplitudes involving these new particles grow with energy, breaking tree-level perturbative unitarity. This growth with energy is controlled if the heavy vector is a gauge boson from a spontaneously broken symmetry, once one includes the Higgs mechanism responsible for the heavy mass generation. However, we will remain agnostic about the origin of the massive vector.
The kinetic terms for these fields can be written as 
\al{\label{eq:vectors_kinetic_term}
\La_\mathcal{B} &\supset  
\frac{1}{2} \left( 
    \partial_\mu \mathcal{B}_\nu \partial^\nu \mathcal{B}^\mu 
    -\partial_\mu \mathcal{B}_\nu \partial^\mu \mathcal{B}^\nu 
    + M^2 \mathcal{B}_\mu \mathcal{B}^\mu
\right)\,,\\
\La_{\mathcal{B}_1} &\supset
    D_\mu \mathcal{B}_1^{\dagger\nu} D_\nu \mathcal{B}_1^{\mu} 
    -D_\mu \mathcal{B}_{1\nu}^{\dagger} D^\mu \mathcal{B}_1^{\nu} 
    + M^2 \mathcal{B}_{1\mu}^{\dagger} \mathcal{B}_1^{\mu}\,,\\
\La_\mathcal{W} &\supset
\frac{1}{2} \left( 
    D_\mu \mathcal{W}_\nu D^\nu \mathcal{W}^\mu 
    -D_\mu \mathcal{W}_\nu D^\mu \mathcal{W}^\nu 
    + M^2 \mathcal{W}_\mu \mathcal{W}^\mu
\right)\,,\\
\La_{\mathcal{W}_1} &\supset
D_\mu \mathcal{W}_1^{\dagger \nu a} D_\nu \mathcal{W}_1^{\mu a} 
-D_\mu \mathcal{W}_{1\nu}^{\dagger a} D^\mu \mathcal{W}_1^{\nu a} 
+ M^2 \mathcal{W}_{1\mu}^{\dagger a} \mathcal{W}_1^{\mu a}\,,
}
where $M$ is the mass parameter. Out of the four models above, the Abelian SM-singlet vector $\cal B$ is a special case. Using the Stueckelberg approach~\cite{Stueckelberg:1938hvi,Stueckelberg:1938zz}, one can show that this model is renormalizable despite the mass term~\cite{Pauli:1941zz,Ruegg:2003ps}. For the remaining three models, which all involve vectors charged under the SM gauge group, the Stueckelberg approach cannot account for their mass in a renormalizable theory.

For these charged vectors an extra gauge-invariant term should also be included~\cite{Ferrara:1992yc,Henning:2014wua,Brandeis1970}:
\al{\label{eq:additional_interactions_vector}
&\La_{\mathcal{B}_1} \supset 
-i\,g'\,k_{\B_1}\,
\mathcal{B}_1^{\dagger\mu} \mathcal{B}_1^\nu B_{\mu\nu}\,,\\
&\La_{\mathcal{W}} \supset 
-\frac12\,g\,k_{\W}\,\epsilon^{abc} 
\mathcal{W}^{\mu a} \mathcal{W}^{\nu a} W^c_{\mu\nu}\,,\\
&\La_{\mathcal{W}_1} \supset
    -i\,g'\,k_{\W_1,1}\,\mathcal{W}_{1\mu}^{\dagger a} \mathcal{W}_{1\nu}^{a} B^{\mu\nu}
    -g\,k_{\W_1,2}\,\epsilon^{abc} 
    \mathcal{W}_{1\mu}^{\dagger a} \mathcal{W}_{1\nu}^{b} W^{\mu\nu c}\,,
}
where we introduced the coefficients $k_{\B_1},~k_{\W},~k_{\W_1,1},~k_{\W_1,2}$. In order to respect tree-level unitarity, these coefficients need to be fixed to one. Indeed, it can be explicitly verified that this is necessary to cancel the quadratic energy growth in the scattering processes $X \mathcal{X}\rightarrow X \mathcal{X}$, where $X$ denotes the SM massless gauge bosons and $\mathcal{X}$ the heavy vectors. Note that one cannot expect a Higgs mechanism to control the energy growth in this amplitude since that Higgs would not couple to the massless (unbroken) gauge bosons. However, a Higgs mechanism underlying the mass generation of $\mathcal{X}$ would be necessary to tame the quadratic energy growth in a $\mathcal{X} \mathcal{X}\rightarrow \mathcal{X} \mathcal{X}$. Furthermore, a quartic term for the heavy vector would be needed to cancel the quartic energy growth. 
All of this discussion is analogous to the form of the low energy interactions in the SM, 
when $\mathcal{X}$ is taken to be the charged $W^\pm$ and $X$ the photon. Vertices such as \eqref{eq:additional_interactions_vector} for the $W$ boson are generated through EWSB.
We do not include either the Higgs mechanism or the quartic term in the considered UV models, 
because they do not affect our calculations. The latter cannot contribute in tree-level matching while the former could in principle be accounted for by the single-field scalar models we include.

In addition to the argument based on perturbative unitarity, Ref.~\cite{Henning:2014wua} proved in a purely algebraic way that $k_\mathcal{X}=1$ for massive vector fields in theories with spontaneously broken symmetries, regardless of the details of symmetry breaking. Other arguments based on perturbative renormalizability have also been put forward in the context of QCD, to predict $k_\mathcal{\rho}=1$ in the coupling of the $\rho$-meson to the photon~\cite{Feuillat:2000ch,Djukanovic:2005ag,Djukanovic:2013mka}. 
Furthermore, the result $k_\mathcal{X}=1$ has been observed in 
Refs.~\cite{Barbieri:2015yvd,Biggio:2016wyy}, where the explicit computation to the dipole moment was performed in extensions of the SM with heavy vectors. There, only when $k_\mathcal{X}$ was unity would a divergence for the dipole process, which had no counterterm to absorb it, vanish. 
From yet another perspective, the choice $k_{\cal X}=1$ coincides with the so-called ``minimal coupling'' on-shell interaction of Ref.~\cite{Arkani-Hamed:2017jhn}. This is defined as the amplitude that recovers the configuration of particles with opposite helicities in the high-energy limit.

\newcommand{\vlinex}[1]{\phantom{\begin{matrix} x\\[#1]x \end{matrix}}}
\newlength{\Oldarrayrulewidth}
\newcommand{\Cline}[2]{%
  \noalign{\global\setlength{\Oldarrayrulewidth}{\arrayrulewidth}}%
  \noalign{\global\setlength{\arrayrulewidth}{#1}}\cline{#2}%
  \noalign{\global\setlength{\arrayrulewidth}{\Oldarrayrulewidth}}}
\newcolumntype{?}{!{\vrule width 1.5pt}}

{\renewcommand{\arraystretch}{1.7}
\begin{sidewaystable}[p] 
\begin{center}
\vspace{-1cm}
\begin{minipage}{2\textheight} 
\resizebox{0.5\textwidth}{!}{
\begin{tabular}{ c?c?c|c|c?c|c|c|c? } 
\multicolumn{2}{c}{}&\multicolumn{3}{c?}{\textbf{Scalar extensions}}
&\multicolumn{4}{c}{\textbf{Vector extensions}}
 \\[3mm]\Cline{1.5pt}{2-9}
                             &&$S \sim(1,1,0)$ 
                             & $\Xi\sim(1,3,0)$ 
                            & $\Xi_1\sim(1,3,1)$ 
                            & $\mathcal{B}^\mu\sim(1,1,0)$ 
                            & $\mathcal{B}^\mu_{1}\sim(1,1,1)$ 
                            & $\mathcal{W}^\mu\sim(1,3,0)$ 
                            & $\mathcal{W}^\mu_{1}\sim(1,3,1)$ 
                            \\\Cline{1.5pt}{2-9}
dim.\ 4 \hspace{2mm}&
$|\phi|^4\,*$    &0& 0 & 0 & 0& 0 &$0$&$0$
\\\Cline{1.5pt}{2-9}
\multirow{3}{*}{dim.\ 6 \hspace{-1mm} 
$\left\{ \vlinex{6mm}\right.$ \hspace{-5mm}}&
$\O_{\phi}$     &$4 {\muH^2}$  &  $4\lambda -16 \lambda \muH^2 
$  & $8 \lambda - 16 \lambda \muH^2$ 
    &$-\frac12g^2 \muH^2$ 
    &$-2\lambda +\frac14({g'}^2 k_{\B_1}-g^2-10g_{\B_1}^2) \muH^2$
    &$-\lambda -\frac18(2{g'}^2+10g_\W^2-g^2 k_\W) \muH^2$
    &$-\frac12\lambda -\frac1{32}(3g_{\W_1}^2+2{g'}^2k_{\W_1,1}^2+2g^2k_{\W_1,2}^2)\muH^2$
\\ \cline{2-9}
&$\O_{\phi\square}$     &$-\frac12+2\muH^2$  &  $\frac12 -2\muH^2$  & 
    $ 2-8\muH^2$
 & $-\frac12$&$-\frac12$&$-\frac38$&$-\frac18$    
    \\ \cline{2-9} 
&$\O_{\phi D}$           &&$-2+8\muH^2$  & $4-16\muH^2$ 
 &$-2$ & 1&&$-\frac14$
\\\Cline{1.5pt}{2-9}
\multirow{1}{*}{$\phi^8$ \hspace{2mm}}&
$\O_{\phi^8}$    &$8 \lambda^2$  &  $8\lambda^2$  &
    &$g^2\lambda$ 
    &$\frac12 (2g_{\B_1}^2+g^2-{g'}^2 k_{\B_1}) \lambda$
    &$\frac14(2{g'}^2+2g_\W^2-g^2 k_\W)\lambda$
    &$-\frac1{16}(2{g'}^2k_{\W_1,1}+2g^2k_{\W_1,2}-g_{\W_1}^2)\lambda$
\\\cline{2-9}
\multirow{2}{*}{$\phi^6D^2$ \hspace{-1mm} 
$\left\{ \vlinex{-0.8mm} \right.$ \hspace{-8mm}}&$\O_{\phi^6}^{(1)}$    &$-8 \lambda $&  & $32\lambda$ 
    &$-2g^2$ 
    &$\frac14 (2g_{\B_1}^2+3{g'}^2k_{\B_1}-6g^2)$
    &$-\frac18(6{g'}^2+2g_\W^2-5g^2k_\W)$
    &$-\frac1{32}(g_{\W_1}^2+6{g'}^2k_{\W_1,1}+8g^2k_{\W_1,2})$
\\\cline{2-9}
&$\O_{\phi^6}^{(2)}$    &&$-8\lambda $ & $16\lambda$ 
    &$-\frac12 g^2$ 
    &$\frac14(4g_{\B_1}^2+2{g'}^2k_{\B_1}+g^2)$
    &$-\frac{1}{2}{g'}^2$
    &$-\frac1{16}(g_{\W_1}^2+2{g'}^2k_{\W_1,1}+g^2k_{\W_1,2})$
\\\cline{2-9}
\multirow{3}{*}{$\phi^4D^4$ \hspace{-1mm} 
$\left\{ \vlinex{6mm}\right.$ \hspace{-8mm}}&$\O_{\phi^4}^{(1)}$    &&$4$  & 
    &$-2$ &2&$\frac12$&
\\\cline{2-9}
&$\O_{\phi^4}^{(2)}$    &&  & 8 
    &2 &&$\frac12$&
\\\cline{2-9}
&$\O_{\phi^4}^{(3)}$    &$2$&$-2$  &  
    & &$-2$&$-1$& 
\\\Cline{1.5pt}{2-9}
\multirow{2}{*}{$X\phi^4D^2$ \hspace{-1mm} 
$\left\{ \vlinex{-0.8mm}\right.$ \hspace{-5mm}}&$g\O_{W\phi^4D^2}^{(1)}$    &\multicolumn{1}{c}{}
                    &\multicolumn{1}{c}{}&
    &$2$ 
    &$2$
    &$-\frac12(1+2k_\W)$&\\\cline{2-2}\cline{6-9}
&$g'\O_{B\phi^4D^2}^{(1)}$ 
        &\multicolumn{1}{c}{}
        &\multicolumn{1}{c}{}&
       &$-2 $&$-2k_{\B_1}$&$\frac32 $& 
\\\Cline{1.5pt}{2-2}\Cline{1.5pt}{6-9}
\multirow{4}{*}{$X^2\phi^4$ \hspace{-1mm} 
$\left\{ \vlinex{14mm}\right.$ \hspace{-8mm}}&$g^2\O_{\phi^4W^2}^{(1)}$ 
        &\multicolumn{3}{c?}{\raisebox{-6mm}{\huge 0}}
    &$\frac14 $&$\frac14$&$-\frac1{16}(1+2k_\W)$
    &$\frac{1}{32}(k_{\W_1,2}-1)$\\[-2.5mm]\cline{2-2}\cline{6-9}
&$g^2\O_{\phi^4W^2}^{(3)}$  
        &\multicolumn{1}{c}{}
        &\multicolumn{1}{c}{}&
        & &&&$\frac{1}{32}(k_{\W_1,2}-1)$\\\cline{2-2}\cline{6-9}
&$g'g\O_{\phi^4WB}^{(1)}$  
        &\multicolumn{3}{c?}{}
        & 
    &$\frac14(1-k_{\B_1})$
    &$\frac18(1-k_\W)$
    &$\frac{1}{16}(k_{\W_1,1}+k_{\W_1,2}-2)$\\\cline{2-2}\cline{6-9}
&${g'}^2\O_{\phi^4B^2}^{(1)}$  
        &\multicolumn{1}{c}{}
        &\multicolumn{1}{c}{}&
        & $-\frac14 $
    &$-\frac14k_{\B_1}$&$\frac3{16} $
    &$\frac1{16}(k_{\W_1,1}-1)$
\\\Cline{1.5pt}{2-9}
\end{tabular}
}
\end{minipage}
\end{center}
\caption{ \label{tab:matching-scalars}
\label{tab:matching-vectors}
Tree-level matching contributions to the SMEFT from single-particle extensions of the SM. The UV models are defined in Appendix~\ref{app:UVmodels}, where we also provide the off-shell matching results (\textit{i.e.}~before field redefinitions).
The higher-dimensional operators are defined in Ref.~\cite{Grzadkowski:2010es} at dimension six and in Ref.~\cite{Murphy:2020rsh} at dimension eight.
We have suppressed overall factors of 
$\kappa^2/M^2$ and $g_\mathcal{X}^2$, where $\kappa$ is the dimensionful Higgs-scalar coupling, $g_\mathcal{X}$ is the vector-Higgs coupling and $M$ is the mass of the heavy particle. Both $\kappa$ and $g_\mathcal{X}$ are taken to be real. 
In the vector extensions, $\h_\mathcal{X}$ is the interaction with gauge bosons of Eq.~\eqref{eq:additional_interactions_vector}.
Powers of $1/M$ can be reconstructed by dimensional analysis.
We dropped all terms of $O(1/M^8)$ and $O(1/M^6)$ for the scalar and vector extensions, respectively. 
Empty entries are zero and operators that are omitted do not receive matching contributions from any of the considered models (in the limit of zero SM Yukawa couplings). 
\\
* The contributions to the renormalizable $|\phi|^4$ operator have been absorbed into the other operators through a redefinition of $\lambda$. No other SM couplings are redefined at this stage.}
\end{sidewaystable}
}

In the matching calculations we find that dimension eight is the leading order at which the couplings $k_\mathcal{X}$ have an impact. By keeping the $k_\mathcal{X}$ as free parameters, we will explore the significance of taking the limit $k_\mathcal{X}=1$ in interpreting the EFT results.

We performed the tree-level off-shell matching calculation using Feynman diagrams.
The full results in the Green's basis of Refs.~\cite{Gherardi:2020det,Carmona:2021xtq, Chala:2021cgt} have been provided in Appendix~\ref{app:UVmodels}. 
The results in the Warsaw basis at dimension six and Murphy's basis at dimension eight, 
obtained through field redefinitions, 
are presented in Table~\ref{tab:matching-scalars}. 
We work up to dimension eight, which translates to order $\kappa^2/M^6$ for the scalar extensions, where $\kappa$ is the dimensionful Higgs-scalar coupling, and up to $1/M^4$ in the vector boson extensions. 
We have not included contributions from field redefinitions to operators with fermions.
These can be reproduced from the off-shell matching results, given in Table~\ref{tab:offShellMatching}.
Following Ref.~\cite{Ellis:2023zim}, we redefine $\lambda$ such that any matching contribution to $|\phi|^4$ is absorbed into corrections to other operators.

Matching results at dimension six for heavy scalars and vectors exist in Refs.~\cite{delAguila:2010mx,deBlas:2014mba,deBlas:2017xtg}. 
In addition, the real scalar singlet $S$~\cite{Ellis:2023zim},
the real scalar triplet 
$\Xi$~\cite{Corbett:2021eux},
the complex scalar triplet $\Xi_1$~\cite{Banerjee:2022thk},
and the real Abelian vector $\B$~\cite{Hays:2020scx}
have previously been matched to the SMEFT up to dimension eight. 
Partial results also exist for
the real vector boson triplet $\W$~\cite{Criado:2018sdb} and the complex $SU(2)_L$ singlet vector $\B_1$~\cite{Chala:2021wpj}, but these do not include the interaction 
of Eq.~\eqref{eq:additional_interactions_vector} and not all EFT operators considered here. To the best of our knowledge, the tree-level matching results of the complex vector boson triplet $\W_1$ in this section are new.

Substituting our matching results from Table~\ref{tab:matching-scalars} in the expressions for $\Amp[h\gamma\gamma]^{(8)}$ and $\Amp[h\gamma Z]^{(8)}$ (Eqs.~\eqref{eq:h_gammagamma} and \eqref{eq:h_gammaZ}) sets them to zero.
In the case of scalars, this follows trivially, as the classes $X\phi^4D^2$ and $X^2\phi^4$ are not generated. 
For models with heavy vectors, the cancellation is non-trivial and arises from the correlated way in which the operators from the relevant classes are generated. Following the previous discussion on the universal nature of the heavy vector interactions in Eq.~\eqref{eq:additional_interactions_vector}, we remark that 
$h\to\gamma Z$ is actually generated at tree level when 
$k_{\W_1,1}\neq k_{\W_1,2}$, 
which would be consistent with gauge invariance. 
However, taking into account the restriction from perturbative unitarity, $k_{\W_1,1}=k_{\W_1,2}=1$, the $h\to\gamma Z$ amplitude is not generated at tree level. 
This follows from the the fact that $k_{\W_1,1}=k_{\W_1,2}$ is the only way in which the interactions from Eq.~\eqref{eq:additional_interactions_vector} can be written as a commutator of covariant derivatives:
\begin{equation}
    -i\,g'\,\mathcal{W}_{1\mu}^{\dagger a} \mathcal{W}_{1\nu}^{a} B^{\mu\nu}
    -g\,\epsilon^{abc} 
    \mathcal{W}_{1\mu}^{\dagger a} \mathcal{W}_{1\nu}^{b} W^{\mu\nu c}
    =
    -\mathcal{W}_{1\mu}^{\dagger a} 
    \,[D^\mu,D^\nu]\,
    \mathcal{W}_{1\nu}^{a}\,.
\end{equation}

Remarkably, we also find that $\mathcal{W}_1 \sim (1,3,1)$ does not generate any dimension-eight operator with four Higgs fields when $k_{\W_1,1}=k_{\W_1,2}=1$, even though those at dimension six are generated.
It is relevant to mention that it was previously noted that the heavy vector-Higgs interaction that we assume is never generated in a Yang--Mills theory in this model~\cite{Fonseca:2022wtz}.

\section{Tree-level generated operators mixing into Higgs decays}\label{sec:running}

In the previous section, we explicitly confirmed that the EFT operators associated to the Higgs decay processes
$h\to gg$, $h\to\gamma\gamma$ and $h\to\gamma Z$ are generated only at loop level. It is therefore relevant to consider whether their one-loop running can be triggered by operators that are generated at tree level. 
As discussed in Section~\ref{sec:tlo_llo}, the non-renormalization theorem of Ref.~\cite{Cheung:2015aba} does not exclude such mixing effects at dimension eight, in contrast to dimension six.

In Ref.~\cite{Elias-Miro:2013gya} an alternative argument was proposed to rule out mixing into the operators for $h\to gg$ and $h\to\gamma\gamma$ from a TLO at dimension six. This argument is based on Higgs low-energy theorems, which relate amplitudes with different numbers of external Higgs particles
in the limit of vanishing Higgs momentum~\cite{Ellis:1975ap,Shifman:1979eb,Vainshtein:1980ea}. Because of this limit, the low-energy 
theorems of Refs.~\cite{Ellis:1975ap,Shifman:1979eb,Vainshtein:1980ea} are thus unable to fully capture the effects at higher orders in $p_h^2/M^2$ or $m_h^2/M^2$, where $p_h$ and $m_h$ are the Higgs momentum and mass, respectively, and $M$ is the mass of the heavy particle inside the loop (with $m_h \ll M$). Said in another way, the argument of Ref.~\cite{Elias-Miro:2013gya} cannot be directly applied to dimension eight and beyond.
We will therefore explicitly study the RGEs of the decays in the SMEFT at dimension eight.

The SMEFT contributions to the considered Higgs decays up to dimension eight involve the following terms:
\be
\label{eq:h_xx_full}
\Amp^{\mathrm{SMEFT}}[hVV] = \Amp[hVV]^{(6)} +\Amp[hVV]^{(6\times 6)} +\Amp[hVV]^{(8)}\,,
\ee
where $VV=\{gg,\,\gamma\gamma,\,\gamma Z\}$. 
These amplitudes have been defined in Section~\ref{sec:higgsdecays}. 
We need to consider the one-loop RG contribution to all of these terms, which can be triggered by a single insertion of a dimension-six coefficient, $c^{(6)}$; by two dimension-six coefficients, ${c^{(6)}\times c^{(6)}}$; or by a dimension-eight coupling, $c^{(8)}$. 
This structure is schematically depicted in Table~\ref{table:ren_structure}, where we only consider the renormalization triggered by potentially be tree-level generated couplings.

Various entries in Table~\ref{table:ren_structure} are zero at the orders we consider, which simplifies the analysis below.
The zeros in the first row, related to the dimension-six part of the amplitude, follow from the arguments of the previous paragraphs (and from the explicit results of Ref.~\cite{Jenkins:2013wua}) for the insertion of $c^{(6)}$ and from the explicit results of 
Refs.~\cite{Chala:2021pll,AccettulliHuber:2021uoa} for $c^{(6)}\times c^{(6)}$. 
The running of $\Amp[hVV]^{(6\times 6)}$, defined in Eqs.~\eqref{eq:h_gg}, \eqref{eq:h_gammagamma} and \eqref{eq:h_gammaZ}, is also not triggered by TLOs through the insertion of $c^{(6)}$ (at the one-loop order of the full expression).
Insertions of $c^{(6)}\times c^{(6)}$ or $c^{(8)}$ in $\Amp[hVV]^{(6\times 6)}$
are proportional to $\muH^2/\Lambda^6$, which we neglect here, because of the additional $1/\Lambda^2$ factor. 
Lastly, the dimension-eight amplitude cannot receive contributions from a single insertion of $c^{(6)}$, based on the power counting in 
$1/\Lambda$.
but higher order contributions are possible and need to be studied in more detail.

In the remainder of this section, we will consider the dimension-eight RGEs calculated in Refs.~\cite{AccettulliHuber:2021uoa,Chala:2021pll,DasBakshi:2022mwk}  in combination with the explicit matching results to compute the non-zero terms in Table~\ref{table:ren_structure} and assess 
whether TLOs mix into the considered Higgs decay processes. All RGEs presented in this section contain only the part triggered by potentially tree-level generated operators.
(The dimension-eight RGEs including terms that are triggered by LLOs are computed in~\cite{Helset:2022pde,Assi:2023zid}.)
Towards the end of the section, we translate our results to an alternative basis that emphasizes reoccurring patterns. This basis makes the distinction between TLOs and LLOs and their mixing structure more transparent.

\renewcommand{\arraystretch}{1.4}
\begin{table}[]

\centering
\begin{tabular}{@{}cccc@{}}
\toprule
        & \hspace{2mm}$c^{(6)}$\hspace{1mm}\phantom{.} & $ c^{(6)}{\times}c^{(6)}$  & $c^{(8)}$ \\ \midrule

\multicolumn{1}{l|}{$16\pi^2\mu\frac{\dd \Amp[h\to VV]^{(6)}}{\dd \mu}$}       & 0  &   0        & $\times$     \\ 
\multicolumn{1}{l|}{$16\pi^2\mu\frac{\dd \Amp[h\to VV]^{(6\times6)}}{\dd \mu}$} & 0     & $O(\muH^2/\Lambda^6)$       & $O(\muH^2/\Lambda^6)$      \\
\multicolumn{1}{l|}{$16\pi^2\mu\frac{\dd \Amp[h\to VV]^{(8)}}{\dd \mu}$}       & $\emptyset$     & $\times$      & $\times$      \\ \bottomrule

\end{tabular}
\caption{
\label{table:ren_structure}
Schematic depiction of the renormalization of the different contributions to $h\to VV$, with $VV=\{gg,\,\gamma\gamma,\,\gamma Z\}$ through insertions of potentially tree-level generated Wilson coefficients. A ``0'' means that there is no contribution (in explicit results~\cite{Chala:2021pll}), whereas ``$\emptyset$'' means that the respective contribution cannot exist on the basis of dimensional analysis.
Contributions at $O(\muH^2/\Lambda^6)$ are ignored in our work, because they correspond to higher-order effects. Finally, entries labeled with a ``$\times$'' will be the ones under scrutiny in Sections \ref{sec:h_gluglu}, \ref{sec:h_gammagamma} and \ref{sec:h_gammaZ}.   }
\end{table}

\subsection{\texorpdfstring{$h\to gg$}{h -> gg}}\label{sec:h_gluglu}

To study the renormalization of $h\rightarrow g g$, described by the amplitudes defined in Eq.~\eqref{eq:h_gg}, we only need to consider double insertions of dimension-six effective couplings and single insertions of dimension-eight ones. 
The former do not renormalize the amplitudes relevant for the gluon decay according to the results of Refs.~\cite{AccettulliHuber:2021uoa,Chala:2021pll}. We are therefore left with the dimension-eight induced RGEs, 
which read~\cite{DasBakshi:2022mwk}
\al{\label{eq:hgluglu_rge}
16\pi^2\mu\frac{\dd}{\dd \mu}\left( \frac{\Amp[h g g]}{v^3/\Lambda^4}\right) &= 12 \,\frac{\muH^2}{v^2} \, C_{\phi^4 G^2}^{(1)} + \left( 4\gamma_\phi - 3 {g'}^2 - 9 g^2 -14 g_s^2 + 48 \lambda \right) C_{\phi^4 G^2}^{(1)}  \\
 &\qquad + 2 g_s \left( \mathrm{Tr}\left[ C^\dagger_{qdG\phi^3} Y_d^{\phantom{\dagger}} + Y_d^\dagger C_{qdG\phi^3}^{\phantom{\dagger}} + C^\dagger_{quG\phi^3} Y_u^{\phantom{\dagger}} + Y_u^\dagger C_{quG\phi^3}^{\phantom{\dagger}}  \right]\right)\,,
}
where $g_s$ is the gauge coupling of $SU(3)_c$, $\gamma_\phi = 
\mathrm{Tr}[
3 Y_u^{\phantom{\dagger}} Y^\dagger_u
+ 3 Y_d^{\phantom{\dagger}} Y^\dagger_d
+ Y_e^{\phantom{\dagger}} Y^\dagger_e]$, 
where $Y_{u,d,e}$ are the SM Yukawa couplings of the up, down and electron fields, respectively (we follow the convention of Ref.~\cite{Chala:2021pll}) and $C_{q\psi G\phi^3}$ are the WCs of dimension-eight dipole operators 
$\mathcal{O}_{q\psi G\phi^3} 
= \overline{q}\, \sigma_{\mu\nu} T^A \, 
\psi \Phi \, (\phi^\dagger \phi) \, G^{A,\mu\nu}$, with 
$\psi\Phi = u\,\tilde{\phi}$ or $d\,\phi$, where $\tilde \phi \equiv i\sigma^2 \phi^*$.
Note that the first term in Eq.~\eqref{eq:hgluglu_rge}, proportional to $\muH^2$, corresponds to the renormalization of the dimension-six part of the amplitude, $\A^{(6)}$, triggered by a dimension-eight coefficient.

With Eq.~\eqref{eq:hgluglu_rge}, one can now explore whether TLOs mix into this loop-level process. 
Equation~\eqref{eq:hgluglu_rge} receives contributions from $C_{\phi^4 G^2}^{(1)}$ and $C_{q\psi G\phi^3}$. We have previously argued that the former is a loop-level generated coefficient, because the Higgs decay to two gluons is a loop-level process. We also argue that the dimension-eight dipole operators are loop-level generated, like their dimension-six counterparts, thereby refining the classification of Refs.~\cite{Craig:2019wmo, Murphy:2020rsh}.
The reason is that a contribution to the dipole moment of light fermions cannot exist at tree level in any weakly coupled renormalizable theory, and $C_{q\psi G\phi^3}$ is the only operator that contributes to the dipole kinematic structure at tree level (in the basis of Ref.~\cite{Murphy:2020rsh}) -- see for example Ref.~\cite{Hamoudou:2022tdn} for the matching conditions to the low-energy EFT dipole operator. We therefore conclude that the running of $h\to gg$ is not affected at one loop by TLOs.

Another perspective on the fact that $h\to gg$ does not run at one loop in the EFT comes from explicit full-theory calculations in generic UV extensions. For example, the amplitudes of Ref.~\cite{Shifman:1979eb} 
are free of logarithms in the heavy mass expansion, regardless of the SM extension. 
This implies that the mixing of tree-level generated operators into the loop-level $h\to gg$ decay will be absent at one loop at all mass dimensions.

To illustrate that the dipole operators are not generated at tree level, we work out a concrete matching example. We consider a toy UV model with two vector-like quarks (VLQs), $Q_u$ and $Q_q$, with the same quantum numbers as the 
right-handed singlet up-quark ($u$) and the quark doublet ($q$), respectively.
The Lagrangian we consider has the following extra interactions,
\be
\La_\text{SM+VLQ} \supset 
\left( y_{Q_u} \bar q\, \tilde \phi\, Q_u + y_{Q_q} \bar Q_q\, \tilde \phi\, u + y_{QqQu} \bar Q_q\, \tilde \phi\, Q_u +h.c.\right),
\ee
where we consider degenerate masses for simplicity. We matched this model to the SMEFT up to dimension eight at tree level using \verb!Matchete!~\cite{Fuentes-Martin:2022jrf}. 
This simple extension is instructive because it generates a tree-level Feynmann diagram with the same external fields as the dipole operator. However, once we project the results from \verb!Matchete! into Murphy's basis~\cite{Murphy:2020rsh}, we observe that the contribution to $C_{q\psi G\phi^3}$ cancels. 

To see this cancellation more explicitly, we can start from the result from \verb!Matchete!,
\al{\label{eq:VLQ_mathcing}
\La_\text{SMEFT}^\text{VLQ} & \supset \frac{1}{M_Q^4}|\phi|^2D_\mu \bar q\,\tilde\phi\left(2 y_{Q_u}^{\phantom{*}} y_{Q_qQ_u}^* + Y_u y_{Q_q}^{\phantom{*}}\right)D^\mu u  \\
& \qquad + \frac{g_s}{M_Q^4}|\phi|^2 \bar q\,\tilde\phi\left(2 y_{Q_u}^{\phantom{*}} y_{Q_qQ_u}^* + Y_u y_{Q_q}^{\phantom{*}}\right)\sigma_{\mu\nu}G^{\mu\nu} u,
}
where we ignored other operators not relevant for the discussion.
Projecting the first operator to Murphy's basis involves using integration by parts,
\be
|\phi|^2 D_\mu \bar q\,\tilde\phi D^\mu u = -\frac{1}{2}\left(|\phi|^2 D^2 \bar q\,\tilde\phi u + |\phi|^2 \bar q\,\tilde\phi D^2 u\right)+\cdots,
\ee
where the ellipsis denotes irrelevant operators with derivatives on the Higgs fields. 
The dipole operators can then be obtained using the identity $D^2 \psi=\slashed{D}^2\psi+g_s\sigma^{\mu\nu}G_{\mu\nu}\psi$  (ignoring other gauge bosons). It thus follows that the tree-level contribution to $C_{q\psi G\phi^3}$ is cancelled in this example. Said in another way, the combination of terms in Eq.~\eqref{eq:VLQ_mathcing} does not contribute to the on-shell three-point dipole vertex.

\subsection{\texorpdfstring{$h\to \gamma \gamma$}{h -> gamma gamma}}
\label{sec:h_gammagamma}

Let us now turn our attention to the di-photon Higgs decay. Unlike in $h\to g g$, a non-trivial result for the running of this amplitude can be obtained without Yukawa couplings.
We therefore consider the limit of vanishing Yukawa couplings.\footnote{Fermionic operators do not contribute in the limit of vanishing Yukawas. The presence of fermions in the theory does affect the RGE of the SM charges and the wavefunction renormalization of the bosonic effective operators. These contributions are included in the RGEs of Refs.~\cite{AccettulliHuber:2021uoa,Chala:2021pll,DasBakshi:2022mwk} that we used. We checked, however, that their combined effect is zero.} 
The result for the renormalization of $h\to \gamma\gamma$ is~\cite{AccettulliHuber:2021uoa,Chala:2021pll,DasBakshi:2022mwk}
{\allowdisplaybreaks
\begin{align}
\label{eq:rgehgammagamma}
16\pi^2\mu\frac{\dd }{\dd \mu} 
    \left(
\frac{\Amp\left[h\gamma\gamma\right]}{v^3/\Lambda^4}
    \right)
    &= 
    -3e^2{g'}^2\Bigg( 
        \frac{C_{\phi^4W^2}^{(1)}}{g^2}
        +\frac{C_{\phi^4W^2}^{(3)}}{g^2}
        -\frac{C_{\phi^4WB}^{(1)}}{g'g}
        +\frac{C_{\phi^4B^2}^{(1)}}{{g'}^2}
    \Bigg)
    \nonumber\\[2mm]
    &\hspace{10mm}+e^2 g^2 \Bigg( 
        -9\frac{C_{\phi^4W^2}^{(1)}}{g^2}
        +3\frac{C_{\phi^4W^2}^{(3)}}{g^2}
        +3\frac{C_{\phi^4WB}^{(1)}}{g'g}
        \nonumber\\&\hspace{10mm}\qquad \qquad
        -9\frac{C_{\phi^4B^2}^{(1)}}{{g'}^2}
        +\frac{3}{2}\frac{C_{W\phi^4D^2}^{(1)}}{g}
        +\frac{3}{2}\frac{C_{B\phi^4D^2}^{(1)}}{g'}
    \Bigg)
    \nonumber\\[2mm]
    &\hspace{10mm}+e^2 \lambda \Bigg(
        36\frac{C_{\phi^4W^2}^{(1)}}{g^2}
        +28\frac{C_{\phi^4W^2}^{(3)}}{g^2}
        -32\frac{C_{\phi^4WB}^{(1)}}{g'g}
        \nonumber\\&\hspace{10mm}\qquad\qquad 
        +36\frac{C_{\phi^4B^2}^{(1)}}{{g'}^2}
        -\frac{C_{W\phi^4D^2}^{(1)}}{g}
        -\frac{C_{B\phi^4D^2}^{(1)}}{g'}
    \Bigg)
    \,,
\end{align}
which} includes contributions from the running of the couplings,
$16\pi^2\mu\dd g'/\dd \mu = 41{g'}^3/6$,\, 
$16\pi^2\mu\dd g/\dd \mu = -19g^3/6$,
besides that of the WCs. 
We note that double insertion of dimension-six coefficients contribute to the RGEs of the individual WCs that appear in the $h\to \gamma\gamma$ amplitude, Eq.~\eqref{eq:h_gammagamma}, but these cancel even before imposing any matching condition on the WCs.

Inspection of Eq.~\eqref{eq:rgehgammagamma} suggests that TLOs could renormalize $h\to \gamma\gamma$ due to the dependence on operators of the classes $X^2\phi^4$ and $X\phi^4 D^2$, which are potentially tree-level generated. 
However, as we have seen in Section~\ref{sec:matching}, some linear combination of TLOs in these two classes are actually LLOs. 
We therefore directly substitute the WCs by the corresponding tree-level matching results in Table~\ref{tab:matching-scalars}, which accounts for \textit{all} renormalizable and weakly coupled UV completions that can give a non-zero tree-level contribution to fully bosonic operators with four Higgs bosons. 

The matching results are
\begin{align}\label{eq:gammagammamixingRes}
    16\pi^2\mu\frac{\dd }{\dd \mu} 
    \left(
\frac{\Amp\left[h\gamma\gamma\right]}{v^3/\Lambda^4}
    \right)
    &= e^2 \left\{ 
    \begin{array}{cl} 
    g_{\mathcal{B}_1}^2 (\lambda-\frac32g^2 )(\h_{\mathcal{B}_1}-1), \quad\phantom{.}      
                &       \mathcal{B}_1 \sim (1,1,1)\\
        \frac12
        g_{\mathcal{W}}^2
        (\lambda-\frac32g^2) (\h_{\mathcal{W} }-1), \quad\phantom{.} 
                &\mathcal{W} \sim (1,3,0)\\
    \frac14
    g_{\mathcal{W}_1}^2
    (\lambda-\frac32g^2 )(\h_{\mathcal{W}_1,1}-1), \quad\phantom{.} 
                &\mathcal{W}_1 \sim (1,3,1) \end{array}
    \right.\,,
\end{align}
where the tree-level matching conditions of the scalar extensions and $\mathcal{B}\sim(1,1,0)$ set Eq.~\eqref{eq:rgehgammagamma} to zero. 
The scalar extensions do not generate any of the WCs in Eq.~\eqref{eq:rgehgammagamma}, while the heavy Abelian vector model generates them in a correlated way that sets the overall expression to zero.
For the remaining vector extensions, which are charged under the SM gauge group, the result is proportional to $(k_{\mathcal{X}} - 1)$, with $k_{\mathcal{X}}$ defined in Eq.~\eqref{eq:additional_interactions_vector}. As discussed in Section~\ref{sec:matching}, $k_{\mathcal{X}} = 1$ is necessary to ensure tree-level perturbative unitarity and the renormalizability of the heavy vector theory. 
We also explicitly verified that the amplitude $h\to \gamma \gamma$ with a heavy vector inside the loop generates a divergence proportional to $(k_{\mathcal{X}} - 1)$.
Under our assumptions of a weakly coupled and renormalizable theory, \ie $k_{\mathcal{X}} = 1$, the conclusion is therefore that TLOs do not mix into $h\to\gamma\gamma$. 

Similarly to the non-renormalization of the $h\to gg$ decay amplitude, this is also implied by the full one-loop results for the $h\to\gamma\gamma$ amplitude in generic SM extensions~\cite{Shifman:1979eb}, where the heavy mass expansion does not produce any logarithmic dependence.
(Reference~\cite{Shifman:1979eb} assumes $k_{\mathcal{X}}=1$ for the heavy vector extension.)

\subsection{\texorpdfstring{$h\to \gamma Z$}{h -> gamma Z}}
\label{sec:h_gammaZ}
In contrast to the previous case, the RGE of $\mathcal{A}[h\gamma Z]$ (Eq.~\eqref{eq:h_gammaZ}) is induced by fermionic operators, even in the limit of vanishing Yukawa couplings.
These, however, have only two Higgs fields, which means that they are not generated at tree level in the considered UV scenarios with vanishing Yukawa matrices.
We therefore again neglect fermionic operators from the outset. 
In this case, the RGE for $h\to \gamma Z$ is given by
{\allowdisplaybreaks
\begin{align}
\label{eq:rgehgammaZ}
16\pi^2\mu\frac{\dd }{\dd \mu} 
    \left(
\frac{\Amp\left[h\gamma Z \right]}{v^3/\Lambda^4}
    \right)&= 
 \frac{{g'}^3 g}{6}\Bigg(
    23\frac{C^{(1)}_{\phi^4W^2}}{g^2} 
    +23\frac{C^{(3)}_{\phi^4W^2}}{g^2}
    +12\frac{C_{\phi WB}}{g'g}
    -47\frac{C_{\phi^4 B^2}}{{g'}^2} 
        \nonumber\\&\hspace{3cm}
    -\frac{9}{2}\frac{C^{(1)}_{ W \phi^4 D^2}}{g}
    +\frac{17}{4}\frac{C_{B\phi^4 D^2}}{g'}
    \Bigg)
    \nonumber\\[1mm]&\quad
    +\frac{g'g^3}{6}\Bigg(
    -49\frac{C^{(1)}_{\phi^4W^2}}{g^2} 
    +47\frac{C^{(3)}_{\phi^4W^2}}{g^2}
    -24\frac{C_{\phi WB}}{g'g}
    +73\frac{C_{\phi^4 B^2}}{{g'}^2} 
        \nonumber\\&\hspace{2cm}
    +13\frac{C^{(1)}_{ W \phi^4 D^2}}{g}
    -\frac{27}{4}\frac{C_{B\phi^4 D^2}}{g'}
    +\frac{9}{4}C^{(1)}_{\phi^4}
    -\frac{9}{4}C^{(2)}_{\phi^4}
    \Bigg)
    \nonumber\\[1mm]&\quad
    +g'g\lambda\Bigg(
72 \frac{C^{(1)}_{\phi^4W^2}}{g^2}
    +56 \frac{C^{(3)}_{\phi^4 W^2}}{g^2} 
    -72\frac{C_{\phi^4 B^2}}{{g'}^2}
    -9 \frac{C^{(1)}_{ W \phi^4 D^2}}{g} 
    + 9\frac{C_{B\phi^4 D^2}}{g'} 
    \Bigg)
 \nonumber\\[1mm]&\quad
 + 16\pi^2\mu\frac{\dd}{\dd \mu}\bigg(\frac{g^2-{g'}^2}{g' g}  
    \frac{\mathcal{A}[h\gamma\gamma]}{v^3/\Lambda^4}\bigg)\,.
\end{align}}

The running of $h\to \gamma Z$ receives contributions from WCs of the $\phi^4D^4$ class, which can be generated at tree level, for example by scalar extensions of the SM. 
The relevant linear combination responsible for the $h\to \gamma Z$ renormalization is $C^{(1)}_{\phi^4}-C^{(2)}_{\phi^4}$, which is non-zero for the two $SU(2)_L$ triplet scalar models considered; no cancellation happens with contributions from other operators, as the classes $X\phi^4D^2$ or $X^2\phi^4$ are not generated in these models (see Table~\ref{tab:matching-scalars}). As for the heavy vectors, we observe that for $k_{\mathcal{X}}=1$, $\mathcal{B}_1$ and $\mathcal{W}$ also give a non-zero result to Eq.~\eqref{eq:rgehgammaZ},
\begin{align}
\label{eq:gammaZresults}
    16\pi^2\mu\frac{\dd }{\dd \mu} 
    \left(
\frac{\Amp\left[h\gamma Z \right]}{v^3/\Lambda^4}
    \right)
     &=
        g'g^3
    \left\{ \begin{array}{cl}
        \frac32\frac{\kappa_{\Xi}^2}{M^2}, \quad\phantom{.} & \Xi \sim (1,3,0),\\
        -3\frac{\,|\kappa_{\Xi_1}|^2}{M^2},      \quad\phantom{.}  & \Xi_1\sim (1,3,1),\\
        \frac94 g_{\mathcal{B}_1}^2,  \quad\phantom{.}  & \mathcal{B}_1 \sim (1,1,1),~\h_{\mathcal{B}_1}=1,\\
        -\frac98 g_{\mathcal{W}}^2, \quad\phantom{.}  & \mathcal{W} \sim (1,3,0),~\h_{\mathcal{W}}=1
        \end{array}
    \right. \,.
\end{align}
This result proves through explicit calculation that TLOs can mix into the decay $h\to \gamma Z$. 

We will explore this effect in more detail in 
Section~\ref{sec:FullModel}, where we reproduce the logarithm in a full-theory calculation (with the real scalar triplet UV model, $\Xi$) and study its phenomenological impact.

\subsection{An alternative operator basis}
\label{sec:abasis}
The above results have been presented in terms of the basis in Eq.~\eqref{eq:La_EFT}. We are, however, interested in the specific linear combinations of Wilson coefficients that contribute to the $h\to\gamma\gamma$ and $h\to\gamma Z$ processes. In addition, \emph{all} weakly coupled UV models generate the EFT operators in a correlated way, namely
$C_{W\phi^4D^2}^{(1)}=-C_{B\phi^4D^2}^{(1)}=8\,C_{W^2\phi^4}^{(1)}=-8\,C_{B^2\phi^4}^{(1)}$. To make this more transparent, we consider a change of basis of the dimension-eight operators with field-strength tensors, defined through the following full-rank matrix,
\begin{equation}\label{eq:newBasisCouplings}
    \begin{pmatrix}
        C_{h\gamma\gamma}\\[1.6mm]C_{h\gamma Z}\\[1.6mm]
        C_3\\[1.6mm]C_4\\[1.6mm]C_5\\[1.6mm]C_\text{TLO}
    \end{pmatrix} 
    \equiv 
    \left(
\begin{array}{cccccc}
 \frac{1}{2} & \frac{1}{2} & -\frac{1}{2} & \frac{1}{2} & 0 & 0 \\[1.5mm]
 \frac{1}{2 } & \frac{1}{2} & 0 & -\frac{1}{2} & -\frac{1}{16} & \frac{1}{16} \\[1.5mm]
 0 & -1 & \frac{1}{2 } & 0 & -\frac{1}{8 } & -\frac{1}{8 } \\[1.5mm]
 0 & {1} & 0 & 0 & 0 & 0 \\[1.5mm]
 0 & 0 & 0 & 0 & \frac{1}{8 } & \frac{1}{8 } \\[1.5mm]
 0 & 0 & 0 & 0 & \frac{1}{16 } & -\frac{1}{16 } \\
\end{array}
\right)
\begin{pmatrix}
    \frac{1}{g^2}\,C_{\phi^4W^2}^{(1)}\\[1.5mm]
    \frac{1}{g^2}\,C_{\phi^4W^2}^{(3)}\\[1.5mm]
    \frac{1}{g'g}\,C_{\phi^4WB}^{(1)}\\[1.5mm]
    \frac{1}{{g'}^2}\,C_{\phi^4B^2}^{(1)}\\[1.5mm]
    \frac{1}{g}\,C_{W\phi^4D^2}^{(1)}\\[1.5mm]
    \frac{1}{g'}\,C_{B\phi^4D^2}^{(1)}
\end{pmatrix}\,.
\end{equation}\\[-5mm]

\renewcommand{\arraystretch}{1.4}
\begin{table}[t]
\centering
\resizebox{1\textwidth}{!}{
\begin{tabular}{ ?c?c|c|c?c|c|c|c? } 
\multicolumn{1}{c}{}&\multicolumn{3}{c?}{\textbf{Scalar extensions}}
&\multicolumn{4}{c}{\textbf{Vector extensions}}\\
 \Cline{1.5pt}{1-8}
    &$S$ 
    & $\Xi$ 
    & $\Xi_1$ 
    & $\mathcal{B}^\mu$ 
    & $\mathcal{B}^\mu_{1}$ 
    & $\mathcal{W}^\mu$ 
    & $\mathcal{W}^\mu_{1}$ \\[-2mm]
    &$(1,1,0)$ 
     & $(1,3,0)$ 
    & $(1,3,1)$ 
    & $(1,1,0)$ 
    & $(1,1,1)$ 
    & $(1,3,0)$ 
    & $(1,3,1)$
 \\\Cline{1.5pt}{1-8}
$\O_{\phi^4}^{(1)}$    &&4&&$-2$ &2&$1/2$&\\\hline
$\O_{\phi^4}^{(2)}$    &&&8&2 &&$1/2$&\\\hline
$\O_{\phi^4}^{(3)}$    &2&$-2$&& &$-2$&$-1$& \\\Cline{1.5pt}{1-8}
$C_{h\gamma\gamma}$ &\multicolumn{1}{c}{}&\multicolumn{1}{c}{}&
    & 
    &
    &
    &\\ 
    \cline{1-1}\cline{5-8}
$C_{h\gamma Z}$ &\multicolumn{1}{c}{}&\multicolumn{1}{c}{}&
    & 
    &
    &
    &$\frac{1}{32}(k_{\W_1,2}-k_{\W_1,1})$\\ \cline{1-1}\cline{5-8}
$C_3$ &\multicolumn{3}{c?}{\raisebox{-6mm}{\LARGE 0}}
    & 
    &$\frac18 (k_{\B_1}-1)$
    &$\frac1{16} (k_\W-1)$
    &$\frac1{32} (k_{\W_1,1}-1)$\\[-3.8mm] \cline{1-1}\cline{5-8}
$C_4$ &\multicolumn{1}{c}{}&\multicolumn{1}{c}{}&
    &
    &
    &
    &$\frac1{32} (k_{\W_1,2}-1)$\\ \cline{1-1}\cline{5-8}
$C_5$ &\multicolumn{1}{c}{}&\multicolumn{1}{c}{}&
    &
    &$\frac14 (1-k_{\B_1})$
    &$\frac18 (1-k_\W)$
    &\\ \cline{1-1}\cline{5-8}
$C_\text{TLO}$ &\multicolumn{1}{c}{}&\multicolumn{1}{c}{}&
    & $\frac14 $
    &$\frac14$
    &$-3/{16} $
    &\\ \Cline{1.5pt}{1-8}
\end{tabular}
}
\caption{
The same as Table~\ref{tab:matching-scalars}, restricted to dimension-eight operators with exactly four Higgs fields. The operators with field-strength tensors have been translated to the basis defined in Eqs.~\eqref{eq:newBasisCouplings} and~\eqref{eq:alternativeBasis}, which makes the correlations in the matching results more transparent. 
In particular, it manifests that there is only one tree-level generated operator with field strengths when the UV is renormalizable $(k_\mathcal{X}=1)$.
We suppressed an overall factor of
$\kappa^2/M^6$ for the scalar extensions and $g_\mathcal{X}^2/M^4$ for the vector extensions.}
\label{tab:matching-vectors_newbasis}
\end{table} 

\begin{figure}[t]
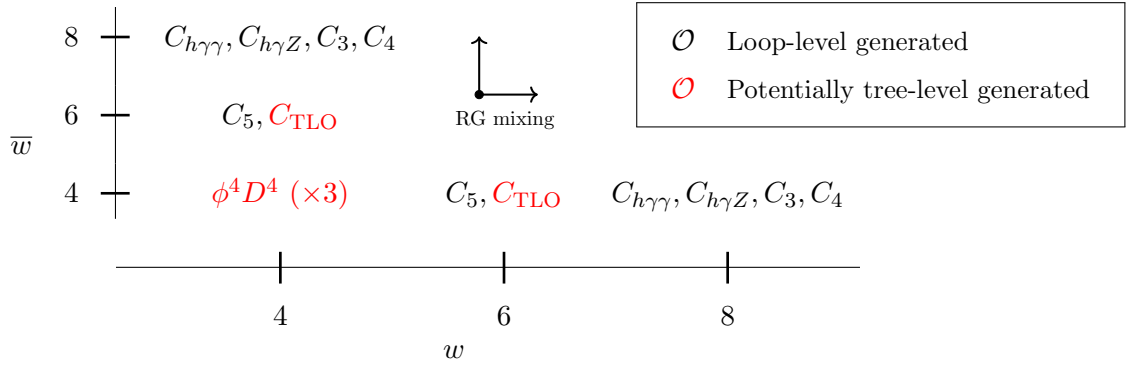

\begin{minipage}{0.55\textwidth}
\centering 
\begin{tabular}{ c c|cccc } 
    \multirow{5}{*}{$\overline w$}&
    8\hr &&$C_{h\gamma\gamma}, C_{h\gamma Z}, C_3, C_4$&&\\[3mm]
    &6\hr&&$C_5, \TLO{C_\text{TLO}}$&&\\[3mm]
    &4\hr&&$\TLO{\phi^4D^4~(\times3)}$
            &$C_5, \TLO{C_\text{TLO}}$
            &$C_{h\gamma\gamma}, C_{h\gamma Z}, C_3, C_4$\\[-1mm]
    &\multicolumn{1}{c}{}&&&&\\\cline{3-6}\\[-12.5mm]
    &\multicolumn{1}{c}{}&
                         &$\mat{\vr \\\\[-7mm]4}$
                         &$\mat{\vr \\\\[-7mm]6}$
                         &$\mat{\vr \\\\[-7mm]8}$\\[-3mm]
    &\multicolumn{5}{c}{$w$}\\[-4.5cm]
    \multicolumn{5}{c}{\hspace{5.4cm} \raisebox{1.3cm}{\arrowpicture}}\\[1.6cm]
\end{tabular}
\end{minipage}
\begin{minipage}{0.35\textwidth} 
\centering
\vspace{-3.3cm}
\fbox{
\begin{tabular}{l l}
\textcolor{black}{$\O$} 
    & {\small Loop-level generated}\\[-1mm]
\textcolor{red}{$\O$} & 
    {\small Potentially tree-level generated}
    \\[0.5mm]
\end{tabular} }
\end{minipage}
\caption{
The same as Fig.~\ref{fig:tabledim8}, for the specific operator basis defined in Eq.~\eqref{eq:newBasisCouplings} and~\eqref{eq:alternativeBasis}. The operators in red are tree-level generated, while LLOs are written in black. There are three operators in the $\phi^4D^4$ class, which can all be generated at tree level.
}
\label{fig:tabledim8_newbasis}
\end{figure}

The first two operators in this basis are more directly related to the physical decay channels, such that only one or two EFT parameters will be constrained from these observables. The dimension-eight amplitudes of Eq.~\eqref{eq:h_gammaZ} are
\begin{align}
\frac{\Amp\left[h\gamma\gamma\right]^{(8)}}{v^3/\Lambda^4} & =
    2 e^2 C_{h\gamma\gamma} \,,
\nn\\
\frac{\Amp\left[h\gamma Z\right]^{(8)}}{v^3/\Lambda^4} & =
    2e^2\,\frac{g^2-{g'}^2}{g'g}\,C_{h\gamma\gamma}
    + 2g' g \,C_{h\gamma Z}\,.
\end{align}
Furthermore, all couplings but $C_\text{TLO}$ have been chosen such that they are not generated at tree level in any of the considered UV models (for $k_{\B_1}=k_\W=k_{\W_1,1}=k_{\W_1,2}=1$). We present the matching results in the new basis in Table~\ref{tab:matching-vectors_newbasis} and Fig.~\ref{fig:tabledim8_newbasis}.
Finally, $C_3$ has been chosen as the only linear combination of Wilson coefficients that mixes into $C_{h\gamma\gamma}$ (besides $C_{h\gamma\gamma}$ itself). The RGE of $\Amp\left[h\gamma\gamma\right]$ 
\eqref{eq:rgehgammagamma} can thus be rewritten as
\begin{align}
\label{eq:rgehgammagamma-basis2}
    16\pi^2\mu\frac{\dd }{\dd \mu} 
    \left(
\frac{\Amp\left[h\gamma\gamma\right]}{v^3/\Lambda^4}
    \right)
    &= 
    -6e^2{g'}^2C_{h\gamma\gamma}     
    -6e^2 g^2 \left( 
        3C_{h\gamma\gamma} + 
        2C_3
    \right)
    +8e^2 \lambda \left(
        9C_{h\gamma\gamma} 
        + C_3
    \right)
    \,.
\end{align}
This provides an alternative perspective on the fact that $C_3$ is loop-level generated (for $k_{\B_1}=k_\W=k_{\W_1,1}=k_{\W_1,2}=1$), besides our explicit matching results.

The RGE of $\Amp\left[h\gamma Z\right]$ \eqref{eq:rgehgammaZ} becomes
\begin{align}
\label{eq:rgehgammaZ-basis2}
    16\pi^2\mu\frac{\dd }{\dd \mu} 
    \left(
\frac{\Amp\left[h\gamma Z\right]}{v^3/\Lambda^4}
    \right)
    &= 
        g'g^3\left(
        6 C_\text{TLO} 
        +\frac{3}{8}C_{\phi^4}^{(1)}
        -\frac{3}{8}C_{\phi^4}^{(2)}
        \right)
    \nn\\&\hspace{-3cm}
    +g'g^3\left(
        4C_{h\gamma\gamma} 
        -\frac{61}{3} C_{h\gamma Z} 
        -4C_3 +12 C_4 +\frac{1}{6} C_5
        \right)
    +16g'g\lambda \, (9C_{h\gamma Z} - C_4)
    \nonumber \\&\hspace{-3cm}
    +{g'}^3g\left(-4C_{h\gamma\gamma} + \frac{35}{3}C_{h\gamma Z} -\frac{1}{6}C_5\right)
    + 16\pi^2\mu\frac{\dd }{\dd\mu} \bigg(\frac{g^2-{g'}^2}{g' g}  
    \frac{\mathcal{A}[h\gamma\gamma]}{v^3/\Lambda^4}\bigg)
    \,,
\end{align}
where the second and third lines are necessarily two-loop effects when the EFT is related to any weakly coupled renormalizable UV theory.
Note that even though $C_\text{TLO}$, $C_{\phi^4}^{(1)}$ and $C_{\phi^4}^{(2)}$ are generated by $\mathcal{B} \sim (1,1,0)$, their effect cancels and no running contribution to $\Amp[h\gamma Z]$ is generated in this model from TLOs. One can also understand this from the full model perspective, since $\mathcal{B}$ is a singlet.

Besides the mixing of TLOs into 
$\mathcal{A}[h\gamma Z]$,
we computed explicitly that $C_3$, $C_4$, $C_5$ 
in Eq.~\eqref{eq:newBasisCouplings} 
also receive RG mixing contributions from tree-level generated parameters. That is, all LLOs (in this subset of operators) except the operator associated to $C_{h\gamma\gamma}$ are renormalized by TLOs. 
The tree-level generated coefficient $C_{\mathrm{TLO}}$ is renormalized by TLOs and $C_5$. 
This is consistent with the classification in Fig.~\ref{fig:tabledim8_newbasis}, which portrays the non-renormalization result of Ref.~\cite{Cheung:2015aba}.
Remarkably, 
mixing at quadratic order in the dimension-six (potentially) tree-level generated WCs does not affect any of the loop-level generated parameters in Fig.~\ref{fig:tabledim8_newbasis}. In contrast, $C_\text{TLO}$ and the $\phi^4D^4$ operators do have such terms in their RGEs.

In terms of operators, the redefinition of Eq.~\eqref{eq:newBasisCouplings} implies that the dimension-eight part of Eq.~\eqref{eq:La_EFT} is rewritten as 
{\allowdisplaybreaks
\begin{align}
    \label{eq:alternativeBasis}
    \La_\text{SMEFT} &\supset 
    C_{h\gamma\gamma} \left( 
    g^2\O_{\phi^4W^2}^{(1)}
    +{g'}^2 \O_{\phi^4B^2}^{(1)}
    \right) +
    C_{h\gamma Z}\left( 
    g^2\O_{\phi^4W^2}^{(1)}
    -{g'}^2 \O_{\phi^4B^2}^{(1)}
    \right)
    \\&\hspace{-1cm}
    -4\,C_3 \, (\phi^\dagger \phi) (\phi^\dagger 
        \left[D_\mu,D_\nu\right]\left[D^\mu,D^\nu\right]\phi) 
    \ -\ 4\,C_4\, 
    (\phi^\dagger \left[D_\mu,D_\nu\right]\phi) 
    (\phi^\dagger \left[D^\mu,D^\nu\right]\phi)
    \nn\\&\hspace{-1cm}
    -4\,C_5 \,
    \left(
        (\phi^\dagger \phi) (\phi^\dagger 
        \left[D_\mu,D_\nu\right]\left[D^\mu,D^\nu\right]\phi)
        +2(\phi^\dagger \phi) 
        (D_\mu\phi^\dagger \left[D^\mu,D^\nu\right]D_\nu \phi)
    \right)
    \nn\\&\hspace{-1cm}
    +2i\,C_\text{TLO}
    \Big(
        (\phi^\dagger \phi) 
        \big(\phi^\dagger 
        (g W_{\mu\nu} - g'B_{\mu\nu})
        \left[D^\mu,D^\nu\right]\phi\big)
        +4(\phi^\dagger \phi)\big(D^\mu\phi^\dagger 
        (g W_{\mu\nu} - g'B_{\mu\nu})
        D^\nu\phi\big)
    \Big)\,,\nn
\end{align}
where} $W_{\mu\nu} = W^a_{\mu\nu}\sigma^a$.

\section{A top-down look at \texorpdfstring{$h\rightarrow \gamma Z$}{h -> gamma Z}}\label{sec:FullModel}

Having studied the Higgs decays from the EFT perspective, we now analyse in more depth specific UV scenarios which can generate the logarithmic contribution at dimension eight. This is a particularly relevant question nowadays, as evidence for $h\to \gamma Z$ has been found by ATLAS and CMS~\cite{ATLAS:2023wqy,ATLAS:2023yqk}.
Regardless of whether a BSM signal is found, or the SM proves to be enough to describe this process, important bounds for heavy physics scenarios can be achieved through the SMEFT parameterization, making it imperative to understand the effect of higher-order terms in these processes. 
We will therefore study if these contributions can be quantitatively important and if they should be taken into account in phenomenological studies.

We begin this section by computing the amplitude for $h\to \gamma Z$ in the UV model with a neutral $SU(2)_L$ triplet scalar
and we perform the heavy mass expansion to relate to the EFT results. 
In Section~\ref{sec:cust}, we make the connection between the logarithmic enhancement of the $h\to\gamma Z$ decay at dimension eight and custodial symmetry breaking in scalar extensions of the SM. We point out that this correlation is broken in UV models with heavy vector bosons.
Finally, we extend the analysis from amplitudes to observables in Section~\ref{sec:5.outlook}, and we conclude that there exist observables in which the renormalization of the $h\to\gamma Z$ decay has a significant impact.

\subsection{A UV theory calculation of \texorpdfstring{$h\rightarrow \gamma Z$}{h -> gamma Z}}\label{sec:triplet_model}

Let us consider the SM extended by an $SU(2)_L$ triplet scalar $\Xi$ with zero hypercharge, which we matched to the SMEFT in Section~\ref{sec:matching}. 
The Lagrangian for $\Xi$ is introduced in Eq.~\eqref{eq:TripletLag}, which we also state here for clarity:
\be
\La_\Xi = \frac{1}{2}D_\mu\Xi^a D^\mu \Xi^a - \frac{1}{2}M^2\Xi^a\Xi^a - \kappa_\Xi \Xi^a \phi^\dagger \sigma^a \phi\,.
\ee
The scalar potential is chosen such that  no extra sources of EWSB are present and the model can be projected onto the SMEFT, as studied in detail in Ref.~\cite{Cohen:2020xca}. The analysis of more general cases when this restriction is lifted would require matching onto HEFT.

After EWSB, both the Higgs and the triplet acquire a vev and will therefore mix due to the interaction term proportional to $\kappa_\Xi$. 
The mixing between neutral (charged) scalars is parametrized by the angle $\theta_0$ ($\theta_+$), and we name the physical scalars $h$, $\xi^0$ and $\xi^\pm$, corresponding to the SM-like Higgs and novel neutral and charged scalars, respectively. For more details of EWSB in this model, we refer the reader to Appendix~\ref{app:Triplet}.

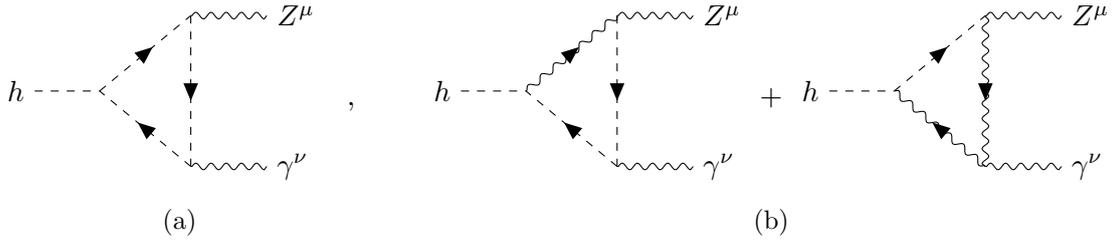
\begin{figure}[t]
    \centering
\begin{subfigure}{0.35\textwidth}
        \centering
\adjustbox{valign=m}{\begin{tikzpicture}
\begin{feynman}
\vertex (Ml) {$h$};
\vertex[right = 1.1cm of Ml] (Mc);
\vertex[right = 1.2cm of Mc] (Mr);
\vertex[above = 1 of Mr] (Va);
\vertex[below = 1 of Mr] (Vb);
\vertex[right = 1. cm of Va] (VZ) {$Z^\mu$};
\vertex[right = 1. cm of Vb] (Vg) {$\gamma^\nu$};
\diagram*{
(Ml) -- [scalar] (Mc),
(Mc) -- [charged scalar] (Va) -- [photon] (VZ),
(Va) -- [charged scalar] (Vb) -- [photon] (Vg),
(Vb) -- [charged scalar] (Mc)
};
\end{feynman}
\end{tikzpicture}}\  \hspace*{0.5mm} , \hspace*{-0.5mm}
\caption{}
        \label{eq:I_scalar}
    \end{subfigure}
    \hfill
    \begin{subfigure}{0.62\textwidth}
        \centering
\adjustbox{valign=m}{\begin{tikzpicture}
\begin{feynman}
\vertex (Ml) {$h$};
\vertex[right = 1.1cm of Ml] (Mc);
\vertex[right = 1.2cm of Mc] (Mr);
\vertex[above = 1 of Mr] (Va);
\vertex[below = 1 of Mr] (Vb);
\vertex[right = 1. cm of Va] (VZ) {$Z^\mu$};
\vertex[right = 1. cm of Vb] (Vg) {$\gamma^\nu$};
\diagram*{
(Ml) -- [scalar] (Mc),
(Mc) -- [charged boson] (Va) -- [photon] (VZ),
(Va) -- [charged scalar] (Vb) -- [photon] (Vg),
(Vb) -- [charged scalar] (Mc)
};
\end{feynman}
\end{tikzpicture}}\ +
\adjustbox{valign=m}{\begin{tikzpicture}
\begin{feynman}
\vertex (Ml) {$h$};
\vertex[right = 1.1cm of Ml] (Mc);
\vertex[right = 1.2cm of Mc] (Mr);
\vertex[above = 1 of Mr] (Va);
\vertex[below = 1 of Mr] (Vb);
\vertex[right = 1. cm of Va] (VZ) {$Z^\mu$};
\vertex[right = 1. cm of Vb] (Vg) {$\gamma^\nu$};
\diagram*{
(Ml) -- [scalar,] (Mc),
(Mc) -- [charged scalar] (Va) -- [photon] (VZ),
(Va) -- [charged boson] (Vb) -- [photon] (Vg),
(Vb) -- [charged boson] (Mc)
};
\end{feynman}
\end{tikzpicture}}
        \caption{}
        \label{eq:I_mixed}
    \end{subfigure}
    \caption{
    Feynman diagrams relevant to the computation of $\Amp_{H,\text{scalar}}$ (a) and $\Amp_{H,\text{mixed}}$ (b) in 
    Eqs.~\hyperref[eq:I_full_expanded_scalar]{(\ref*{eq:I_full_expanded_scalar}, }\hyperref[eq:I_full_expanded_mixed]{ \ref*{eq:I_full_expanded_mixed})}.   
    Diagrams that contribute only to the Lorentz structure $\eta^{\mu\nu}$ have been omitted, because the full integral can be reconstructed from the Lorentz structure $p_Z^\nu p_\gamma^\mu$, see~\cite{Hue:2017cph,Fontes:2014xva} and 
    Appendix~\ref{app:loopresults}. 
    Diagrams with reversed arrows should also be included.
    }
    \label{fig:1loopDiagrams}
\end{figure}

The direct contributions to the amplitude of $h\to \gamma Z$ from loops with additional charged scalar particles have previously been computed in~\cite{Degrande:2017naf,Hue:2017cph}.
We split the amplitude according to the type of diagrams that contribute,
\be\label{eq:I}
i \Amp_H = i \Amp_{H,\text{scalar}} + i\Amp_{H,\text{mixed}},
\ee
where $\Amp_{H,\text{scalar}}$ only contains pure scalar $\xi^{\pm}$ loops and $\Amp_{H,\text{mixed}}$ involves mixed loops of scalars and $W^{\pm}$ bosons, see Fig.~\ref{fig:1loopDiagrams}. 
The full results are reported in Appendix~\ref{app:loopresults}. 
The subscript $H$ is included in $i\Amp_H$ to emphasize that we only include diagrams that have at least one heavy particle in the loop; other contributions with only SM particles in the loop and \textit{indirect} contributions from modifications of SM parameters are not included. We come back to the latter in Section~\ref{sec:5.outlook} and discuss them in more detail in Appendices~\ref{app:EFT} and \ref{app:Triplet}. 

In the limit that the mass of the triplet, which is given by $M$ at leading order in $\kappa_\Xi$, is much larger than the EW scale, the amplitudes in Eq.~\eqref{eq:I} yield 
\begin{align}
\label{eq:I_full_expanded_scalar}
\Amp_{H,\text{scalar}} & = \frac{e^2 \cot \theta_W}{48\pi^2} \left(\frac{\kappa_\Xi}{v}\right)^2\frac{v^3}{ M^4}\left[1+\frac{m_Z^2+2m_h^2}{15 M^2}\right]+\order{\frac{1}{M^8}},\\
\Amp_{H,\text{mixed}} & =  -\frac{e^2}{16\pi^2 s_{2W}}\left(\frac{\kappa_\Xi}{v}\right)^2\frac{v^3}{M^4}\left[1 + \frac{2m_Z^2 - 5m_h^2 - 54 m_W^2 + 54m_W^2\log\left(\frac{M^2}{m_W^2}\right)}{9 M^2}\right] \nonumber\\
&\qquad\qquad\qquad\qquad\qquad\qquad\qquad\qquad\qquad\qquad\qquad\qquad+\order{\frac{1}{M^8}}, \label{eq:I_full_expanded_mixed}
\end{align}
where we have stripped off the common factor of $2(p_Z^\nu p_\gamma^\mu - (p_Z\cdot p_\gamma)\eta^{\mu\nu})$, with $p_{\gamma,Z}$ being the 4\nobreakdash-momenta of the photon and the $Z$ boson. In these expressions, we have omitted terms of order $\order{\kappa_\Xi^4}$, which do not come with a logarithm.
It is worth mentioning that the contribution to $h\to \gamma\gamma$ can be obtained by substituting the $Z$ boson in $i\Amp_{H,\text{scalar}}$ by another photon, \textit{i.e.}~taking $m_Z\to 0$ and $e^2\cot\theta_W\to e^2$. 
Mixed diagrams (Fig.~\ref{eq:I_mixed}) do not contribute to $h\to \gamma\gamma$, since the electromagnetic current is diagonal in the fields.

When translating the results to an EFT expansion, terms proportional to 
$\kappa_\Xi^2M^{-4}$ and
$\kappa_\Xi^2M^{-6}$ correspond to mass dimension six and eight, respectively. 
The logarithm in the expansion of $\Amp_{H,\text{mixed}}$, given by
\be\label{eq:Dim8_log_triplet}
\Amp_{H,\text{log}}^{(8)}=- \frac{3g'g^3}{64\pi^2}\left(\frac{\kappa_\Xi}{v}\right)^2\frac{v^5}{M^6}\,\log\!\left(\frac{M^2}{m_W^2}\right),
\ee
then follows from the tree-level matching result in Eq.~\eqref{eq:gammaZresults}, setting $\Lambda = M$ and $\mu=m_W$, while the identification of the other terms in 
Eqs.~\hyperref[eq:I_full_expanded_scalar]{(\ref*{eq:I_full_expanded_scalar}, }\hyperref[eq:I_full_expanded_mixed]{ \ref*{eq:I_full_expanded_mixed})}
in an EFT computation would require a full one-loop matching as discussed in the introduction.
In the scalar triplet model, the absence of running contributions to $h\to\gamma\gamma$  \eqref{eq:gammagammamixingRes} follows from the absence of mixed diagrams in that case.

\begin{figure}[t]
\centering
\includegraphics[width=1.0\textwidth]{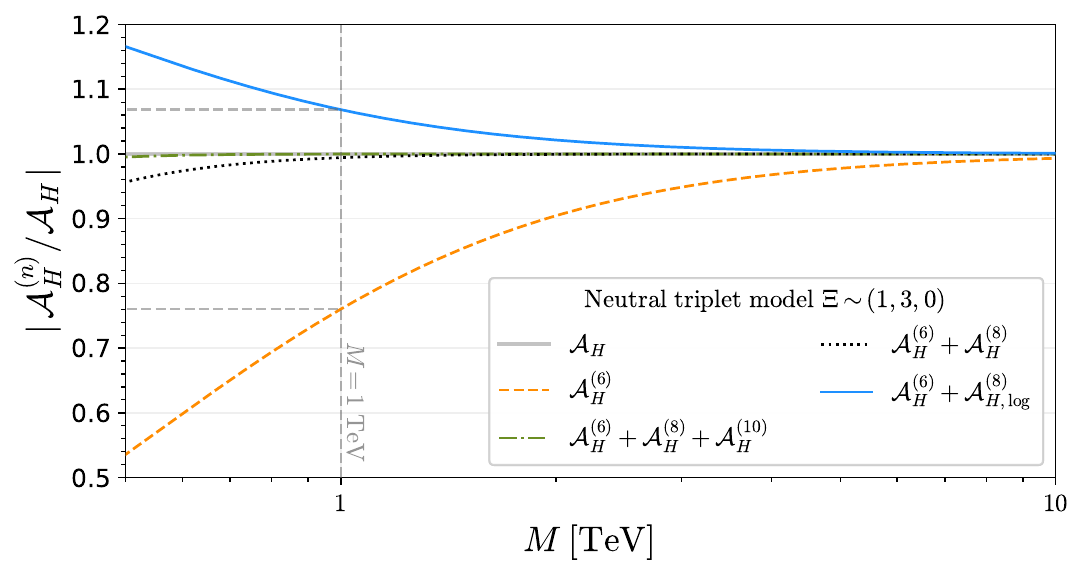}
\caption{Plot of the amplitude $\Amp_H$, expanded up to different orders in the triplet mass $M$, normalized by the result in Eq.~\eqref{eq:I}. We emphasize that $\Amp_H$ contains only loop diagrams with at least one heavy particle inside the loop, while the full amplitude is discussed in Section~\ref{sec:5.outlook}.
}\label{fig:Triplet}
\end{figure}

What remains is a quantitative measure of the importance of this logarithm. In order to see how relevant the effect in Eq.~\eqref{eq:Dim8_log_triplet} is compared to the contribution at dimension six, we expand the amplitude $\Amp_H$ as
\be\label{eq:I_expansion}
\Amp_H = \Amp^{(6)}_H + \Amp^{(8)}_H + \Amp^{(10)}_H + \cdots,
\ee
where $\Amp^{(n)}_H$ corresponds to the EFT expansion at dimension $n$. We then plot this amplitude expanded up to several different orders in Fig.~\ref{fig:Triplet}, normalized by the full result without performing the mass expansion,
obtained from Eqs.~\eqref{eq:I_scalar_result} and \eqref{eq:I_mixed_result}. The grey line is the full result, 
which is equal to one given our normalization.
The dashed orange, dotted black and dash-dotted green curves are, respectively, the results truncated up to dimension six, eight and ten. 
In solid blue we present the dimension-six piece augmented by the logarithmic contribution at dimension eight in Eq.~\eqref{eq:Dim8_log_triplet}. The addition of the logarithm to $\Amp^{(6)}_H$ brings the expansion much closer to the full result, although overshooting it by a relatively small amount. 

We highlight in the plot the value for which $M=1$ TeV with the vertical gray dashed line. For this value of the mass, 
$\Amp^{(6)}_H$ is off by approximately 24\% from the full result, while $\Amp^{(6)}_H+\Amp^{(8)}_{H,\text{log}}$ is only 7\% larger than the full amplitude. Adding the non-logarithmic part at dimension eight, which has the opposite sign, brings the result closer to the full theory. We stress that, in this case, the dimension-eight contribution to the amplitude is dominated by the logarithm, which itself corresponds to 31\% of the full result (in magnitude) versus the 7\% of the non-logarithmic part.  These results indicate that truncating the EFT at dimension six might lead to inaccurate results.Besides providing already a result close to the full theory, the further advantage of adding only the dimension eight logarithm $\Amp^{(8)}_{H,\text{log}}$ lies in the fact that it is easily computed from the tree-level matching results in Table~\ref{tab:matching-scalars} and the RGEs in Eqs.~\eqref{eq:rgehgammagamma} and \eqref{eq:rgehgammaZ}. The complete contributions at dimension six and eight, on the other hand, require a full one-loop matching calculation.

\subsection{Custodial symmetry and UV patterns}\label{sec:cust}

An immediate concern regarding the neutral  scalar triplet model introduced in the previous section, is that the relevant coupling, $\kappa_\Xi$, breaks custodial symmetry at tree level by contributing to the $W$ boson mass without affecting the $Z$ boson mass (see Eq.~\eqref{eq:mW_triplet_model}), resulting in large contributions to $\delta\rho = \rho -1 = \left( \frac{m_W^2}{c_W^2 m_Z^2} - 1\right)$~\cite{deBlas:2022hdk,Paul:2022dds}.

In order to avoid bounds from custodial symmetry breaking while preserving a sizable logarithm, we can consider more elaborate UV models. The remaining scalar extension that can generate TLO mixing into the loop-level process $h\to \gamma Z$ is the complex triplet, $\Xi_1\sim(1,3,1)$, which also breaks custodial symmetry at tree level. This can be seen from the matching results in Table~\ref{tab:matching-scalars}, where the dimension-six WC $C_{\phi D}$, which parameterizes deviations of the $\rho$ parameter from unity,\footnote{Custodial breaking effects will also occur at dimension eight and higher~\cite{Corbett:2021eux}. For simplicity, when considering relations among UV couplings that would follow from custodial symmetry, we impose only that $C_{\phi D}$ is zero after tree-level matching, but higher order corrections to these relations will also exist.} 
is generated by both of these scalar extensions. One could try to restore custodial symmetry following the Georgi--Machacek model~\cite{Georgi:1985nv,Chanowitz:1985ug}, by considering an extension with both the neutral and complex triplets such that the contribution to $\rho$ vanishes, which implies $C_{\phi D} = 0$ at tree level in the EFT. However, the linear combination of dimension-eight WCs responsible for TLO mixing into LLO for the scalar extensions, $C_{\phi^4}^{(1)}-C_{\phi^4}^{(2)}$ (see Eq.~\eqref{eq:rgehgammaZ}), would equally vanish in this case. This follows from the results in Table~\ref{tab:matching-scalars}, where we observe that the generation of $C_{\phi^4}^{(1)}-C_{\phi^4}^{(2)}$ is correlated with that of $C_{\phi D}$ for scalar extensions.

Alternatively to the explicit matching results, this correlation can also be understood from the fact that in the Georgi--Machacek model, given the custodial transformations of the introduced scalars, one cannot draw the mixed Feynman diagram in Fig.~\ref{eq:I_mixed} -- the one responsible for the logarithmic behaviour. A mixed diagram exists in this model, akin to Fig.~\ref{eq:I_mixed},  but only where the external scalar is the fiveplet-custodial scalar, $H_0^5\to\gamma Z$~\cite{Degrande:2017naf}, not for the process of interest with a physical Higgs. That is, couplings needed to obtain the diagram of Fig.~\ref{eq:I_mixed}, \ie the vertices $\xi^+ W^- Z$ and $\xi^+ W^- h$, do not exist in the Georgi--Machacek model for the same $\xi^+$ and the Higgs. We remark that custodial symmetry was also found to offer an extra suppression to the contributions to $h\to\gamma Z$ in the context of Composite Higgs models~\cite{Azatov:2013ura}. 

In order to avoid this correlation between the generation of the dimension-eight RGEs and custodial symmetry breaking, we can consider heavy vector extensions for which this connection is not observed. For example, from the results of Eq.~\eqref{eq:gammaZresults} and Table~\ref{tab:matching-vectors}, we observe that, while $\mathcal{B}\sim(1,1,0)$ does not generate the TLO mixing into LLO effect, it generates $C_{\phi D}$ in a way that can cancel the contribution from $\Xi_1$. 
To achieve $\delta \rho=0$ at tree level, we can thus take a UV model with $\Xi_1$ and $\mathcal{B}$ and choose their couplings in Eqs.~\eqref{eq:ChargedTripletLag} and \eqref{eq:DarkPhotonLag} to satisfy $g_\mathcal{B}^2=2\,(\kappa_{\Xi_1}/M_{\Xi_1})^2$.
This model generates a logarithmic contribution to the $h\to \gamma Z$ decay, while custodial bounds are expected to be respected. 

Another suitable combination of fields which could achieve the logarithmic enhancement in the $h\to\gamma Z$ decay while respecting custodial bounds is the charged vector singlet $\mathcal{B}_1\sim(1,1,1)$ and the neutral scalar triplet of the previous section, with the couplings fixed by setting $g_{\mathcal{B}_{1}}^2=2(\kappa_{\Xi}/M_{\Xi})^2$. This relation between couplings allows for the cancellation of the contribution to $C_{\phi D}$ at tree level. Unlike the previous example, $\mathcal{B}_1\sim(1,1,1)$ also induces TLO mixing into LLO, but enforcing $C_{\phi D}=0$ does not cancel this effect, 
because both particles contribute to the logarithm with the same sign.
We have used Eq.~\eqref{eq:gammaZresults} and Table~\ref{tab:matching-vectors} to observe these patterns.

Let us finally remark that there is one remaining single-field extension capable of generating TLO mixing into LLO: the heavy neutral triplet $\mathcal{W}\sim(1,3,0)$. This model is the only one that does not generate $C_{\phi D}$ at tree level.%
    \footnote{This model could generate $C_{\phi D}$ if one allows for complex couplings~\cite{deBlas:2017xtg}, but we restrict our analysis to real couplings in this work.} A complete study of this heavy vector would need a complete model~\cite{Dekens:2021bro} including the details of the spontaneous symmetry breaking from which $\mathcal{W}$ obtains a mass and is therefore left for future work.

\begin{figure}[t]
\centering
\begin{subfigure}[t]{0.48\textwidth}
\centering
\includegraphics[width=\textwidth]{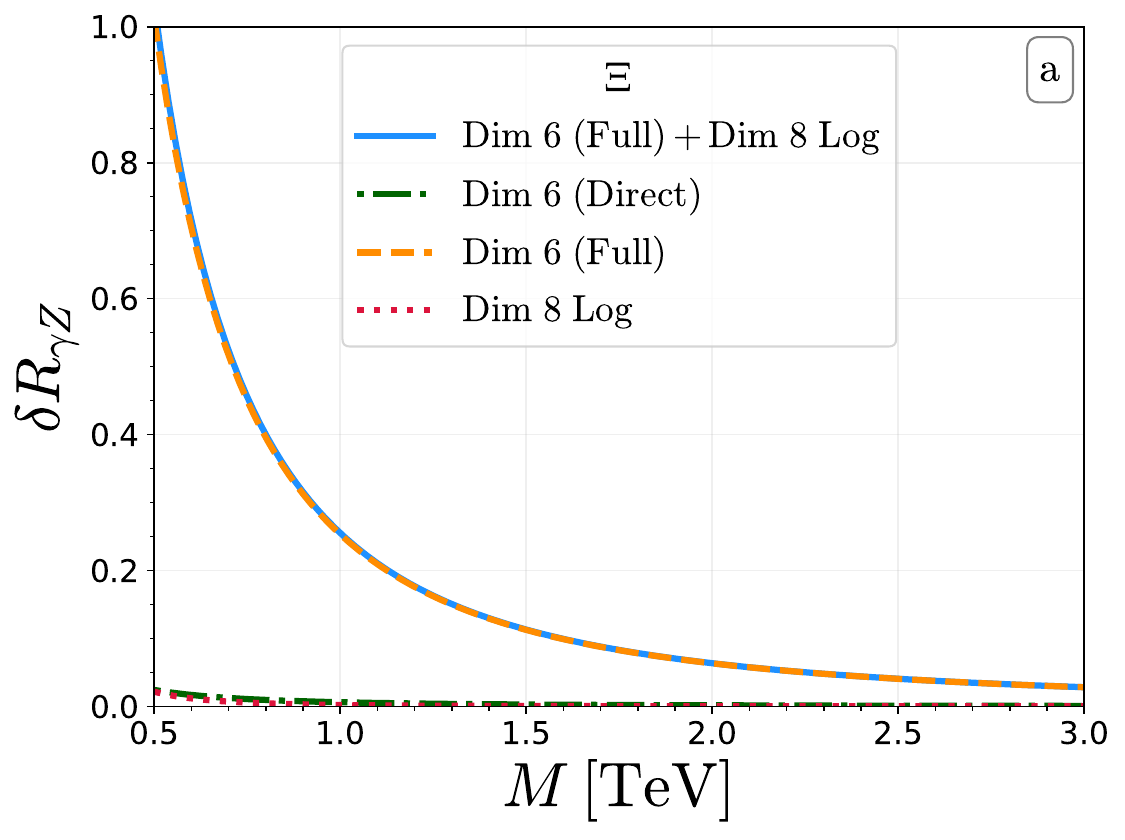}
{\phantomsubcaption\label{fig:sec53_a}}
\end{subfigure}
\begin{subfigure}[t]{0.49\textwidth}
\centering
\includegraphics[width=\textwidth]{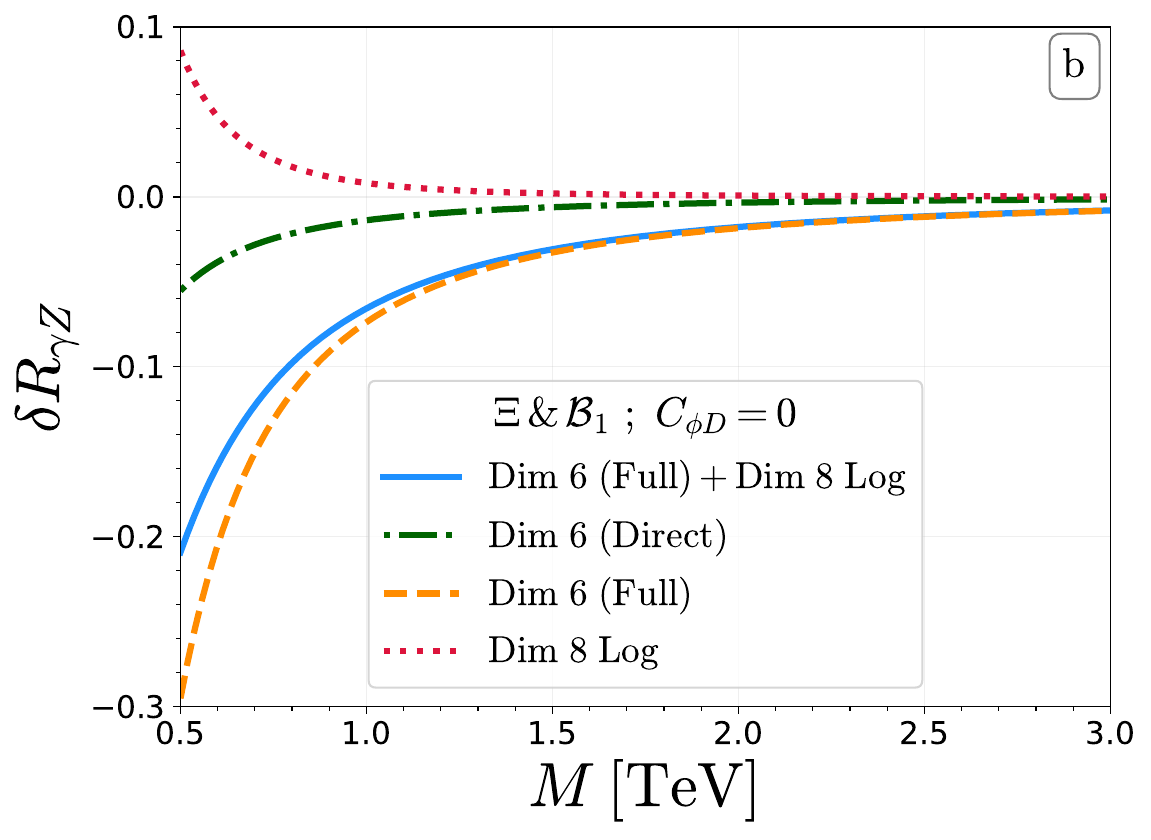}
{\phantomsubcaption\label{fig:sec53_b}}
\end{subfigure}
\begin{subfigure}[t]{0.48\textwidth}
\centering
\includegraphics[width=\textwidth]{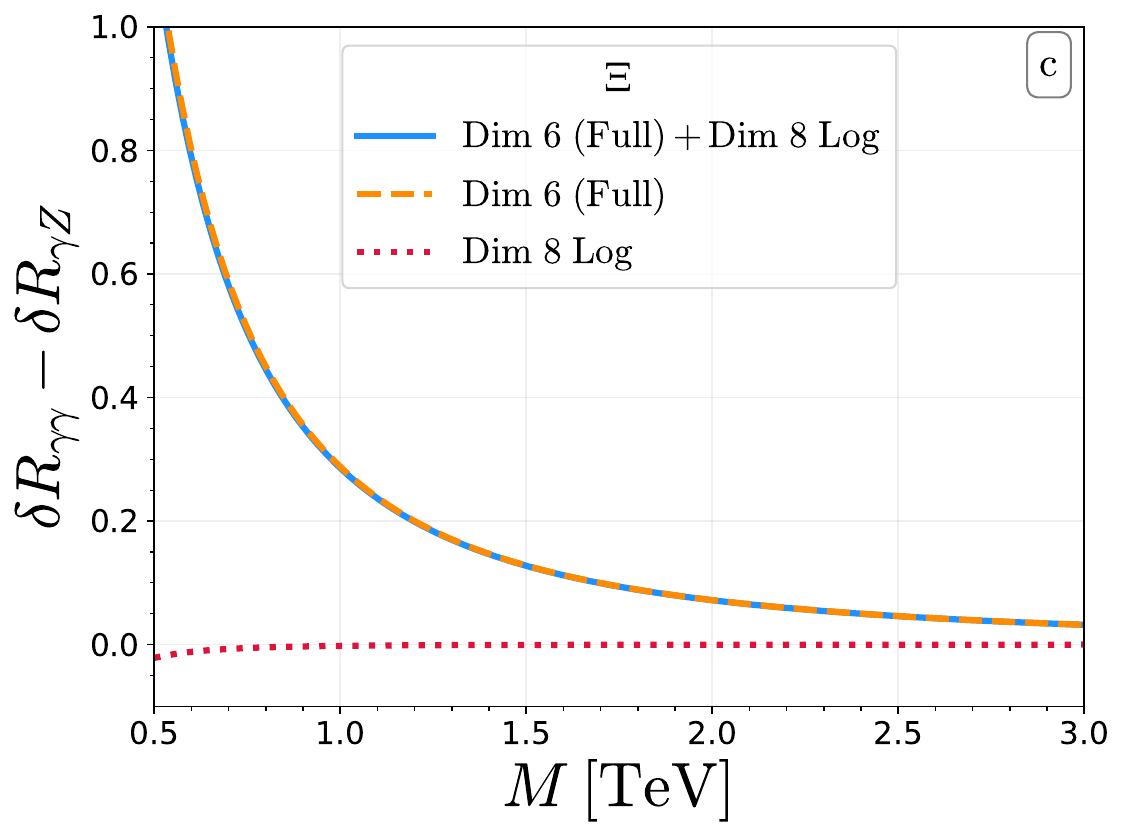}
{\phantomsubcaption\label{fig:sec53_c}}
\end{subfigure}
\begin{subfigure}[t]{0.49\textwidth}
\centering
\includegraphics[width=\textwidth]{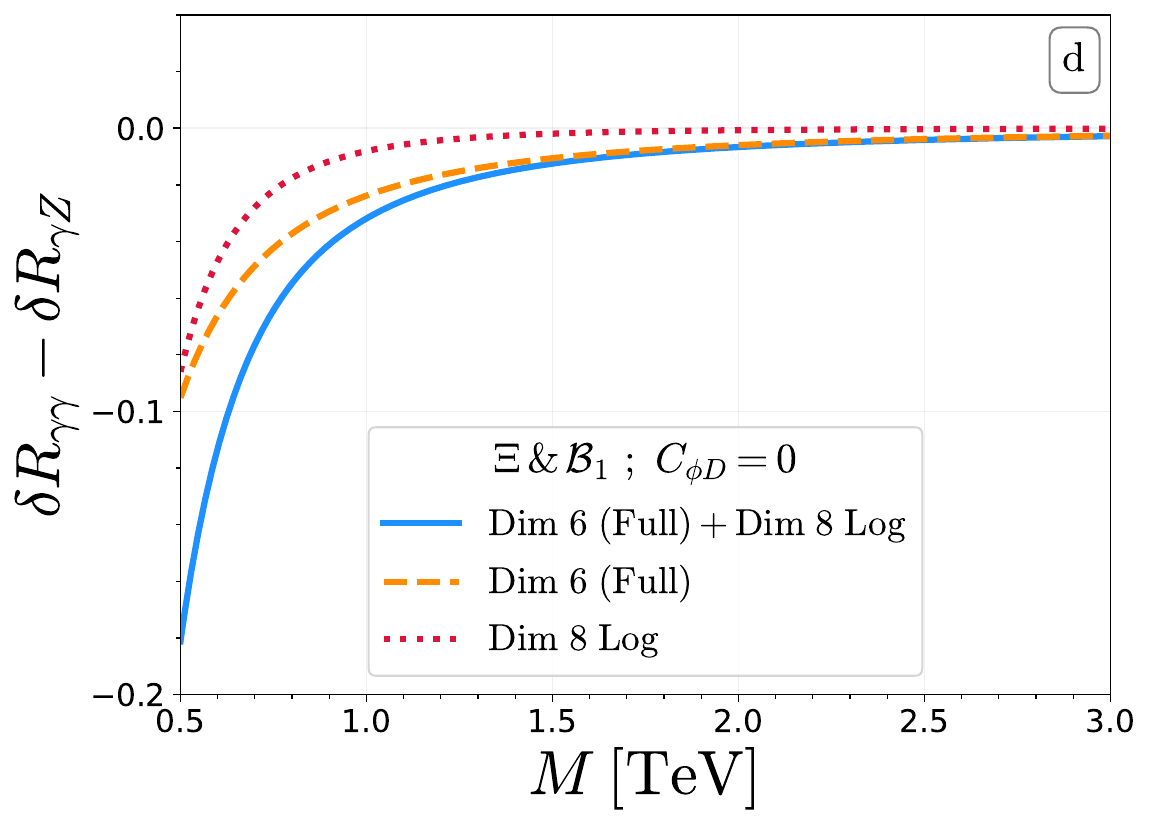}
{\phantomsubcaption\label{fig:sec53_d}}
\end{subfigure}
\caption{Evaluation of the deviations from the SM value of the $h\to \gamma Z$ decay width (panels \textbf{a} and \textbf{b}) and of the leading term in the expansion of the ratio introduced in Eq.~\eqref{eq:ratio} (panels \textbf{c} and \textbf{d}) as a function of the heavy mass $M$. Panels \textbf{a} and \textbf{c} show the results for the scalar triplet model, while panels \textbf{b} and \textbf{d} are for the scalar triplet plus the charged $SU(2)_L$ singlet vector in the custodial limit, $C_{\phi D}=0$ (further assumptions made during the matching calculation are explained in the text). The dot-dashed green lines represent only the direct contributions at dimension six, resulting from $X^2\phi^2$ operators; the dashed orange lines contain all dimension six terms, including also indirect contributions such as those arising from redefining SM parameters; the dotted red lines correspond to contribution from the dimension-eight logarithm; the solid blue lines are the sum of all of these contributions.
}\label{fig:Sec53}
\end{figure}

\subsection{Observable RG mixing effects}\label{sec:5.outlook}
Let us finally estimate the size of the contributions from dimension-eight RGEs to observables, paying close attention to models that avoid constraints from custodial symmetry breaking.
By considering only loop diagrams that involve the heavy scalar, Fig.~\ref{fig:Triplet} pointed to the relevance of the dimension-eight logarithmic contribution relative to the dimension-six one. 
However, a different picture emerges upon including all contributions to $h\to \gamma Z$, in particular those coming from tree-level generated operators which redefine SM parameters and fields. (We exemplified the origin of such indirect effects in Appendix~\ref{app:EFT}.)
Figure~\ref{fig:sec53_a} shows 
$\delta \mathcal{R}_{\gamma Z}$, which is defined as the deviation from the SM prediction of the $h\to\gamma Z$ decay width  (normalized by the SM value). We computed $\delta \mathcal{R}_{\gamma Z}$ using the numerical values obtained in Refs.~\cite[Eq.~4.10]{Dedes:2019bew} 
and~\cite[Eq.~4.16]{Hays:2020scx} (the former for the dimension-six piece and the latter for the tree-level contribution from dimension-eight WCs) trading the arbitrary WCs by matching to the neutral  scalar triplet model. 
We find that the effect of the dimension-eight RGE is around 
$1\%$ or less
of the full dimension-six contribution for 
${M_{\Xi}=1\,\mathrm{TeV}}$. 
Figure~\ref{fig:sec53_a} 
emphasizes the importance of  the indirect contributions at dimension six, which follow from a tree-level matching computation affecting the SM loop-level amplitude.
These dominate over both the dimension-eight logarithm but also over the dimension-six direct contributions (which follow from a one-loop matching computation to the $X^2\phi^2$ class of operators). 

One could conjure more complicated models to cancel these indirect dimension-six effects. However, a more interesting approach is to take a step back and return to a model-independent bottom-up approach, and consider an observable for which these indirect contributions partially cancel. 
To this end, we propose the ratio between the decay width of $h\to \gamma\gamma$ and $h\to \gamma Z$. In the SMEFT, considering again the numerical results from Refs.~\cite{Dedes:2019bew,Hays:2020scx}, including all significant dimension-six effects up to one-loop and only the logarithmic part at dimension eight.%
    \footnote{For these estimates, we have not considered non-logarithmic one-loop contributions at dimension eight, neither from dimension-six squared WCs~\cite{Hays:2020scx} nor from dimension-eight coefficients. We have also neglected dimension-six WCs which will not be generated by the considered UV extensions or 
    which are suppressed by Yukawa couplings (other than the top Yukawa)
    in their matching conditions, such as $C_{d\phi,33}$. Note that $\order{\delta\mathcal{R}^2}$ terms can also be numerically important, especially for the $C_{\phi D}^2$ term, but we dropped them here because they consist of two-loop corrections and in custodial-symmetric models which we will consider they are suppressed.}
This gives
{\allowdisplaybreaks
\begin{align}
\label{eq:ratio}
\frac{1 + \delta {\cal R}_{\gamma \gamma} }{ 1 + \delta {\cal R}_{\gamma Z}}  & = 1+\delta {\cal R}_{\gamma \gamma} - \delta {\cal R}_{\gamma Z}+\order{\delta\mathcal{R}^2}\nonumber\\
&= 1 - \left(\frac{1~\text{TeV}}{\Lambda}\right)^2 \left(0.12 C_{\phi D} - 0.02 C_{u\phi,33} + 0.049 \bar C_{\phi B} - 0.002 \bar C_{\phi W} - 0.024 \bar C_{\phi W B}\right)\nonumber\\
& + 0.0007 \left(\frac{1~\text{TeV}}{\Lambda}\right)^4 \left( 6C_{\mathrm{TLO}}+\frac38 C_{\phi^4}^{(1)} - \frac38 C_{\phi^4}^{(2)}  \right)\log\!\left(\frac {m_h}{\Lambda} \right)+\order{\delta\mathcal{R}^2},
\end{align}
where} for the barred coefficients $\bar C_{\phi X}$ we have factored out the loop factor and the corresponding powers of gauge couplings, \textit{i.e.}~$C_{\phi X}= g_X^2\bar C_{\phi X}/(16\pi^2)$, with $g_X^2 = g^2,{g'}^2,g'g$ for $X=W,B,WB$, respectively. In the second line we have already substituted the dimension-eight WCs responsible for the direct contribution to the $h\to\gamma Z$ decay by their RGEs, including only the terms proportional to TLOs,
following Eq.~\eqref{eq:rgehgammaZ-basis2}; $C_{\mathrm{TLO}}$ is the only potentially tree-level generated coefficient introduced in the new basis defined in Eq.~\eqref{eq:newBasisCouplings}. 

The ratio defined in Eq.~\eqref{eq:ratio} is still very sensitive to custodial symmetry breaking, as $C_{\phi D}$ is the WC with the largest numerical pre-factor. 
Indeed, in Fig.~\ref{fig:sec53_c} we show that this ratio is not yet sensitive enough to dimension-eight effects for the $\Xi$ extension, with the result still being  completely dominated by the dimension-six contributions. A UV direction which could result in a larger relative contribution of the dimension-eight logarithm corresponds to models where custodial symmetry is preserved at tree level, which can be achieved without canceling the dimension-eight RGE effect, as we discussed in the previous section.

Focusing on these custodial symmetric UV scenarios, we can consider a specific scenario of the previous section, a custodial model with $\mathcal{B}_1$ and $\Xi$. We choose $M=M_\Xi=M_{\B_1}$ for simplicity. Following from the tree-level and loop-level matching computations from Refs.~\cite{deBlas:2017xtg,Guedes:2023azv} -- the latter only for the scalar -- the dimension-six contributions to Eq.~\eqref{eq:ratio} cancel, up to $\bar C_{\phi X}$, resulting in
\be
\label{eq:ratioUVmodel}
\frac{1 + \delta {\cal R}_{\gamma \gamma} }{ 1 + \delta {\cal R}_{\gamma Z}}  \approx 1 -0.024 \left(\frac{1~\text{TeV}}{M}\right)^2 \left(\frac{\kappa_\Xi}{M}\right)^2  + 0.004\left(\frac{1~\text{TeV}}{M}\right)^4\left(\frac{\kappa_\Xi}{M}\right)^2 \log\!\left(\frac{m_h}{M}\right) + \order{\delta\mathcal{R}^2},
\ee
where we used $g_{\mathcal{B}_{1}}^2=2(\kappa_{\Xi}/M)^2$ to obtain $C_{\phi D}=0$ and have further assumed that $\bar C_{\phi X}= \kappa^2_\Xi/M^2$. 
The latter assumption
was made since there is no one-loop matching dictionary for heavy vectors. 
We remark, however, that this assumption would be conservative in the neutral scalar triplet model, in which
the $\bar C_{\phi X}$ WCs are suppressed by an additional numerical factor of 1/16~\cite{Guedes:2023azv}.
A complete theory behind the mass generation of the heavy vector would be required to perform this one-loop matching. 

Under the assumption of $\bar C_{\phi X}= \kappa^2_\Xi/M^2$, Figs.~\ref{fig:sec53_b} and \ref{fig:sec53_d} show how much more sensitive the new observable can be to dimension-eight effects in the custodial limit. At this stage, it is reasonable to suggest that the dimension-eight contribution can provide a relevant effect: for particles with $M=1$ TeV, the dimension-eight part corresponds to $\approx 25\%$ of the full result. Given the extreme precision with which $\delta \mathcal{R}_{\gamma\gamma}$ and $\delta \mathcal{R}_{\gamma Z}$ are expected to be measured at the FCC~\cite{deBlas:2019rxi}, the ratio of these observables will be able to probe dimension-eight effects, and it can thereby help to distinguish new physics models.

\section{Conclusion}\label{sec:conclusion}

We have studied the radiative Higgs decays and their renormalization in the SMEFT, assuming weakly coupled renormalizable UV physics. 
Under this assumption, the Higgs decay amplitudes $h\to gg$, $h\to\gamma\gamma$ and $h\to \gamma Z$ are loop-level processes, which translates to loop-suppression factors in some of the EFT parameters.
At subleading order in the EFT expansion (\textit{i.e.}~mass dimension eight), we concluded that the renormalization group equations may lead to sizable -- logarithmically enhanced -- contributions to observables related to the $h\to\gamma Z$ decay in various models.
This therefore provides a promising avenue for future more detailed phenomenological studies.

The logarithmic enhancement of $h\to\gamma Z$ is the result of renormalization group
mixing of operators that can be generated by matching at tree level into operators responsible for the loop-level decay amplitude.
This one-loop effect is absent at any mass dimension in the other Higgs decays ($h\to gg$ and $h\to \gamma \gamma$), 
which follows from results in generic UV completions, as well as for $h\to\gamma Z$ 
at dimension six. 
We established the renormalization of $h\to\gamma Z$ with explicit tree-level matching calculations, supplemented by the known one-loop RGEs from the literature. 
For simplicity, we worked in the absence of Yukawa couplings and CP violating operators, and we chose heavy field extensions in which the SM gauge symmetry is not broken in the UV, such that we can match onto the SMEFT.
Under these assumptions, we considered all bosonic weakly coupled renormalizable single-particle UV extensions in which the mixing of tree-level generated operators into the Higgs decays could be expected. 
We confirmed this effect in four of the models.
Unless cancellations occur, such as in the Georgi--Machacek model, we generically expect this effect also in multi-particle extensions whenever loop diagrams exist with both heavy and light particles in the loop (similar to those in Fig.~\ref{eq:I_mixed}).

Our matching conditions for the scalar extensions and the singlet vector extension in Tables~\ref{tab:matching-scalars} and~\ref{tab:offShellMatching} agree with existing results in the literature. For the charged heavy vectors, we obtained new results by including an additional interaction -- the magnetic dipole term -- between heavy vectors and gauge bosons, defined in Eq.~\eqref{eq:additional_interactions_vector}.
It was found that this interaction plays a crucial role in the consistency of our  analysis: 
the vanishing of the $h\to\gamma Z$ decay amplitude at tree level requires its coefficient (which we left free in the matching calculations) to be fixed to the value
that also respects perturbative unitarity in the full theory. The results thus indicate that a relatively large $h\to\gamma Z$ amplitude at low energies 
 -- resulting from a tree-level matching contribution --
could signal the breakdown of perturbative unitarity in the UV.
In addition, the non-renormalization of the $h\to\gamma\gamma$ amplitude also requires the UV (with heavy vectors) to respect tree-level perturbative unitarity. 
We find that dimension eight is the leading mass dimension at which the additional interaction affects matching calculations at tree level.

In a generic dimension-eight basis (such as the basis of Ref.~\cite{Murphy:2020rsh}),
both the one-loop suppression of the Higgs decays and the non-renormalization of $h\to \gamma \gamma$ are encoded through intricate cancellations between multiple WCs that are generated in a correlated way in tree-level matching calculations. 
The verification of such relations therefore constitutes a valuable consistency check on matching calculations, operator basis transformations and the RGEs. 
We made the correlations between the WCs more transparent by proposing an alternative operator basis.
This basis illuminates that there is only one CP-even tree-level generated linear combination of WCs of all operators in the $X^2 \phi^4$ and $X\phi^4D^2$ classes. All other directions in the parameter space correspond to loop-level processes in the UV, such as the Higgs decays. This is consistent with, but refines the classification of Ref.~\cite{Craig:2019wmo}.
The improved classification could motivate the use of a restricted set of EFT parameters in phenomenological studies,
instead of naively constraining all WCs from a purely bottom-up approach.

With the complete set of models which result in 
tree-level generated operators mixing into the $h\to \gamma Z$ decay (disregarding the Yukawa couplings),
we studied the relation between UV symmetries and the occurance of this new renormalization effect.
We observed that for scalar extensions, the logarithmic enhancement is correlated with custodial symmetry breaking in the UV, leading to important constraints. To break this correspondence, heavy vector particles have to be included.

Our first study on the phenomenological impact of the renormalization of the logarithmically enhanced $h\to\gamma Z$ amplitude motivates future research in this direction. The ratio of observables in Eq.~\eqref{eq:ratioUVmodel} enhances the relative importance of the dimension-eight logarithmic contribution to a potentially observable level. 
It would be useful to reconsider this quantitative prediction in a UV model that includes the mass-generation mechanism of the heavy vector boson 
$\mathcal{B}_1$. 
Furthermore, for a consistent treatment it would also be necessary to include the non-logarithmic one-loop matching contributions from integrating out the heavy vector fields.

Finally, we have assumed the absence of 
CP-violating and fermionic effects. We do not expect that these effects lead to drastically different conclusions, but it would be worthwhile to complete our study by including all possible corrections to the Higgs decays. 
We leave this, and a systematic study of the UV scenarios which could maximize the relevance of the dimension-eight contributions for future work.

\section*{Acknowledgments}

We thank
Mikael Chala,
Mehmet A.\ G\"um\"u\c{s},
Minyuan Jiang,
Aneesh Manohar, Matheus Martines, 
Julie Pag\`es,
Maria Ramos,
and Jos\'e Santiago for useful discussions. 
The work of C.G. and G.G. is supported by the Deutsche Forschungsgemeinschaft under Germany’s Excellence Strategy EXC 2121 “Quantum Universe” – 390833306, as well as
by the grant 491245950.
J.R.N.\ is supported by the Deutsche Forschungsgemeinschaft (DFG, German Research Foundation) -- Projektnummer 417533893/GRK2575 ``Rethinking Quantum Field Theory''.
G.M.S. acknowledges financial support from "Fundação de Amparo à Pesquisa do Estado de São Paulo" (FAPESP) under contracts 2020/14713-2 and 2022/07360-1.
This project has received funding/support from the European Union’s Horizon 2020 research and innovation programme under the Marie Sk\l{}odowska-Curie grant agreement No 860881-HIDDeN.
This project has received funding from the European Union’s Horizon Europe research and innovation programme under the Marie Sk{\l}odowska-Curie Staff Exchange grant agreement No 101086085 -- ASYMMETRY.

\appendix

\section{EFT contributions from redefining SM fields and parameters}\label{app:EFT}

In this appendix, we exemplify the indirect effects that result from redefinitions of SM parameters after spontaneous symmetry breaking, due to the presence of higher-dimensional operators.
Such effects are complementary to the direct effects considered in Section~\ref{sec:higgsdecays}, and they are relevant to the determination of the full one-loop decay rates considered in Section~\ref{sec:5.outlook}. In that section, we use the results from Refs.~\cite{Dedes:2019bew,Hays:2020scx}.
Moreover, the same redefinitions also affect the coefficients of EFT operators, leading to the terms at quadratic order in the dimension-six WCs given in Eqs.~\eqref{eq:h_gg}, \eqref{eq:h_gammagamma} and \eqref{eq:h_gammaZ}.

Let us consider the subset of operators in 
Eq.~\eqref{eq:La_EFT} that contain two field-strength tensors that are either $B_{\mu\nu}$ or $W^a_{\mu\nu}$. The redefinitions of gluon fields and associated couplings are considerably simpler.
(We consider pure-Higgs operators, which generate a redefinition of the Higgs field, in Eq.~\eqref{eq:R_h} below.) The considered Lagrangian is
\al{\label{eq:La_EFT_KT}
\La_{X\phi} & = -\frac{1}{4}B_{\mu\nu}B^{\mu\nu}-\frac{1}{4}W^a_{\mu\nu}W^{a\mu\nu}\\
&+\frac{C_{\phi W}}{\Lambda^2}|\phi|^2W^a_{\mu\nu}W^{a\mu\nu} + \frac{C_{\phi B}}{\Lambda^2}|\phi|^2B_{\mu\nu}B^{\mu\nu} + \frac{C_{\phi WB}}{\Lambda^2}(\phi^\dagger \sigma^a\phi)B_{\mu\nu}W^{a\mu\nu}\\
& + \frac{C_{\phi^4 W^2}^{(1)}}{\Lambda^4}|\phi|^4W^a_{\mu\nu}W^{a\mu\nu} + \frac{C_{\phi^4 W^2}^{(3)}}{\Lambda^4}(\phi^\dagger \sigma^a\phi)(\phi^\dagger \sigma^b\phi)W^a_{\mu\nu}W^{b\mu\nu}\\
& + \frac{C_{\phi^4 B^2}^{(1)}}{\Lambda^4}|\phi|^4B_{\mu\nu}B^{\mu\nu} + \frac{C_{\phi^4 WB}^{(1)}}{\Lambda^4}|\phi|^2(\phi^\dagger\sigma^a\phi)B_{\mu\nu}W^{a\mu\nu}.
}
After the Higgs takes a vev, all effective operators above contribute to the kinetic term of the gauge bosons,
\al{
\La_{X\phi} & \supset -\frac{1}{4}\left[1-\frac{2v^2C_{\phi B}}{\Lambda^2}-\frac{v^4C_{\phi^4B^2}^{(1)}}{\Lambda^4}\right] B_{\mu\nu} B^{\mu\nu}\\
&-\frac{1}{4}\left[1-\frac{2v^2C_{\phi W}}{\Lambda^2}-\frac{v^4C_{\phi^4W^2}^{(1)}}{\Lambda^4}\right] W^{1,2}_{\mu\nu}W^{1,2\mu\nu}\\
&-\frac{1}{4}\left[1-\frac{2v^2C_{\phi W}}{\Lambda^2}-\frac{v^4}{\Lambda^4}\left(C_{\phi^4W^2}^{(1)}+C_{\phi^4W^2}^{(3)}\right)\right] W^{3}_{\mu\nu}W^{3\mu\nu} \\
&-\left[\frac{v^2C_{\phi WB}}{2\Lambda^2}+\frac{v^4C_{\phi^4 WB}^{(1)}}{4\Lambda^4}\right]W^3_{\mu\nu}B^{\mu\nu}.
}
Note that the kinetic terms of $W^{3}$ and $W^{1,2}$ are corrected by effective operators in different ways at dimension eight.
The first step to go back to a canonical kinetic term, whilst preserving the form of the covariant derivatives, is to redefine the fields and gauge couplings as
\al{
B_\mu  &\to B_\mu\left[1-2 \frac{v^2}{\Lambda^2}C_{\phi B}- \frac{v^4}{\Lambda^4}C_{\phi^4 B^2}^{(1)}\right]^{-1/2}, \\
g' &\to g'\left[1-2 \frac{v^2}{\Lambda^2}C_{\phi B}- \frac{v^4}{\Lambda^4}C_{\phi^4 B^2}^{(1)}\right]^{1/2},\\
W^{3}_\mu  &\to W^{3}_\mu\left[1-2 \frac{v^2}{\Lambda^2}C_{\phi W}- \frac{v^4}{\Lambda^4}(C_{\phi^4 W^2}^{(1)}+C_{\phi^4 W^2}^{(3)})\right]^{-1/2},\\
g  &\to g\left[1-2 \frac{v^2}{\Lambda^2}C_{\phi W}- \frac{v^4}{\Lambda^4}(C_{\phi^4 W^2}^{(1)}+C_{\phi^4 W^2}^{(3)})\right]^{1/2},
\label{eq:V_gV_red}
}
where we use the notation of $g'$ and $g$ for the $U(1)_Y$ and $SU(2)_L$ gauge couplings. Similar redefinitions would apply for operators with gluons.
After these redefinitions, the kinetic term becomes
\al{
\La_{X\phi}&\supset  -\frac{1}{4}\left(B_{\mu\nu}~W^3_{\mu\nu}\right)\begin{pmatrix} 1 & -\s \\ -\s & 1 \end{pmatrix}\begin{pmatrix} B^{\mu\nu} \\ W^{3\mu\nu} \end{pmatrix},
}
with
\be
\s = -\frac{v^2}{\Lambda^2}\left[C_{\phi WB}\left(1+\frac{v^2}{\Lambda^2}(C_{\phi W}+C_{\phi B})\right)+\frac{v^2}{2\Lambda^2}C_{\phi^4 WB}^{(1)}\right].
\ee
In order to diagonalize the kinetic term above we ($i$) perform the ``usual'' weak rotation with the angles defined in terms of $g',~g$ and then ($ii$) perform a non-orthogonal rotation to remove the kinetic-mixing. For step ($i$) we the write
\be\label{eq:weak_angle}
\begin{pmatrix} B_\mu \\ W^3_\mu \end{pmatrix} = \begin{pmatrix} c_W & -s_W \\ s_W & c_W \end{pmatrix}\begin{pmatrix} \tilde A_\mu \\ \tilde Z_\mu \end{pmatrix},\quad 
c_W \equiv \frac{g}{\sqrt{g^2+{g'}^2}},\quad s_W \equiv \frac{g'}{\sqrt{g^2+{g'}^2}}.
\ee
Therefore,
\al{
\La_{X\phi} &\supset -\frac{1}{4}\left(B^{\mu\nu}~W^{3\mu\nu}\right)\begin{pmatrix} 1 & -\s \\ -\s & 1 \end{pmatrix}\begin{pmatrix} B_{\mu\nu} \\ W^3_{\mu\nu}\end{pmatrix}\\
& = -\frac{1}{4}\left(\tilde A^{\mu\nu}~\tilde Z^{\mu\nu}\right)\begin{pmatrix} 1-s_{2W} \s & -\s c_{2W} \\ -\s c_{2W} & 1+s_{2W} \s \end{pmatrix}\begin{pmatrix} \tilde A_{\mu\nu} \\ \tilde Z_{\mu\nu} \end{pmatrix}.
}
For step ($ii$) we use
\be\label{eq:non_orth_rot}
\begin{pmatrix} \tilde A_\mu \\ \tilde Z_\mu \end{pmatrix} = \begin{pmatrix} 1+\frac{s_{2W}}{2}\s + \frac{3s_{2W}^2}{8}\s^2 & c_{2W}\s\left(1+\frac{s_{2W}}{2}\s\right)  \\ 0 & 1-\frac{s_{2W}}{2}\s +\frac{\s^2}{2}\left(s_W^4+s_W^2c_W^2+c_W^4\right) \end{pmatrix}\begin{pmatrix} A_\mu \\ Z_\mu \end{pmatrix},
\ee
and consequently
\al{
\La_{X\phi} &\supset  -\frac{1}{4}\left(\tilde A_{\mu\nu}~\tilde Z_{\mu\nu}\right)\begin{pmatrix} 1-s_{2W} \s & -\s c_{2W} \\ -\s c_{2W} & 1+s_{2W} \s \end{pmatrix}\begin{pmatrix} \tilde A^{\mu\nu} \\ \tilde Z^{\mu\nu} \end{pmatrix}\\
 & = -\frac{1}{4}A_{\mu\nu}A^{\mu\nu} - \frac{1}{4} Z_{\mu\nu}Z^{\mu\nu},
}
which is finally in canonical form. The total rotation is the product of the transformations in Eqs.~\eqref{eq:weak_angle} and \eqref{eq:non_orth_rot}:
\be\label{eq:full_weak_rot}
\begin{pmatrix} B_\mu \\ W^3_\mu \end{pmatrix}=\begin{pmatrix} c_W\left(1+\frac{s_{2W}}{2} \s + \frac{3s_{2W}^2}{8}\s^2\right) & \ \ -s_W + \s c_W^3 + \frac{s_W}{2}\s^2\left(c_{2W}-\frac{3}{4}s_{2W}^2\right) \\ s_W\left(1+\frac{s_{2W}}{2} \s + \frac{3s_{2W}^2}{8}\s^2\right) & \ \ c_W-\s s_W^3 +\frac{c_W}{2}\s^2\left(c_{2W}+\frac{3s_{2W}^2}{4}\right) \end{pmatrix}\begin{pmatrix} A_\mu \\ Z_\mu \end{pmatrix}.
\ee
After this rotation, the mass matrix becomes diagonal with eigenvalues
\be
m_A^2=0,\quad m_Z^2 = \frac{v^2}{4}\left({g'}^2+g^2\right)\left[1 - s_{2W} \s + \s^2 \right].
\ee
This implies the following form of the effective coupling,
\be\label{eq:gZ}
g_Z^2 = \left({g'}^2+g^2\right)\left[1 - s_{2W} \s + \s^2 \right].
\ee
With that, we have rewritten the original Lagrangian in terms of the physical photon and $Z$ boson fields, $A_\mu$ and $Z_\mu$.
Let us reemphasize that the $Z$ boson mass receives further contributions from operators that were not part of Eq.~\eqref{eq:La_EFT_KT}, such as the ones considered in Eq.~\eqref{eq:La_phi} below.

The transformation in Eq.~\eqref{eq:full_weak_rot} modifies the structure of the covariant derivative. In order to put it in the same form as in the SM, that is
\be
D_\mu \supset ie\Q A_\mu + ig_Z Z_\mu (T^3-\tilde s_W^2 \Q),
\ee
with $\Q$, $T^3$ the $U(1)_{\rm EM}$ and third $SU(2)_L$ generators, we define the physical electric charge $e$ as
\be\label{eq:e_charge}
e = \e\left(1+\frac{s_{2W}^{\phantom{2}}}{2}\s + \frac{3s_{2W}^2}{8}\s^2\right),
\ee
where $\e=g' c_W = gs_W$. In addition, $\tilde s_W$ is defined by
\be\label{eq:sWtilde}
\tilde s_W^2 = s_W^2\left[1- \s\left(\frac{c_W}{s_W}-s_{2W}^{\phantom{2}}\right) - 2 \s^2c_W^2 c_{2W}^{\phantom{2}}\right].
\ee

In addition to redefinitions that arise from canonically normalizing the gauge boson kinetic terms, there are further indirect effects that arise from redefinitions of the Higgs field. Let us therefore consider the operators that contain only Higgs doublets (and derivatives),
\al{\label{eq:La_phi}
\La_\phi & = \frac{C_\phi}{\Lambda^2}|\phi|^6 + \frac{C_{\phi D}}{\Lambda^2}|\phi^\dagger D_\mu \phi |^2 + \frac{C_{\phi\Box}}{\Lambda^2}|\phi|^2\Box |\phi|^2\\
&+\frac{C_{\phi^8}}{\Lambda^4}|\phi|^8+\frac{C_{\phi^6}^{(1)}}{\Lambda^4}|\phi|^4D_\mu\phi^\dagger D^\mu \phi +\frac{C_{\phi^6}^{(2)}}{\Lambda^4}|\phi|^2 (D_\mu\phi^\dagger \sigma^a D^\mu \phi)(\phi^\dagger \sigma^a \phi).
}
After the Higgs takes a vev,  $\phi = (0,v+\tilde h)/\sqrt{2}$, its kinetic term becomes
\be
\La_\phi \supset \frac{1}{2}(\del \tilde h)^2\left\{1 - 2\frac{v^2}{\Lambda^2}\left[C_{\phi\Box} - \frac{C_{\phi D}}{4}\right] + \frac{v^4}{4\Lambda^4}\left[C_{\phi^6}^{(1)}+C_{\phi^6}^{(2)}\right]\right\},
\ee
which is normalized by
\be\label{eq:R_h}
\tilde h =h\left\{1 - 2\frac{v^2}{\Lambda^2}\left[C_{\phi\Box} - \frac{C_{\phi D}}{4}\right] + \frac{v^4}{4\Lambda^4}\left[C_{\phi^6}^{(1)}+C_{\phi^6}^{(2)}\right]\right\}^{-1/2},
\ee
where $h$ is the canonically normalized Higgs field.
The decays $h\to \gamma\gamma$ and $h\to \gamma Z$ can then be computed by keeping terms with one Higgs in Eq.~\eqref{eq:La_EFT} and after performing the transformations in Eqs.~\eqref{eq:V_gV_red}, \eqref{eq:full_weak_rot}, \eqref{eq:e_charge} and \eqref{eq:R_h}. The result of all these steps is given in Eqs.~\eqref{eq:h_gammagamma} and \eqref{eq:h_gammaZ} up to $\order{1/\Lambda^4}$.

\section{Weakly coupled UV models and off-shell matching results}
\label{app:UVmodels}

In this appendix we list the single-particle extensions to the SM considered in Section~\ref{sec:higgsdecays}. For each model, we integrate out the heavy particles at tree level and keep contributions up to dimension eight. We  match these results to the Green's basis of Refs.~\cite{Gherardi:2020det,Carmona:2021xtq,Chala:2021cgt},
resulting in Table~\ref{tab:offShellMatching}. Afterwards, we reduce them to the physical basis in Refs.~\cite{Grzadkowski:2010es,Murphy:2020rsh}. The matching results in the physical basis are presented in Table~\ref{tab:matching-scalars}.

In the Lagrangian for each UV model, $M$ denotes the mass of the heavy particle, $\kappa_{S},~\kappa_{\Xi},~\kappa_{\Xi_1}$ are dimensionful couplings for scalar extensions and $g_{\B},~g_{\B_1},~g_{\W},~g_{\W_1}$ are dimensionless couplings in the case of vector extensions. 
We have assumed that all these couplings are real.
Quantum numbers are presented with respect to the SM gauge group $SU(3)_c\times SU(2)_L\times U(1)_Y$.

\begin{itemize}
\item \textbf{Real scalar singlet} $S\sim (1,1,0)$
\be
\label{eq:SingletLag}
\La_S = \frac{1}{2}\del_\mu S\del^\mu S - \frac{1}{2}M^2S^2 - \kappa_S S|\phi|^2.
\ee
The SM with an additional singlet scalar field is one of the simplest extensions of the SM, therefore being the subject of many studies. We refer the reader to Refs.~\cite{Silveira:1985rk,Patt:2006fw,OConnell:2006rsp,Englert:2011yb} and references therein.

\item \textbf{Real scalar triplet} $\Xi\sim (1,3,0)$
\be
\label{eq:TripletLag}
\La_\Xi = \frac{1}{2}D_\mu \Xi^a D^\mu \Xi^a - \frac{1}{2}M^2\Xi^a\Xi^a - \kappa_\Xi\Xi^a \phi^\dagger\sigma^a\phi.
\ee
This model, first introduced in Ref.~\cite{Ross:1975fq}, has a rich phenomenology (see for instance Refs.~\cite{Gunion:1989ci,Lynn:1990zk,Blank:1997qa,Forshaw:2003kh,Chen:2006pb,Chankowski:2006hs,Chivukula:2007koj,Bandyopadhyay:2020otm,Cheng:2022hbo,FileviezPerez:2008bj,Chiang:2020rcv,Ellis:2023zim,Corbett:2021eux,Corbett:2017ieo,Krauss:2018orw,Khan:2016sxm,FileviezPerez:2022lxp,Ashanujjaman:2023etj}). 
We study this model in more detail in Section~\ref{sec:5.outlook} and Appendix~\ref{app:Triplet}.

\item \textbf{Complex scalar triplet} $\Xi_1\sim (1,3,1)$
\be\label{eq:ChargedTripletLag}
\La_{\Xi_1} = D_\mu \Xi^{a\dagger}_1 D^\mu \Xi^a_1 - M^2\Xi^{a\dagger}_1\Xi^a_1 -\left( \kappa_{\Xi_1}\Xi^a_1\phi^\dagger\sigma^a\tilde \phi + h.c.\right),
\ee
with $\tilde \phi = i\sigma^2\phi^*$. The complex triplet has been employed, for example, in the Georgi--Machacek model~\cite{Georgi:1985nv,Chanowitz:1985ug} and also in neutrino mass models~\cite{Arhrib:2011uy}.

\item \textbf{Real Abelian vector} $\B\sim(1,1,0)$
\be\label{eq:DarkPhotonLag}
\La_\B = \frac{1}{2}\left(\del_\mu\B_\nu \del^\nu\B^\mu - \del_\mu\B_\nu\del^\mu\B^\nu 
+ M^2\B_\mu\B^\mu\right) + \left(ig_\B \B^\mu \phi^\dagger D_\mu \phi +h.c.\right).
\ee
The addition of a SM-singlet vector is a well motivated extension of the SM, for it surges out of Abelian extensions of the SM gauge group. This class of models has a very rich phenomenology, see for instance Ref.~\cite{Langacker:2008yv} and references therein.

\item \textbf{Complex $SU(2)_L$ singlet vector} $\B_1\sim (1,1,1)$
\begin{align}
\La_{\B_1} = &~ D_\mu\B_{1\nu}^\dagger D^\nu\B^{\mu}_1 - D_\mu\B_{1\nu}^\dagger D^\mu\B^{\nu}_1 
+ M^2\B_{1\mu}^\dagger\B^\mu_1 + \left(ig_{\B_1} \B_1^\mu \phi^\dagger D_\mu \tilde\phi + h.c. \right)\nonumber\\
&\quad - ig'k_{\B_1}\B_1^{\mu\dagger}\B_1^\nu B_{\mu\nu}.
\end{align}
This model has  previously been studied in, for example, Refs.~\cite{Fonseca:2016jbm,deBlas:2019rxi,Chala:2021wpj} in the context of EFTs.

\item \textbf{Real $SU(2)_L$ triplet vector} $\W\sim(1,3,0)$
\al{
\La_\W = &~ \frac{1}{2}\left(D_\mu\W_\nu^a D^\nu\W^{\mu a} - D_\mu\W_\nu^a D^\mu\W^{\nu a} 
+ M^2\W_\mu^a\W^{\mu a}\right) \\
&+ \left(ig_\W \W^{\mu a} \phi^\dagger \sigma^a D_\mu \phi + h.c.\right) - \frac{1}{2}g\,k_{\W}\epsilon^{abc}\W^{\mu a}\W^{\nu b} W_{\mu\nu}^c.
}
This model was previously studied in the context of the LHC in Ref.~\cite{deBlas:2012qp} and in a more general context in Ref.~\cite{Pappadopulo:2014qza}. It was also used to explain the $(g-2)$ anomaly, for instance in Ref.~\cite{Biggio:2016wyy}, and as a source of neutrino-less double beta decay~\cite{Fonseca:2016jbm}.

\item \textbf{Complex $SU(2)_L$ triplet vector} $\W_1\sim (1,3,1)$
\al{
\La_{\W_1} = &~ D_\mu\W_{1\nu}^{a\dagger} D^\nu\W^{\mu a}_1 - D_\mu\W_{1\nu}^{a\dagger} D^\mu\W^{\nu a}_1 
+ M^2\W_{1\mu}^{a\dagger}\W^{\mu a}_1   \\
& +\left(ig_{\W_1} \W^{\mu a}_1 \phi^\dagger \sigma^a D_\mu \tilde \phi+h.c.\right)\\
& - g\,k_{\W_1,1}\epsilon^{abc}\W_1^{\mu a\dagger }\W_1^{\nu b} W_{\mu\nu}^c- ig'k_{\W_1,2}\W_1^{\mu a\dagger }\W_1^{\nu a} B_{\mu\nu}.
}
This model is much less studied in the literature compared to other models, being mentioned in Refs.~\cite{Fonseca:2016jbm,deBlas:2012qp}, for example. More recently, Ref.~\cite{Fonseca:2022wtz} argued that the coupling of $\W_1$ to two Higgs fields cannot emerge from a Yang--Mills theory.

\end{itemize}

\newcommand{\mc}{\multicolumn{1}{c}{}}

{\renewcommand{\arraystretch}{1.5}
\begin{table}[p] 
\begin{center}
\resizebox{\textwidth}{!}{
\begin{tabular}{ ?c?c|c|c?c|c|c|c? } 
 \multicolumn{1}{c}{}&\multicolumn{3}{c?}{\textbf{Scalar extensions}}
&\multicolumn{4}{c}{\textbf{Vector extensions}}\\
 \Cline{1.5pt}{1-8}
                             &$S$ 
    & $\Xi$ 
    & $\Xi_1$ 
    & $\mathcal{B}^\mu$ 
    & $\mathcal{B}^\mu_{1}$ 
    & $\mathcal{W}^\mu$ 
    & $\mathcal{W}^\mu_{1}$ \\[-2mm]
    &$(1,1,0)$ 
     & $(1,3,0)$ 
    & $(1,3,1)$ 
    & $(1,1,0)$ 
    & $(1,1,1)$ 
    & $(1,3,0)$ 
    & $(1,3,1)$
                            \\\Cline{1.5pt}{1-8}
$|\phi|^4\,*$               &0& 0 & 0 & 0 & 0 &0&0\\\Cline{1.5pt}{1-8}
$\O_{\phi\square}$     &$-$1/2& $-$1/2 &   &$-$1/2 && $-$1/8&\\\hline 
$\O_{\phi D}$           & & $-$2 & 4 &$-$2&1&&$-1/4$\\\hline
$\R_{\phi D}^\prime$    & & 2 & 4 & &$-$1&$-$1/2&$-1/4$\\\Cline{1.5pt}{1-8}
$\O_{\phi^4}^{(1)}$    & &4 &   &$-$2&2&1/2&\\\hline
$\O_{\phi^4}^{(2)}$    & &  & 8 &2&&1/2&\\\hline
$\O_{\phi^4}^{(3)}$    &2&$-$2 & & &$-$2&$-$1&  \\\Cline{1.5pt}{1-8}
$\R_{\phi^4}^{(4)}$    &2&$-$2 &  &\mc&\mc&\mc  &\\\cline{1-4}
$\R_{\phi^4}^{(6)}$    & &4 &  &\mc&\mc&\mc  &\\\cline{1-4}
$\R_{\phi^4}^{(8)}$    &1/2&1/2 &  &
            \multicolumn{4}{c?}{\raisebox{-5mm}{\huge 0}} 
            \\[-2.5mm]\cline{1-4}
$\R_{\phi^4}^{(10)}$    & &2 & 4 &\mc&\mc&\mc  &\\\cline{1-4}
$\R_{\phi^4}^{(11)}$    &1&$-$1 &4 &\mc&\mc&\mc  &\\\cline{1-4}
$\R_{\phi^4}^{(12)}$    & &  &8&\mc&\mc&\mc  &\\\Cline{1.5pt}{1-8}
$g\O_{W\phi^4D^2}^{(1)}$    & \mc&\mc &   &$-$2&4&$-1/2-k_\W$&$-k_{\W_1,2}/2$\\
\cline{1-1}\cline{5-8}
$g\R_{W\phi^4D^2}^{(6)}$    &\mc&\mc&  & &1&$-k_\W/4$&\\\cline{1-1}\cline{5-8}
$g\R_{W\phi^4D^2}^{(7)}$    &\mc&\mc&  &2 &$-$1&&$k_{\W_1,2}/4$\\
\cline{1-1}\cline{5-8}
$g'\O_{B\phi^4D^2}^{(1)}$    &\mc&\mc&  &$-$2 &$-2k_{\B_1}$&3/2 &\\
\cline{1-1}\cline{5-8}
$g'\R_{B\phi^4D^2}^{(3)}$    &\multicolumn{3}{c?}{\huge 0}  &  &$-k_{\B_1}/2$&1/2 &$k_{\W_1,1}/8$\\
\Cline{1.5pt}{1-1}\Cline{1.5pt}{5-8}
$g^2\O_{\phi^4W^2}^{(1)}$    &\mc&\mc&   & &3/8&$-1/16-k_\W/8$&$-1/32$\\
\cline{1-1}\cline{5-8}
$g^2\O_{\phi^4W^2}^{(3)}$    &\mc&\mc&  &$-$1/4 &1/8&&$-1/32$\\
\cline{1-1}\cline{5-8}
$g'g\O_{\phi^4WB}^{(1)}$    &\mc&\mc&   &$-$1/2 &$1/2-k_{\B_1}/4$&$1/8-k_\W/8$&$(k_{\W_1,1}-2)/16$\\
\cline{1-1}\cline{5-8}
${g'}^2\O_{\phi^4B^2}^{(1)}$    &\mc&\mc&   &$-$1/4&$-k_{\B_1}/4$&3/16&$(k_{\W_1,1}-1)/16$\\\Cline{1.5pt}{1-8}
\end{tabular}
}
\end{center}
\caption{\label{tab:offShellMatching}
Tree-level matching contributions to the SMEFT in the Green's basis from single-particle extensions of the SM. 
The higher-dimensional operators are defined in~\cite{Grzadkowski:2010es,Carmona:2021xtq,Gherardi:2020det} at dimension six and in~\cite{Murphy:2020rsh,Chala:2021cgt} at dimension eight.
The operators denoted by $\mathcal{R}$ are chosen to be removed by field redefinitions when we transform to the physical basis.
We have suppressed an overall factor of 
$\kappa^2/M^2$ in the results for the scalar extensions, as well as an overall factor of
$g_\mathcal{X}^2$ in the results for the vector extensions. Powers of $1/M$ can be reconstructed by dimensional analysis.
Empty entries are zero and operators that are omitted do not receive matching contributions from any of the considered models. 
\\
* Any contribution to the renormalizable $|\phi|^4$ operator has been negated by a redefinition of 
$\lambda$, which does not affect any other matching condition.
}
\end{table}
}

\section{Details on the scalar triplet model}\label{app:Triplet}

In this appendix, we review the UV model of Section~\ref{sec:triplet_model} in more detail. 

\subsection{Mixings and diagonalization}

We add to the SM particle content a $SU(2)_L$ triplet of zero hypercharge, $\Xi$, described by the Lagrangian
\be
\La_\Xi = \frac{1}{2}D_\mu \Xi^a D^\mu \Xi^a - \frac{1}{2} M^2 \Xi^a\Xi^a-\kappa_\Xi\Xi^a\phi^\dagger \sigma^a \phi,
\ee
with $ M^2$ and $\kappa_\Xi$ dimensionful parameters.
The covariant derivative is explicitly given by
\be
D_\mu\Xi^c = \left(\partial_\mu \Xi^c  - gW_\mu^a \Xi^b \epsilon^{abc}\right),
\ee
where we use the convention $\Xi = \sigma^a\Xi^a$ for the triplet components, that leads to following kinetic term
\al{\label{eq:triplet_kinetic_term}
\frac{1}{2}D_\mu \Xi^a D^\mu \Xi^a = \frac{1}{2}(\partial_\mu\Xi^c)^2 - g\epsilon^{abc} W_\mu^a \Xi^b \partial_\mu \Xi^c + \frac{g^2}{2}\left[(W_\mu^a)^2(\Xi^b)^2 - (W_\mu^a \Xi^a)^2\right].
}

The potential describing the dynamics of the scalars is written as
\be
V(\phi,\Xi) = -\muH^2|\phi|^2+\lambda|\phi|^4 +\kappa_\Xi \Xi^a\phi^\dagger \sigma^a \phi + \frac{1}{2}M^2\Xi^a \Xi^a.
\ee
In general, both scalars can take a vev. We therefore expand the fields as
\be
\phi=\begin{pmatrix} \phi^+ \\ \frac{v+\phi^0}{\sqrt{2}} \end{pmatrix},\quad \Xi = \begin{pmatrix} v_T + \Xi^0 & \sqrt{2}\Xi^+ \\ \sqrt{2}\Xi^- & -(v_T + \Xi^0) \end{pmatrix},
\ee
where $v,v_T$ are the vevs of $\phi$ and $\Xi$, respectively, and we have defined
\be
\Xi^1 = \frac{\Xi^++\Xi^-}{\sqrt{2}},\quad \Xi^2 = \frac{-\Xi^++\Xi^-}{i\sqrt{2}},\quad \Xi^3 = \Xi^0,
\ee
in terms of the charge eigenstates $\Xi^0,\Xi^\pm$. The associated tadpole equations then read
\al{\label{eq:tadpole_eqs}
\frac{\partial V}{\partial \text{Re} \phi^0}\Big|_{\phi,\Xi = 0} & = v\left(-\muH^2  + \lambda v^2 -\kappa_\Xi v_T\right)=0,\\
\frac{\partial V}{\partial \Xi^0 }\Big|_{\phi,\Xi = 0} & = M^2 v_T - \frac{1}{2}\kappa_\Xi v^2 = 0,
}
that are solved by
\be
v^2 = \frac{2\muH^2M^2}{2M^2\lambda-\kappa_\Xi^2},\quad v_T = \frac{\kappa_\Xi v^2}{2M^2} = \frac{\kappa_\Xi\muH^2}{2M^2\lambda-\kappa_\Xi^2}.
\ee

We now proceed to diagonalize the mass matrices. We start with the neutral sector. The relevant term is
\al{
V& \supset \frac{1}{2}(\text{Re}\phi^0~~\Xi^0)\begin{pmatrix} 2\lambda v^2  & -\kappa_\Xi v\\ -\kappa_\Xi v &  M^2\end{pmatrix}\begin{pmatrix} \text{Re}\phi^0\\ \Xi^0 \end{pmatrix},
}
which is diagonalized by
\be\label{eq:neutral_diagonalization}
U=\begin{pmatrix}c_{\theta_0} & -s_{\theta_0} \\ s_{\theta_0} & c_{\theta_0} \end{pmatrix},\quad s_{2{\theta_0}} = -\frac{2\kappa_\Xi v}{m_h^2 - m_{\xi^0}^2},
\ee
where we use the notation $s_\alpha \equiv \sin\alpha$, $c_\alpha = \cos\alpha$. The physical states $h,\xi^0$ are defined by
\be
\begin{pmatrix} \text{Re}\phi^0 \\ \Xi^0 \end{pmatrix} = U \begin{pmatrix} h \\ \xi^0 \end{pmatrix},
\ee
with masses given by
\be\label{eq:triplet_neutral_masses}
m_{h,\xi^0}^2 = \left(\lambda v^2 + \frac{1}{2}M^2\right)\pm \left(\lambda v^2 - \frac{1}{2}M^2\right)\sqrt{1+\frac{4\kappa_\Xi^2v^2}{(2\lambda v^2 - M^2)^2}}.
\ee
Hence, $h$ is associated to the SM-like Higgs, while $\xi^0$ is the novel neutral scalar state.

The charged sector is a bit more complicated, because $\Xi^\pm$ enters in the definition of the Goldstones. To determine the mass matrix, we first need to identify the gauge-fixing Lagrangian:
\al{
\La_\text{gf} \supset -\frac{1}{\xi_W}\left|\partial^\mu W^+_\mu + ig\xi_W\left(v_T\Xi^+ - \frac{v}{2}\phi^+\right)\right|^2 + g^2\xi_W\left|v_T\Xi^+ - \frac{v}{2}\phi^+\right|^2,
}
where $\xi_W$ is the gauge-fixing parameter for $W_\mu^\pm$. The mass matrix for the charged sector then reads
\al{\label{eq:chagerd_mass_matrix}
V \supset ( \phi^+~~ \Xi^+) \begin{pmatrix} -\kappa_\Xi v_T & \kappa_\Xi v\\ \kappa_\Xi v & M^2 \end{pmatrix}\begin{pmatrix} \phi^- \\ \Xi^- \end{pmatrix}+ g^2\xi_W\left|v_T\Xi^+ - \frac{v}{2}\phi^+\right|^2.
}
Notice that only a particular combination, namely $v_T\Xi^+ - \frac{v}{2}\phi^+$, enters in the gauge fixing, meaning that this is the combination that plays the role of the Goldstone boson for the charged vectors. Then, the natural choice for us is to rotate the charged states to give us this specific combination. Thus,
\al{\label{eq:charged_rotation}
\left\{\begin{array}{ll}G^+ & \equiv c_{\theta_+} \phi^+ - s_{\theta_+} \Xi^+\\
\xi^+ &\equiv s_{\theta_+} \phi^++c_{\theta_+} \Xi^+\end{array}\right. ,  \qquad s_{\theta_+} = \frac{-v_T}{\sqrt{v_T^2+(v/2)^2}},\quad c_{\theta_+} = \frac{-v/2}{\sqrt{v_T^2+(v/2)^2}}.
}
The transformation above not only selects appropriately the Goldstone, but also diagonalizes the mass matrix in Eq.~\eqref{eq:chagerd_mass_matrix}. The complete mass matrix becomes
\be
V  \supset g^2\xi_W\left(v_T^2+\frac{v^2}{4}\right)G^-G^+ + m_{\xi^\pm}^2\xi^+\xi^-,
\ee
with the charged mass given by
\be\label{eq:charged_mass}
m_{\xi^\pm}^2 = c_{\theta_+}^2M^2 +2 s_{\theta_+}^2v_T\kappa_\Xi +2 c_{\theta_+}s_{\theta_+} v \kappa_\Xi.
\ee
Through these rotations we can now consistently take the limit $\xi_W\to \infty$ to go to unitary gauge and completely decouple the Goldstone $G^\pm$.

Lastly, from the kinetic term of the triplet in Eq.~\eqref{eq:triplet_kinetic_term} we obtain a contribution to the $W$ boson mass. Its mass written in term of the Lagrangian parameters is
\be\label{eq:mW_triplet_model}
m_W^2 = \frac{g^2v^2}{4}\left[1+4\left(\frac{v_T}{v}\right)^2\right].
\ee
The mass of the $Z$ boson is given by the SM expression.

\subsection{Computation of the \texorpdfstring{$h\rightarrow \gamma Z$}{h -> gamma Z} amplitude}\label{app:loopresults}

We now report the results obtained for the diagrams in Fig.~\ref{fig:1loopDiagrams},
which give BSM contributions to $h\to \gamma Z$.

\begin{table}[t!]\renewcommand{\arraystretch}{1.5}
\centering
\begin{tabular}{c|c} \hline
Particles & Feynman rule (incoming momenta) \\ \hline \hline
$\gamma_\mu \xi^+(p_+)\xi^-(p_-)$ & $-ie(p_- - p_+)_\mu$ \\
$Z_\mu \xi^+(p_+)\xi^-(p_-)$ & $-ie\cot \theta_W (p_- - p_+)_\mu+\order{\kappa_\Xi}$ \\
$h\xi^+\xi^-$ & $g_{h\xi^+\xi^-}=-s_{\theta_0}s_{\theta_+}^2 \kappa_\Xi -2 c_{\theta_0}s_{\theta_+}c_{\theta_+} (\kappa_\Xi +2 v_T \lambda)$ \\
$\gamma_\mu Z_\nu \xi^+\xi^-$ & $2ie^2\cot \theta_W \eta_{\mu\nu}+\order{\kappa_\Xi}$ \\
$\gamma_\lambda(p_\gamma) W_\mu^+(p_+) W_\nu^-(p_-)$ & $-ie\left[\eta_{\mu\nu}(p_+ - p_-)_\lambda + \eta_{\lambda\nu}(p_- - p_\gamma)_\mu + \eta_{\mu\lambda}(p_\gamma - p_+)_\nu\right]$ \\ \hline
$h(p_h)\xi^\pm(p_\xi) W_\mu^\mp$ & $\pm\frac{ig}{2}(2c_{\theta_+}s_{\theta_0} - s_{\theta_+}c_{\theta_0})(p_h-p_\xi)_\mu$ \\
$\xi^\pm W_\mu^\mp Z_\nu$ & $-\frac{iegv s_{\theta_+}}{s_{2W}}$
\end{tabular}
\caption{Feynman rules relevant for the computation of $h\to \gamma Z$, to leading order in $\kappa_\Xi$ and in unitary gauge. 
}\label{tab:Triplet_FR}
\end{table}

We summarise the relevant Feynman rules in Table~\ref{tab:Triplet_FR}. Using them, we can easily evaluate the diagrams in Fig.~\ref{eq:I_scalar} with only charged scalars in the loop. The full result reads
\al{\label{eq:I_scalar_result}
i\Amp_{H,\text{scalar}} & = \frac{ie^2x_h^2\cot \theta_W g_{h\xi^+\xi^-}}{8\pi^2m_Z^2(x_h-x_Z)^2}\times\\
&\quad~\times\Bigg\{\sqrt{1-4x_h}f(x_h)+ x_Zf(x_h)^2+\frac{x_Z}{x_h} -1- \sqrt{1-4x_Z}f(x_Z)- x_Zf(x_Z)^2\Bigg\},
}
where $x_i\equiv m_{\xi^\pm}^2/m_i^2$ and the function $f$ is defined as
\be
f(x) = \log\left[1-\frac{1}{2x}\left(1-\sqrt{1-4x}\right)\right].
\ee
This result, though in a different notation, agrees with was what previously found in the literature for scalar loops~\cite{Gunion:1989we,Hue:2017cph,Degrande:2017naf}.

The result of the loops in Fig.~\ref{eq:I_mixed} is more involved due to the extra mass scale in the loop, resulting in dilogarithms.
Defining
\be\label{eq:diagrams_gZ_2}
i\Amp_{H,1}^{\mu\nu}\equiv \adjustbox{valign=m}{\begin{tikzpicture}
\begin{feynman}
\vertex (Ml) {$h$};
\vertex[right = 1.75cm of Ml] (Mc);
\vertex[right = 1.2cm of Mc] (Mr);
\vertex[above = 1.2 of Mr] (Va);
\vertex[below = 1.2 of Mr] (Vb);
\vertex[right = 1.5 cm of Va] (VZ) {$Z_\mu$};
\vertex[right = 1.5 cm of Vb] (Vg) {$\gamma_\nu$};
\diagram*{
(Ml) -- [scalar, momentum=$p_h$] (Mc),
(Mc) -- [charged scalar, momentum=$k$] (Va) -- [photon, momentum=$p_Z$] (VZ),
(Va) -- [charged boson, momentum=$q_1$] (Vb) -- [photon, momentum'=$p_\gamma$] (Vg),
(Vb) -- [charged boson, momentum=$q_2$] (Mc)
};
\end{feynman}
\end{tikzpicture}},
\quad
i\Amp_{H,2}^{\mu\nu}\equiv\adjustbox{valign=m}{\begin{tikzpicture}
\begin{feynman}
\vertex (Ml) {$h$};
\vertex[right = 1.75cm of Ml] (Mc);
\vertex[right = 1.2cm of Mc] (Mr);
\vertex[above = 1.2 of Mr] (Va);
\vertex[below = 1.2 of Mr] (Vb);
\vertex[right = 1.5 cm of Va] (VZ) {$Z_\mu$};
\vertex[right = 1.5 cm of Vb] (Vg) {$\gamma_\nu$};
\diagram*{
(Ml) -- [scalar, momentum=$p_h$] (Mc),
(Mc) -- [charged boson, momentum=$k$] (Va) -- [photon, momentum=$p_Z$] (VZ),
(Va) -- [charged scalar, momentum=$q_1$] (Vb) -- [photon, momentum'=$p_\gamma$] (Vg),
(Vb) -- [charged scalar, momentum=$q_2$] (Mc)
};
\end{feynman}
\end{tikzpicture}},
\ee
and using the Feynman rules from Table~\ref{tab:Triplet_FR}, we can build the loop integrals:
\al{\label{eq:I1}
i\Amp_{H,1}^{\mu\nu} & = - 2\times\frac{ig}{2}(2c_{\theta_+}s_{\theta_0} - s_{\theta_+}c_{\theta_0})\times \frac{iegv s_{\theta_+}}{s_{2W}} \times (-ie) \times i^3\\
&\times~ \int\frac{\dd^4k}{(2\pi)^4}\ \frac{(p_h+k)_\beta\eta_{\mu\lambda}\left[\eta_{\rho\delta}(-q_2-q_1)_\nu+\eta_{\nu\rho}(q_1+p_\gamma)_\delta+\eta_{\delta\nu}(q_2-p_\gamma)_\rho\right]}{[k^2-m_{\xi^\pm}^2][q_1^2-m_W^2][q_2^2-m_W^2]}\\
&\times ~ \left(-\eta_{\beta\delta} + \frac{q_{2\beta}q_{2\delta}}{m_W^2}\right)\left(-\eta_{\lambda\rho} + \frac{q_{1\rho}q_{1\lambda}}{m_W^2}\right),
}
\al{\label{eq:I2}
i\Amp_{H,2}^{\mu\nu} & = - 2\times\frac{ig}{2}(2c_{\theta_+}s_{\theta_0} - s_{\theta_+}c_{\theta_0})\times \frac{iegv s_{\theta_+}}{s_{2W}} \times (-ie) \times i^3\\
& \times~ \int\frac{\dd^4k}{(2\pi)^4}\ \frac{(q_2-p_h)_\alpha \eta_{\mu\beta}(q_1+q_2)_\nu}{[k^2-m_W^2][q_1^2-m_{\xi^\pm}^2][q_2^2-m_{\xi^\pm}^2]}\left(-\eta_{\alpha\beta}+\frac{k_\alpha k_\beta}{m_W^2}\right).
}
The factor of 2 in the expressions above come from considering the same diagrams with reversed arrows.
Then, $\Amp_{H,\text{mixed}}$ is given by
\be\label{eq:I_mixed_result}
\Amp_{H,\text{mixed}}=
(\Amp_{H,1}^{\mu\nu}+\Amp_{H,2}^{\mu\nu})\big|_{p^\nu_Zp_\gamma^\mu~\text{piece}},
\ee
where 
we select only the coefficient proportional to $p^\nu_Zp_\gamma^\mu$. 
As explained in Refs.~\cite{Fontes:2014xva,Hue:2017cph}, the Ward identity and on-shell conditions enforce an overall Lorentz structure 
${p_Z^\nu p_\gamma^\mu} - {(p_Z\cdot p_\gamma)\eta^{\mu\nu}}$. This allows the reconstruction of the full integral from only the Lorentz structure $p_Z^\nu p_\gamma^\mu$. This is also the reason why we omitted other diagrams, which only contribute with terms proportional to $\eta^{\mu\nu}$, due to the particular Lorentz structure of the interactions.
The expression obtained from Eq.~\eqref{eq:I_mixed_result} agrees with previous results~\cite{Degrande:2017naf,Hue:2017cph}.

\bibliographystyle{JHEP2}
\bibliography{main}  

\end{document}